\documentclass[acmsmall]{acmart}

\usepackage[english]{babel}
\usepackage{blindtext}
\usepackage{xurl}
\usepackage{tabularx}
\usepackage{enumitem}
\usepackage{hyperref}
\usepackage{url}

\usepackage{caption}
\usepackage{tikz}
\usepackage{amsmath}

\usepackage{subcaption}
\usepackage[export]{adjustbox}

\newcommand{\parab}[1]{\vspace{0.05in}\noindent\textbf{#1}}
\newcommand{\parait}[1]{\vspace{0.05in}\noindent\textit{#1}}

\newcommand{\ashwin}[1]{\textcolor{black}{#1}}

\usepackage{algorithm}
\usepackage{algorithmicx}
\usepackage{algpseudocode}

\acmDOI{}

\acmISBN{}

\begin{document}
\title[Xaminer]{Xaminer: An Internet Cross-Layer Resilience Analysis Tool}

\renewcommand\footnotetextcopyrightpermission[1]{}
\settopmatter{printacmref=false, printccs=false, printfolios=false}

\author{Alagappan Ramanathan}
\orcid{0009-0003-3293-1790}
\affiliation{%
   \institution{University of California, Irvine}
   \country{USA}
}
\email{alagappr@uci.edu}

\author{Rishika Sankaran}
\orcid{0009-0009-3497-2821}
\affiliation{%
   \institution{University of California, Irvine}
   \country{USA}
}
\email{rsankar1@uci.edu}

\author{Sangeetha Abdu Jyothi}
\orcid{0009-0000-0503-4478}
\affiliation{%
   \institution{University of California, Irvine \& VMware Research}
   \country{USA}
}
\email{sangeetha.aj@uci.edu}

\begin{abstract}

A resilient Internet infrastructure is critical in our highly interconnected society. However, the Internet faces several vulnerabilities, ranging from natural disasters to human activities, that can impact the physical layer and, in turn, the higher network layers, such as IP links. In this paper, we introduce Xaminer, the first Internet cross-layer resilience analysis tool, to evaluate the interplay between physical- and network-layer failures. Using a cross-layer Internet map and a failure event model, Xaminer generates a risk profile encompassing a cross-layer impact report, critical infrastructure identification at each layer, and the discovery of trends and patterns under different failure event settings. Xaminer's key strengths lie in its adaptability to diverse disaster scenarios, the ability to assess risks at various granularities, and the capability to generate joint risk profiles for multiple events. We demonstrate Xaminer's capabilities in cross-layer analysis across a spectrum of disaster event models and regions, showcasing its potential role in facilitating well-informed decision-making for resilience planning and deployments.

\end{abstract}

\maketitle

\section{Introduction}

In today's highly interconnected world, a resilient Internet infrastructure is indispensable for global communication, commerce, and daily life. The Internet has complex dependencies across its physical and software components, each playing a vital role in maintaining seamless connectivity. However, this robust system is susceptible to various vulnerabilities and failures.

Natural disasters on land and in the ocean can impact physical infrastructure components such as cables and landing stations. Moreover, human activities, such as marine operations, also pose threats to the Internet infrastructure. Physical infrastructure failure is typically amplified at the network layer due to the common practice of multiple network operators sharing a single cable. While redundancy mechanisms exist, the limited backbone cable infrastructure and extensive infrastructure sharing can amplify the effect on the network layer. For instance, a cable cut in Egypt with the AAE-1 cable resulted in multiple countries and organizations across the globe facing outages and increased latencies~\cite{multiple_providers_aae1_outage}. 

Cross-layer analysis, i.e., analyzing the interplay between physical cable damages and the impact on the higher network layers, is crucial for improving Internet resilience. It allows us to quantify the consequences of physical-layer failures, identify cross-layer failure patterns, and gain insights that are invaluable for bolstering resilience, informed decision-making, and optimized resource allocation.

Prior analysis frameworks~\cite{durairajan_flood, mayer_earthquake, anderson_wildfire,honk_kong, hurricane_1, hurricane_2}, while valuable, focused on a specific Internet infrastructure layer, yielding only a partial view. Broadly, these frameworks can be categorized into two groups. The first category assessed the impact on physical infrastructure, typically resulting from localized disasters~\cite{durairajan_flood, mayer_earthquake, anderson_wildfire}. However, these analyses often overlooked the extensive repercussions of physical infrastructure damage. For instance, the Tohoku earthquake in Japan caused increased latencies at networks in Hong Kong~\cite{honk_kong}, but a framework characterizing the impact of earthquakes on the physical infrastructure in Japan will not be able to identify this impact at the network layer. In contrast, the second category explored network layer impacts using Border Gateway Protocol (BGP) data~\cite{honk_kong, hurricane_1, hurricane_2}, but they were confined to specific regions and past events and lacked predictive capabilities. Due to complex BGP policies implemented by network operators, it is difficult to extend such analysis of past events to make future predictions. Moreover, none of the prior frameworks efficiently quantified the combined impact of multiple disasters and failures while distinguishing each event's individual contribution.

In response to these challenges, in this paper, we present Xaminer, an Internet cross-layer resilience analysis tool. Specifically, Xaminer uses a cross-layer map and a failure event model to first identify the set of vulnerable cable segments in the physical layer and then employs an analysis pipeline to identify the cross-layer impact and extract failure patterns and trends. Xaminer's entire pipeline is designed with modularity, extensibility, and plug-and-play integration as key goals, enabling the use of a wide range of failure models and cross-layer maps.

Resilience analysis in Xaminer is implemented as a multi-stage workflow. First, Xaminer constructs an intermediate representation of the cross-layer map to capture cross-layer effects at various granularities, spanning the physical and network layers. Cable segments, which represent the finest granularity for a failure at the physical layer, serve as the building block to generate the intermediate representation. Next, using the failure event models, Xaminer extracts all cable segments at risk, thus identifying the physical layer impact. Finally, Xaminer's analysis module combines the cross-layer intermediate representation with the list of impacted cable segments to generate a cross-layer risk profile. This includes the cross-layer impact analysis, critical infrastructure identification at each layer, and the discovery of trends and patterns under different failure event settings.

One of Xaminer's unique features is its ability to perform analysis at varied granularities. This functionality allows users to get the cross-layer risk profiles across multiple geographical resolutions, from global to regional to individual cable segments, cables, or landing stations. Another key feature of Xaminer is its ability to create joint risk profiles for multiple events and disasters. Whether it is earthquakes, hurricanes, or other disruptive events, by integrating the assessments of various threats into its joint risk profile, Xaminer evaluates the combined risk of multiple events on both the physical and network layers. These capabilities empower decision-makers to zoom in on specific regions or zoom out for a broader perspective across a single or collection of failure events, thus providing a comprehensive understanding of cross-layer infrastructure dynamics. Hence, these features could prove invaluable for organizations seeking to fortify their Internet infrastructure against a multitude of potential challenges.

Using the cross-layer map of submarine links generated by Nautilus~\cite{nautilus}, we conduct extensive experiments with a broad range of disaster event models, failure probabilities, and regions to demonstrate Xaminer's capabilities. We make a few key observations. First, each disaster type exhibits distinct characteristics for cross-layer impacts globally. The fraction of infrastructure impacted at the physical and IP layers can be highly skewed. For instance, approximately 60\% of submarine cable segments are vulnerable to earthquakes globally. However, since the distribution of IP links in the cross-layer map is highly skewed, only 30\% of total IP links are vulnerable. On the other hand, solar storms predominantly impact the network layer. Although only 15\% of cable segments are vulnerable, these segments host about 60\% of IP links in the dataset. These results indicate that there is a clear imbalance in the impact at the physical and network layers. 

Second, surprisingly, many landlocked countries face a high connectivity risk under disasters affecting submarine cables. For example, Chad, a land-locked African nation, experiences high vulnerability due to its dependence on a small set of neighboring countries for submarine connectivity. Third, we measure connectivity pattern similarities and observe that countries in close proximity and within larger regional blocs tend to share similar network connectivity distribution patterns, regardless of their underlying cable infrastructure. This observation suggests that economic cooperation, political agreements, and shared infrastructure deployments significantly influence a country's network connectivity.

In summary, we make the following contributions:
\vspace{1mm}
\begin{itemize}[leftmargin=*,nolistsep]
    \item We develop Xaminer, an Internet cross-layer resilience analysis tool.
    \item Xaminer is modular and extensible. It supports plug-and-play integration for a broad range of cross-layer maps and failure models.
    \item Xaminer offers high flexibility in cross-layer resilience analysis. It supports varied geographical granularities, a wide range of probabilistic and deterministic failure models, and joint risk profiling across multiple threats.
    \item We demonstrate the capabilities of Xaminer through analysis using a real-world submarine cross-layer map, Nautilus~\cite{nautilus}, with nearly a million links and four popular disaster models~\cite{ungshap,hurricanes_ocha_hdx,elevation_model,solar_storm}.  
\end{itemize}

\section{Background and Related Work}

In this section, we provide an overview of prior research on Internet resilience analysis and cross-layer mapping.

\parab{Internet Resilience.} Previous works on Internet resilience focused on multiple facets including the development of resilient protocols~\cite{resilient_protocols_1, resilient_protocols_2, resilient_protocols_3, resilient_protocols_4, resilient_protocols_5, resilient_protocols_6}, analysis of Internet reachability~\cite{Internet_reachability_1, Internet_reachability_2}, evaluation of network survivability~\cite{network_survivability_1, network_survivability_2, graph_vulnerability_1, graph_vulnerability_2, graph_vulnerability_3, graph_vulnerability_4, graph_vulnerability_5}, analysis of Internet resilience~\cite{resilience_analysis_1, resilience_analysis_2, resilience_analysis_3, resilience_analysis_4, durairajan_flood} and the analysis and detection of network outages~\cite{detecting_outages_1, detecting_outages_2, detecting_outages_3, detecting_outages_4, detecting_outages_5, detecting_outages_6, detecting_outages_7, earthquake_1, hurricane_1}. Xaminer contributes specifically to the analysis of Internet resilience and the identification of outage trends. 

While outages can happen for many reasons, the intensity and frequency of natural disasters that cause Internet outages are increasing due to climate change. This has prompted significant advancements in evaluating the impact of natural disasters on the Internet. For example, a recent study employed forecast models to identify potential failure hotspots for submarine cables (physical layer) due to various climate change factors~\cite{clare_submarine_hotspot}. Previous works on analyzing the impact of failures on the Internet can be broadly categorized into two: those that assess the impact of specific past events primarily by analyzing BGP data (e.g., the Tohoku earthquake~\cite{earthquake_1, earthquake_2, earthquake_4} and Hurricane Sandy~\cite{hurricane_1, hurricane_2}) and those that predict the physical layer impact of specific types of disasters using various failure models (e.g., wildfires~\cite{anderson_wildfire}, floods~\cite{durairajan_flood}, and earthquakes~\cite{mayer_earthquake}). However, the limitation of these prior works is their focus on specific geographic regions or layers of the Internet stack. In contrast, our tool, Xaminer, allows for the use of disaster forecast models to assess Internet resilience across both the physical and network layers, covering geographical regions at varied granularities.

\parab{Cross-Layer Mapping.} \ashwin{To understand failure trends across physical and higher layers of the Internet, a cross-layer map is essential. A cross-layer map provides a mapping between the IP links and the physical cables.} Only a few works, namely iGDB~\cite{igdb} and Nautilus~\cite{nautilus} have presented solutions in the field of cross-layer mapping. iGDB generates a map between terrestrial cable infrastructure and IP links using Thessian polygon segments and IP geolocation, while Nautilus utilizes IP geolocation and AS-Cable owner relationships to map IP links and submarine cables.

\parab{Nautilus.} In our study, we use the cross-layer map generated by Nautilus to gain a comprehensive understanding of the cascading impact of natural disasters on the IP and AS layers. \ashwin{Nautilus leverages RIPE~\cite{ripe_atlas} and CAIDA~\cite{CAIDA} traceroute measurements to extract IP links} and categorize each IP link into three broad categories: definitely submarine, potential submarine, and definitely terrestrial. These categories reflect the confidence level in their association with a submarine cable. \ashwin{To generate the mapping, Nautilus employs a two-pronged approach. Nautilus generates an initial mapping between IP links and cables by using IP geolocation (11 sources), Speed-of-Light validation, and the location of the submarine cable endpoints. Using AS information (IP to AS mapping from 4 sources) and submarine cable owner information, Nautilus further refines its mapping. Finally, Nautilus aggregates the results to generate a comprehensive mapping of IP links to submarine cables, along with a confidence score per mapping.}

\ashwin{In our analysis, we specifically focus on IP links classified as definitely submarine, considering only the top cable prediction when multiple cable predictions exist. We select Nautilus over iGDB due to its superior coverage (9$\times$ better) and fewer cable predictions per IP link ($\approx$ 75\% fewer). Moreover, Nautilus validates its mapping efficacy through three experiments: (i) analysis of past failure events, (ii) targeted traceroute measurements, and (iii) comparison with network operator maps. The results of these experiments reveal that links mapped by Nautilus were present before and after failure events but disappeared during the events. With targeted measurements, only 4\% of the links did not have a corresponding match in Nautilus' cable prediction. Moreover, for 77\% of the links, Nautilus accurately predicts the expected cable as its top prediction, and for an additional 19\% of the links, Nautilus provided a secondary (non-top) cable match. Furthermore, there is a significant overlap between Nautilus cable predictions and network operator maps.}

\section{Design}

\begin{figure*}
  \centering
  \includegraphics[width=\textwidth]{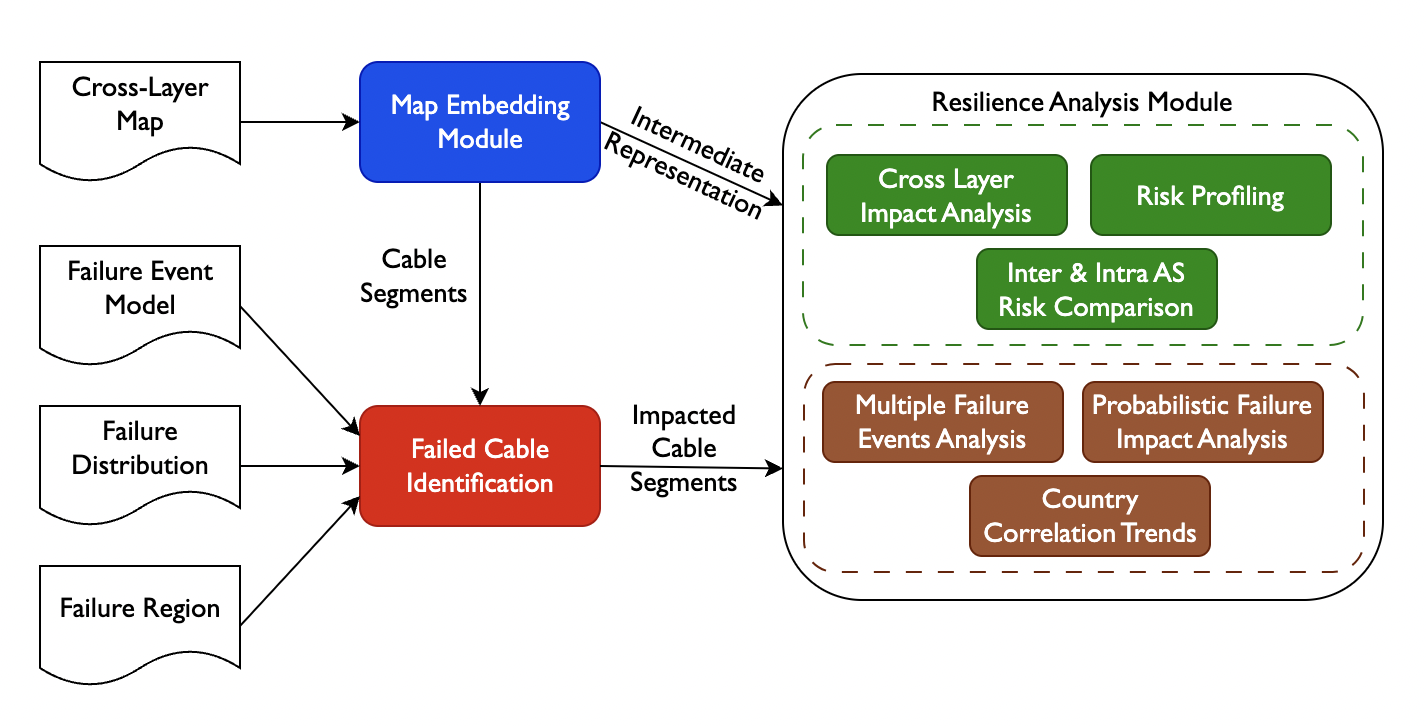}
  \caption{Xaminer System Architecture}
  \label{system_diagram}
\end{figure*}

In this section, we present the design of Xaminer, an Internet cross-layer resilience analysis tool (Figure~\ref{system_diagram}). The high-level workflow of Xaminer is as follows:

\begin{itemize}[leftmargin=*,nolistsep]
    \item First, using the map embedding module, Xaminer generates concise intermediate representations to capture the cross-layer dependencies.
    \item Second, the failed cable identification module uses the failure event information to identify the impacted cable segments in the physical layer.
    \item Finally, the resilience analysis module uses the intermediate representation and the failed cable segments to perform an extensive analysis which includes identifying the cross-layer impact and profiling the risk at various geographical scales.
\end{itemize}

\subsection{Map Embedding Module} \label{map_embedding_module}

To efficiently conduct cross-layer resilience analysis at different granularities, a concise intermediate representation is essential. Without a suitable representation, analyses at varying granularities would require distinct, multi-step pipelines. For example, identifying cable segments connected to an AS would entail extracting AS associated with the IP endpoints, locating IP links tied to these endpoints, and mapping the corresponding cable segments. Hence to avoid complex pipelines and generate a concise representation, Xaminer relies on three sources (detailed in \S~\ref{cross_layer_map_dataset}): (i) a cross-layer map with IP links to cable segment mapping, (ii) an IP to geolocation map, and (iii) an IP to AS Number (ASN) map. 

First, we introduce some key terminologies: (i) Cable Segments (CS), identified as tuples (cable, landing station 1, landing station 2), are used as the building blocks to capture the impact of physical layer failures, (ii) N-Country (NC) represents all IP links (ie., network layer aggregation) with an endpoint within the given country's borders, (iii) P-Country refers to the set of all cable segments with a landing point in the specified country.

Xaminer generates two intermediate maps that are used for the rest of the analysis: (i) the \textit{Cable Segment-N-Country (CS-NC) map} and (ii) the \textit{Cable Segment-AS (CS-AS) map}. \ashwin{The CS-NC map encompasses all IP links that use the specified cable segment and terminates in the given country. Note that the endpoint of an IP link might be farther inland from the corresponding cable endpoint, sometimes even in a different country. Hence, to understand the impact on the IP layer, it is imperative to know the geolocation of IP endpoints traversing a cable segment. To generate the CS-NC map, Xaminer first extracts the country (N-Country) information from IP geolocation sources for each IP link. Next, from the cross-layer map, Xaminer identifies the cable segments associated with the IP link. Finally, Xaminer aggregates all IP links associated with each cable segment for every country.} In short, for a given cable segment and country pair, this map contains all IP links terminating in the given country that use the specified cable segment. \ashwin{The CS-AS map captures all IP links that use the specified cable segment and have an endpoint in the given AS. This map allows us to analyze the impact of physical layer failures at an organization (AS) level. To construct the CS-AS map, for each IP link, Xaminer collects the Autonomous System Number (ASN) information from IP to ASN sources. Similar to the CS-NC map generation, Xaminer leverages the cross-layer map to finally aggregate all IP links associated with each unique combination of cable segments with ASN.} To sum up, for a given cable segment and AS pair, this map contains all IP links with an endpoint in the given AS that uses the specified cable segment.

\ashwin{Encoded as key-value pairs, these intermediate maps (CS-NC and CS-AS) concisely capture cross-layer infrastructure information using cable segments as the building blocks (keys). This concise representation facilitates the analysis of complex failure scenarios involving multiple cable segment failures as a composite of individual cable segment failures. For instance, assessing the impact of failures of all physical cables within a country (P-Country) involves combining the impacts of individual cable segments with a landing point in that country. The use of a simple key-value data structure allows fast value matching for the same keys and enables a seamless combination of results from both maps. We find this to be more efficient than a complex graph data structure. }

\subsection{Failed Cable Identification} \label{failed_submarine_cable_identification}

To identify the cross-layer impact of a failure event, it is critical to first estimate the list of cable segments at the risk of failure. These cable segments represent the key building blocks for the intermediate maps to analyze the cross-layer impact. Hence to identify the cable segments at risk, Xaminer relies on three user-provided inputs: (i) the failure event model, (ii) the failure distribution, and (iii) the failure region. 

The failure event model includes: (i) location-to-impact intensity\ashwin{, stored as a key-value data structure,} which gives the maximum intensity for the disaster that can be potentially experienced at each location \ashwin{(key)} represented as a (latitude, longitude) tuple. (ii) impact thresholds, which represent the limits above/below which the disaster affects the physical infrastructure, and (iii) spatial probing radius, which determines the distance for extrapolating data from the location-to-impact intensity map. The location-to-impact intensity map is sampled at specific spatial resolutions. To determine the impact experienced at a specific location, we employ the spatial probing radius. For the given geolocation, we generate a region around it using the spatial probing radius, identify disaster intensity values for any available points within the region, and assign the specified location's intensity as the maximum value within the spatial probing radius. For example, for a spatial resolution of 0.1$^\circ$ ($\approx$ 100 km$^2$) in the location-to-impact intensity map, a 10 km spatial probing radius helps to extrapolate to all intermediate points.

To address partial failure scenarios (for example, a 5\% failure across locations in Japan), Xaminer relies on the failure distribution input, consisting of (i) the probability of failure, indicating the fraction of locations considered as impacted, and (ii) the sampling strategy, such as random, top-n, or weighted. The sampling strategy defines the mechanism for choosing the fraction of impacted locations. Lastly, the failure region outlines the geographic scope of the analysis. Detailed information on disaster-based event models is given in \S~\ref{disaster_models_dataset}. 

\begin{figure*}
  \centering
  \includegraphics[width=0.9\textwidth]{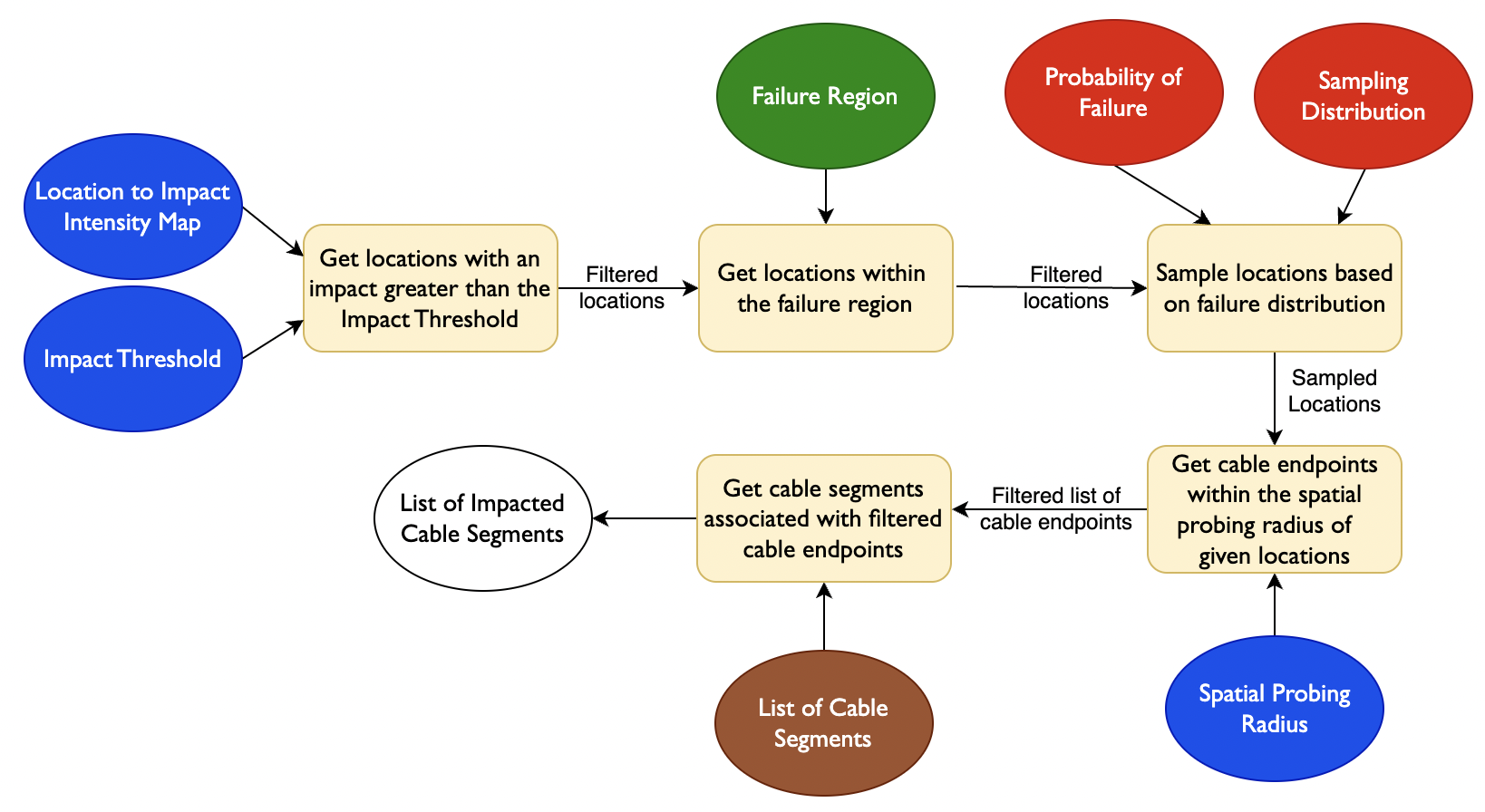}
  \caption{Workflow for identifying the failed cable segments. Blocks in blue represent the inputs from failure event models, red represents the inputs from failure distribution, and yellow represents the processing pipeline.}
  \label{failed_cable_segment_fig}
\end{figure*}

Next, we detail the high-level workflow with an example. A workflow representation of this pipeline is presented in Figure~\ref{failed_cable_segment_fig}. Suppose we want to identify the list of cable segments at risk in Japan due to earthquakes and have configured Xaminer as follows: (i) failure event model as the earthquake model with an impact threshold of 6 PGA (Peak Ground Acceleration)~\cite{pga} and a spatial probing radius of 10 km, (ii) failure distribution as a 5\% failure probability with a weighted sampling strategy prioritizing high-impact locations, and (iii) failure region as Japan. 

Under these settings, the Xaminer pipeline first extracts all locations from the earthquake \ashwin{location-to-impact intensity} model where the impact intensity exceeds PGA value of 6. Then Xaminer filters these pairs to include only those within the geographic scope of Japan, as specified by the failure region. Next, the sampling strategy comes into play, where Xaminer employs the chosen failure probability to sample 5\% of the pairs from the filtered (latitude, longitude) pairs, with a higher probability of selecting pairs with high impact intensity due to the weighted sampling strategy. With these sample pairs, Xaminer identifies the cable landing stations located within a 10 km radius of these pairs. Finally, relying on the list of all cable segments, Xaminer compiles a list of cable segments that are associated with the at-risk landing stations. These segments are the ones that are bound to fail during earthquake events meeting the specified criteria.

\subsection{Resilience Analysis Module}

Using the impacted cable segments from the Failed Cable Identification module and the intermediate representations from the Map Embedding module, the resilience analysis module provides capabilities to perform a wide array of analyses under multiple settings. This module is designed to be extensible to allow users to define custom analysis methods whilst taking advantage of the in-built capabilities. First, we describe the analysis methods and then describe additional capabilities provided by Xaminer.

\subsubsection{Analysis Methods} \label{resilience_analysis_methods}\hfill

\parab{Cross-Layer Impact Analysis.} We evaluate event impact across six key components: cable segments, cables, IP links, IPs, AS links, and ASes. This analysis is crucial for identifying vulnerable infrastructure components and gaining cross-layer insights into cable infrastructure, IP links, and AS connections. To assess cross-layer impact, Xaminer retrieves associated IP links from the CS-NC map using the list of impacted cable segments. Subsequently, using the IP to ASN map, all the corresponding AS links for the IP links are identified. From the IP and AS links information, Xaminer identifies the unique IP and AS endpoints and determines the associated cables from the cable segments. Finally, the impact on each infrastructure component---cable segments, cables, IP links, IPs, AS links, and ASes---is normalized with respect to their respective totals. 

\parab{Risk Profiling.} With risk analysis under a failure event, we quantify the normalized risk profile of a country/AS, i.e., the fraction of IP links associated with the country or AS at the risk of failure. This helps identify countries and ASes facing the most significant impact under the given failure event. \ashwin{Note that this analysis only reveals the affected links and not the fraction of traffic affected since the underlying datasets do not include information on traffic.} Xaminer combines the impacted cable segments and the CS-NC and CS-AS maps to estimate the number of affected IP links associated with each country and AS, respectively. These values are subsequently normalized using the total number of IP links for each country or AS.

\parab{Intra- and Inter-AS Risk Comparison.} In this analysis, we assess the changes in the number of intra- and inter-AS connections per country under the failure event. This analysis, in conjunction with the N-Country risk profile, reveals dependency patterns between a country's risk profile and its intra- and inter-AS connectivity alterations. To identify the risk for all countries under a failure event, Xaminer retrieves all AS links associated with the impacted cable segments in the country. Subsequently, Xaminer calculates the total count of intra- and inter-AS links for each country, providing insights into connectivity shifts following a failure event. Note that an AS link is considered intra-AS when the ASN on both ends of the link is the same.

\subsubsection{Additional Capabilities} \label{additional_capabilities}

To enhance user capabilities for exploring the combined impact of multiple failure events and the effects of varying failure probabilities and sampling methods, Xaminer has two modules:

\parab{Multiple Failure Events Analysis.} While assessing the impact of individual events is crucial, achieving long-term resilience goals often entails understanding the combined effects of multiple events to create a joint risk profile. Xaminer enables this by initially evaluating individual event models to identify affected cable segments. These segments are then aggregated across all events. For example, consider the analysis of two distinct events---earthquakes and hurricanes---in Japan, where the earthquake event identifies impacted segments as \{segment-1, segment-2\}, while the hurricane event identifies \{segment-2, segment-3\}. Aggregating these impacts results in the overall impacted segments, \{segment-1, segment-2, segment-3\}. As Xaminer's analyses are built upon cable segments as fundamental building blocks, this compilation of impacted segments from multiple events facilitates all Xaminer analyses.

\parab{Probabilistic Failure Impact Analysis.} Understanding the impact of failure events under diverse failure models is essential for identifying key vulnerabilities and optimizing resource allocation for infrastructure deployments. Xaminer offers capabilities to explore cross-layer event impacts with user-defined metrics, including impact thresholds, failure probabilities, and sampling strategies. Xaminer systematically analyzes and aggregates these combinations, providing a visual representation. For example, a network operator aiming to enhance network resilience in earthquake-prone regions can leverage Xaminer to assess the impact of various thresholds of earthquake intensity, facilitating informed decisions on infrastructure improvements. 

\parab{Country Correlation Trends.} Assessing and improving the Internet infrastructure resilience requires an understanding of distribution patterns and key trends. Xaminer enables the examination of correlation trends between countries. In particular, we examine how cables terminating in a country influence the connectivity patterns of other countries using it for IP-layer connectivity. Xaminer achieves this by using the CS-NC map to aggregate cable segments with segment endpoints in each country (P-Country). This aggregation produces the PC-NC (P-Country-N-Country) map, which is stored as a key-value data structure. A PC-NC map entry for a country pair $(C_1, C_2)$ corresponds to the set of all IP links with at least one IP endpoint in $C_2$ and using a cable segment with at least one cable endpoint in $C_1$. Using this PC-NC map, Xaminer employs hierarchical clustering with Ward's distance~\cite{ward_algor} to group the correlation distributions of these countries.

\section{Datasets}

Xaminer is a cross-layer resilience analysis tool that relies primarily on (i) a cross-layer map and (ii) disaster models to derive insights into failure patterns. In this section, we present a brief overview of the datasets and their characteristics. It is worth noting that all data sources used by Xaminer can be replaced with equivalent variants, enabling quick integration of the evolving landscape of information in cross-layer mapping and disaster modeling.

\subsection{Cross-Layer Map} \label{cross_layer_map_dataset}

Here, we describe the data sources that form the inputs for the Map Embedding module (\S~\ref{map_embedding_module}).

\parab{Nautilus} Nautilus~\cite{nautilus} is the first cross-layer mapping framework that maps IP links to submarine cables. This framework assigns an IP link to (i) one of the three broad categories: definitely submarine, potential submarine, and definitely terrestrial, and (ii) a list of potential cable segments, each with an associated prediction score. We demonstrate the capabilities of Xaminer using the Nautilus dataset as our input cross-layer map, using the top cable segment prediction for the definitely submarine links, unless otherwise specified. \ashwin{For the definitely submarine category, Nautilus encompasses data for 1.13 million IP links, offering extensive coverage across Asia, Europe, the Americas, and Oceania. However, its coverage is more limited for certain regions in Africa, Central Asia, and South America. Figure~\ref{fig:countries_colored} shows the coverage for all countries.}
 
\parait{Geolocation:} Geolocation is a key parameter that the Map Embedding module (\S~\ref{map_embedding_module}) uses to aggregate the cross-layer statistics at a country level. We use the top geolocation results from Nautilus to maintain consistency with the obtained cable segment predictions. Nautilus incorporates geolocation information from 11 different sources~\cite{ripe_ipmap, maxmind, ip2location, ipinfo, db-ip, ipregistry, ipgeolocation, ipapi, ipapi_co, ipdata, caida_itdk} and applies Speed-of-Light (SoL) validation to filter out any inaccurate inferences. The final geolocation output, aggregated using clustering, has better accuracy compared to any single geolocation source~\cite{nautilus}.

\parait{IP to ASN mapping:} The ASN (Autonomous System Number) information is used to aggregate the cross-layer information at an AS level. Similar to geolocation, any independent IP to ASN mapping could be plugged into Xaminer, but we rely on the top ASN prediction for an IP from Nautilus. Nautilus uses data from four IP to ASN mapping sources~\cite{radb_server, routinator_whois, cymru_whois, caida_as_to_org} and aggregates the results using the maximum agreement (voting) principle.

While all the sources described here use Nautilus and its associated results to maintain consistency, Xaminer supports using alternatives for all these sources.

\subsection{Disaster Models} \label{disaster_models_dataset}

While Xaminer can accept any event model as input, we illustrate its capabilities using four disaster models to identify failed cable segments (\S~\ref{failed_submarine_cable_identification}). For most disasters, we utilize data from various sources and aggregate them at a spatial resolution of 0.1-degree latitude and longitude, roughly equivalent to a grid of 100 km$^2$. The thresholds and spatial probing radius depend on the intensity scales and grid sizes. It's important to note that the parameters chosen for these experiments may not precisely reflect actual intensity. However, Xaminer offers the flexibility to adjust any parameter and rerun analyses as needed. 

\parab{Earthquakes.} To model earthquakes, Xaminer relies on the data provided by the UN-GSHAP program~\cite{ungshap}, to map each grid to a PGA value with a 10\% chance of exceedance in the next 50 years. Xaminer converts the PGA (Peak Ground Acceleration) values to the MMI (Modified Mercalli Intensity) scale using Worden et. al~\cite{worden}. As structural damages occur beyond an MMI value of VI (6), Xaminer sets the threshold to VI (6) and the spatial probing radius as 10 km.

\parab{Hurricanes.} Xaminer relies on the UN-OCHA-HDX~\cite{hurricanes_ocha_hdx} hurricane dataset, to model hurricanes. Xaminer then maps each grid to its highest recorded wind speed (in knots). Since infrastructure damages are observed at wind speeds greater than 64 knots~\cite{saffir_simpson} and the incomplete information for some grids, Xaminer sets the threshold to 64 knots and the spatial probing radius to an extended 50 km. 

\parab{Sea Level Rise.} To model sea level rise, Xaminer relies on the global elevation data from the ETOPO model by NOAA~\cite{elevation_model} and assigns each grid with its corresponding elevation. Xaminer sets the spatial probing radius to 10 km and the default threshold to a 1-meter rise in sea level.

\parab{Solar Storms.} Solar storms, a space weather event tend to affect networking infrastructures within specific latitude thresholds. Hence to model solar storm impact, Xaminer maps all submarine landing stations to their corresponding latitudes. For analysis, Xaminer employs a spatial probing radius of 1 km and sets the default threshold to 50$^\circ$ latitude.
\section{Demonstration of Xaminer Capabilities}

In this section, we present the capabilities of the Xaminer tool, which combines the cross-layer map with failure event models to generate insights about cross-layer failures and identify critical components and trends. In addition to enabling cross-layer analysis,  Xaminer ensures compatibility with a broad range of failure models with varying configurations. To demonstrate these capabilities, we present the following experiments using Xaminer:
\begin{itemize} [leftmargin=*,nolistsep]
    \item An analysis of the cross-layer impact of events with a regional scope.
    \item An assessment of the cross-layer effects of large-scale events with a global scope.
    \item Failure analysis with probabilistic failure models and multiple disasters.
    \item An event-independent cross-layer analysis to identify global patterns and trends.
    \item A comparison of Xaminer results on singular failure events with real-world data.
\end{itemize}
These capabilities are essential for analyses which are critical for improving disaster preparedness, planning responses, assessing the resilience of infrastructure, allocating resources more effectively, implementing better risk mitigation measures, and planning and deploying infrastructure.

\subsection{Regional Impact Assessment}

Identifying the impact of various disasters over a region is essential for assessing infrastructure resilience. To comprehensively assess regional infrastructure resilience, we leverage the cross-layer impact analysis feature detailed in our resilience analysis module (\S~\ref{resilience_analysis_methods}). This feature provides a quantitative measure of overall infrastructure risk across different layers, offering insights into the potential impact of specific events on the region's infrastructure. In our experiments, we explore three diverse regions with distinct disaster models: (i) Earthquakes in Japan, (ii) Hurricanes in the Caribbean, and (iii) Earthquakes in the Pacific Northwest (PNW). The graphical representation of the impact of these disasters on various layers is illustrated in Figure~\ref{fig:regional_maximal_impact}. \ashwin{Details on additional regions can be found in Appendix~\ref{region_maximal_impact_appendix}.}

\begin{figure}[!ht]
  \begin{minipage}[b]{0.48\linewidth}
    \centering
    \includegraphics[width=\columnwidth]{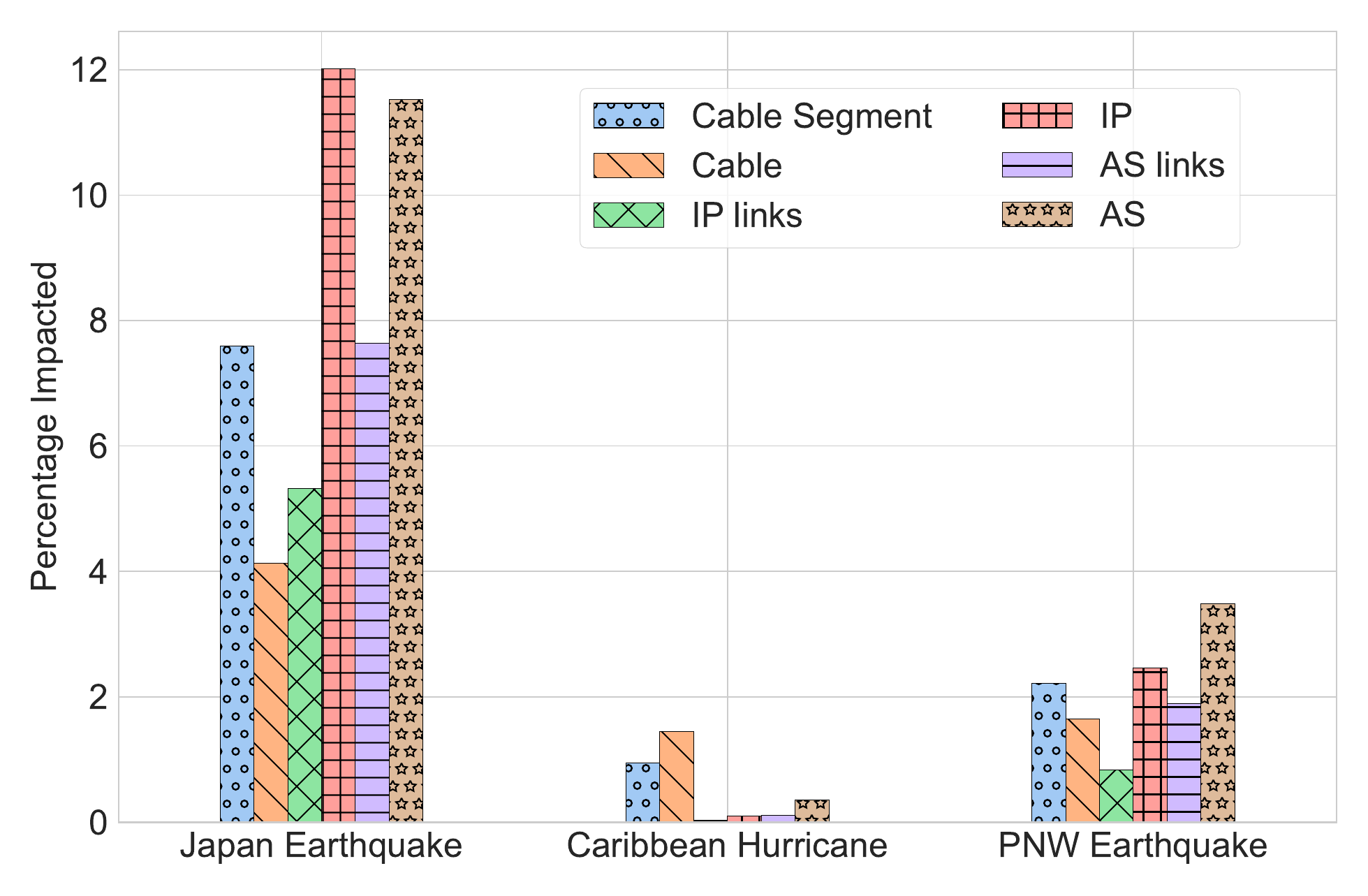}
    \caption{The maximum percentage of infrastructure at risk for various layers due to various disasters at a regional level.}
    \label{fig:regional_maximal_impact}
  \end{minipage}%
  \hfill
  \begin{minipage}[b]{0.48\linewidth}
    \centering
    \includegraphics[width=\columnwidth]{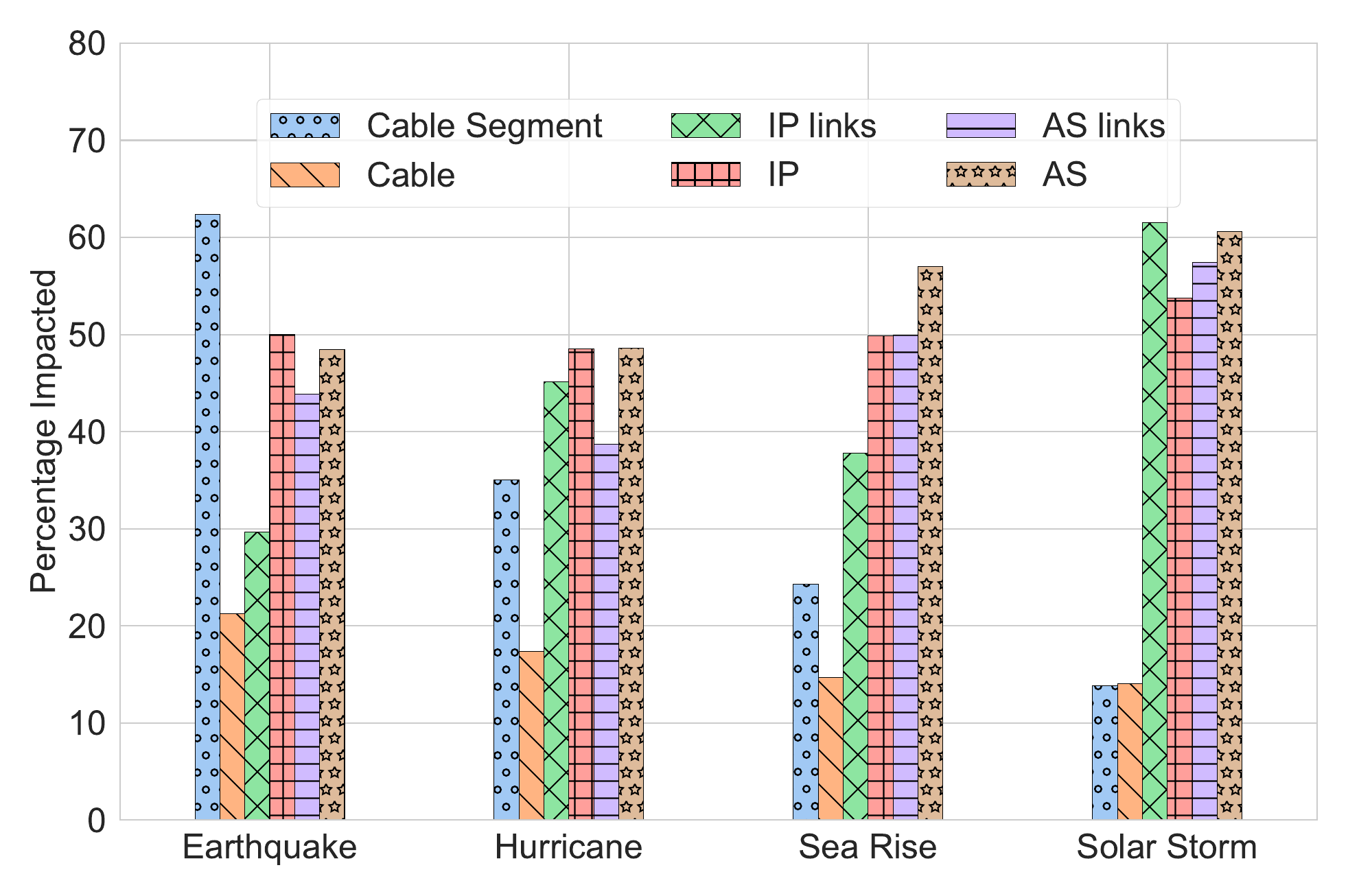}
    \caption{The maximum percentage of infrastructure at risk for various layers due to various disasters globally.}
    \label{fig:disaster_maximal_impact}
  \end{minipage}
\end{figure}

\parait{Insights:} Our analysis, facilitated by the cross-layer map, reveals that each region exhibits unique patterns in response to a particular disaster type. Notably, in Japan, a region with a limited number of cables, the network layer components bear a higher risk compared to the Caribbean, where the physical layer components face maximal susceptibility.

\parait{Capabilities:} This experimental configuration, encompassing regions of diverse compositions, underscores Xaminer's versatility in addressing scenarios that range from individual countries (e.g., Japan) to regional clusters of countries (e.g., the Caribbean) and sub-regions within a country (e.g., the Pacific Northwest).

\subsection{Global Impact Assessment}

To assess the impact of various large-scale events like natural disasters globally, we (i) evaluate the maximum infrastructure at risk across layers, (ii) gauge a country's risk profile, and (iii) assess the potential infrastructure at higher risk.

\subsubsection{Assessing Maximum Infrastructure Risk} ~\label{assessing_max_risk}

Given an event model, understanding the maximum risk profile globally is important and to achieve this, we leverage the cross-layer impact analysis functionality within the resilience analysis module (Section~\ref{resilience_analysis_methods}). This feature quantifies the overall infrastructure risk across different layers, estimating the percentage of infrastructure potentially affected by a specific event.

In our experiment, we adapt the failure event model inputs using the disaster models described in Section~\ref{disaster_models_dataset}. We maintain the default settings for failure distribution (100\% probability of failure) and failure regions (global scope). The resulting impact on various layers of these disasters is depicted in Figure~\ref{fig:disaster_maximal_impact}.

\parait{Insights and Use Cases:} With the given cross-layer map, an imbalance is evident between physical and network layer components. Interestingly, solar storms, with a 50$^\circ$ threshold and a 1 km spatial probing radius, pose the highest network layer risk, despite affecting fewer submarine cables. In contrast, earthquakes, set with a 6 PGA threshold and a 10 km spatial probing radius, present the greatest risk to the submarine infrastructure. These findings hold significance for targeted infrastructure investments. For example, given the higher physical layer risk observed during earthquakes, future investments may focus on enhancing the earthquake resilience of physical cables and landing stations.

\parait{Capability:} This experimental setup, featuring a range of event models with specific thresholds and search radii, showcases Xaminer's versatility in accommodating various failure event models.

\subsubsection{Gauging Country's Risk Profile}

\begin{figure}[ht]
    \centering
    \begin{subfigure}{\textwidth}
        \centering
        \includegraphics[width=0.8\textwidth]{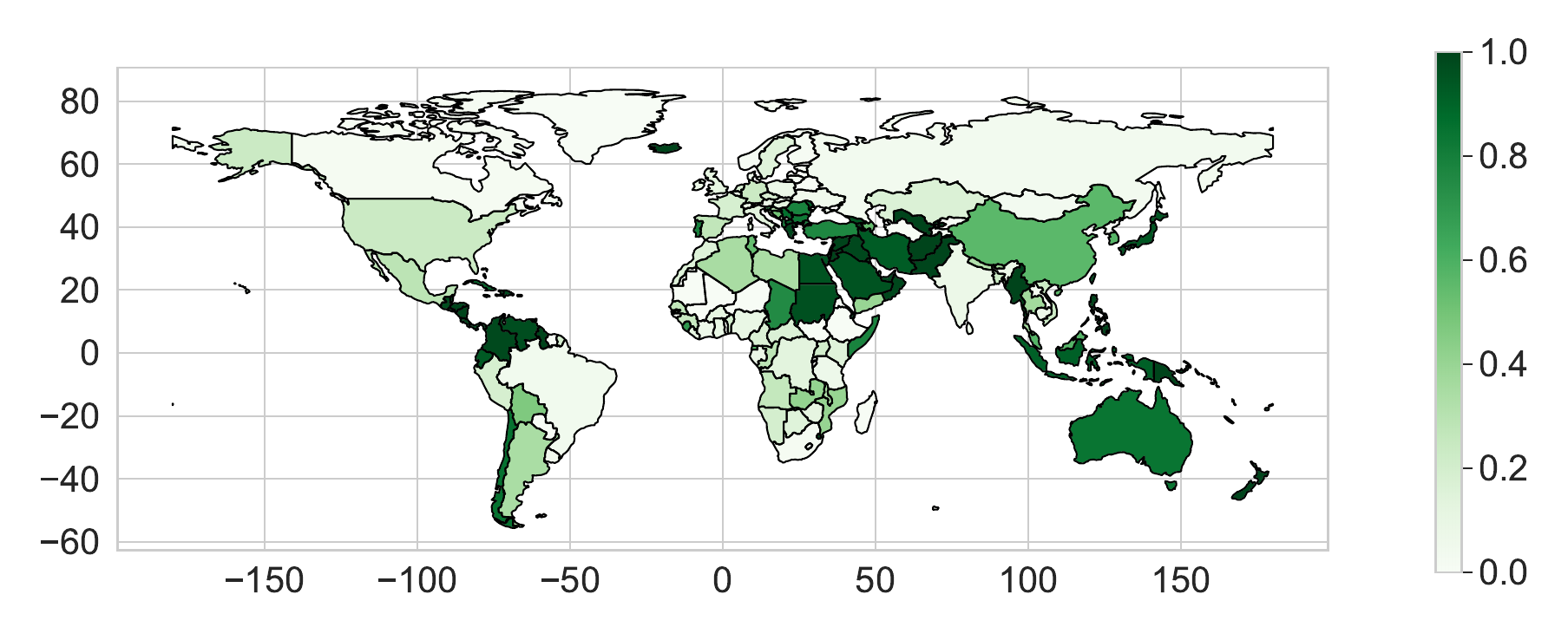}
        \caption{The risk profile for countries due to earthquakes (with a threshold of 6 PGA or higher)}
        \label{fig:earthquake_countries_impact}
    \end{subfigure}

    \begin{subfigure}{\textwidth}
        \centering
        \includegraphics[width=0.8\textwidth]{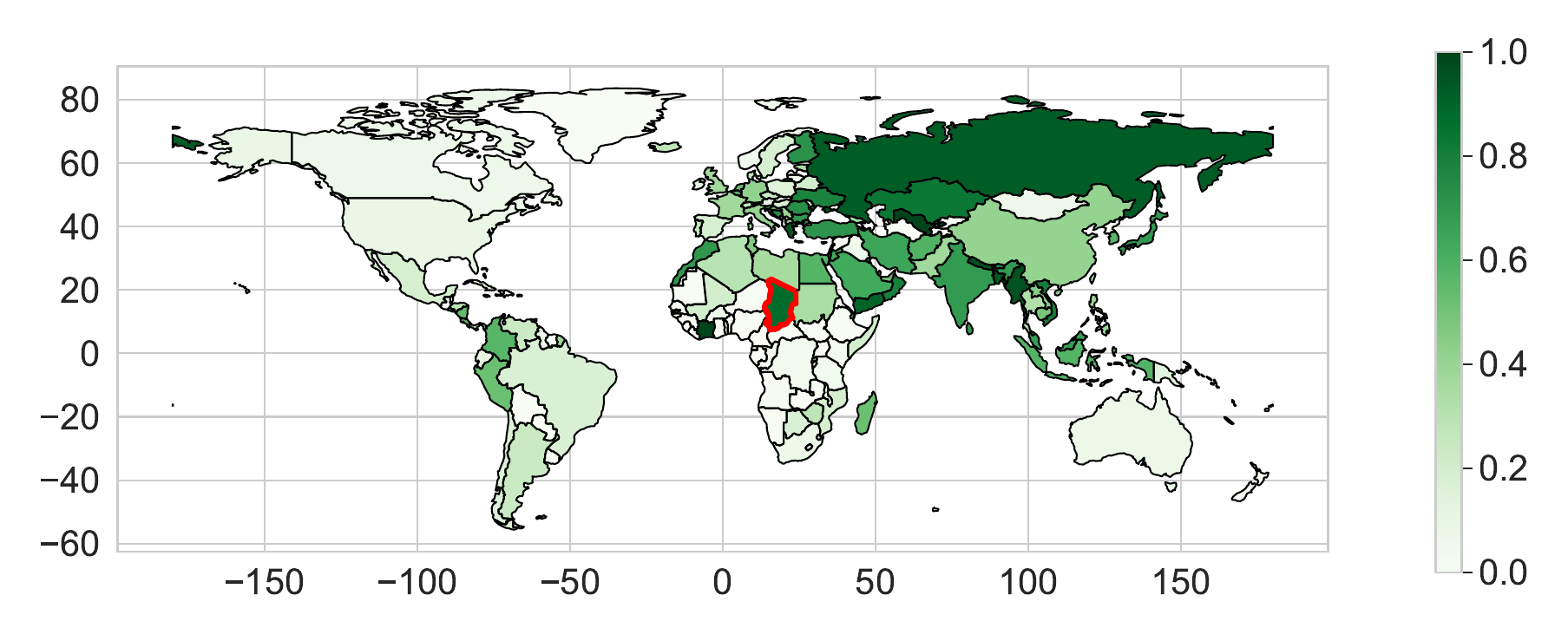}
        \caption{The risk profile for countries due to sea level rise (with a threshold of 1 meter or lower)}
        \label{fig:searise_countries_impact}
    \end{subfigure}
    \label{fig:countries_impact_maximal}
    \caption{The risk profiles for countries due to various disaster settings. The results for each country are normalized based on the number of IPs within the country.}
\end{figure}

While our previous analysis provided insights at a global level, it is also imperative to understand risk profiles at the country level. To achieve this, we:

\textit{Assess Country Risk Profiles:} We determine the normalized impact of a specific failure event on a country, focusing on the fraction of IPs within that country linked to submarine cable infrastructure impacted by the event. This assessment is made possible through the Risk Profiling functionality in the resilience analysis module (Section~\ref{resilience_analysis_methods}). The results for earthquakes and sea rise can be found in Figures~\ref{fig:earthquake_countries_impact} and \ref{fig:searise_countries_impact}. \ashwin{For hurricanes and solar storms results, refer to Appendix~\ref{gauging_risk_profile_appendix}.}

\begin{figure}[ht]
    \centering
    \begin{subfigure}{\textwidth}
        \centering
        \includegraphics[width=0.8\textwidth]{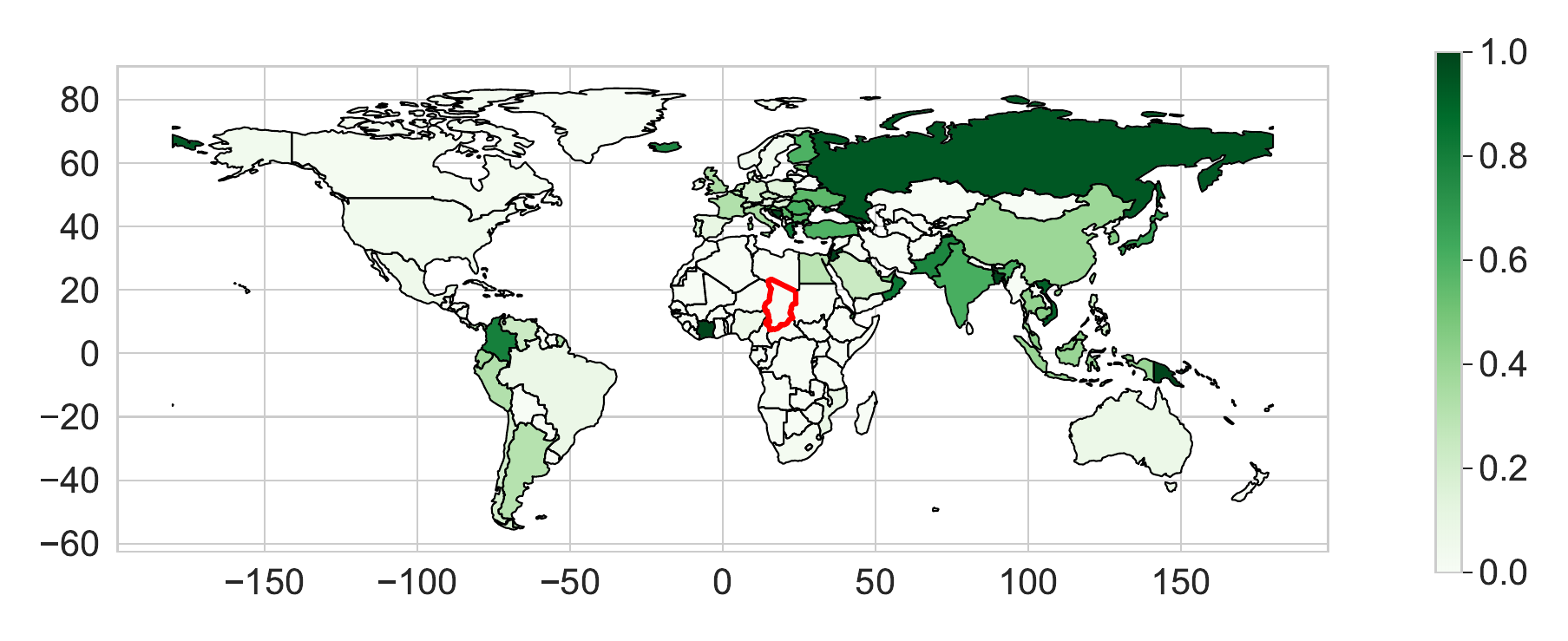}
        \caption{The normalized Intra-AS impact for countries due to sea rise (with a threshold of 1 meter or lower)}
        \label{fig:searise_intra_impact}
    \end{subfigure}

    \begin{subfigure}{\textwidth}
        \centering
        \includegraphics[width=0.8\textwidth]{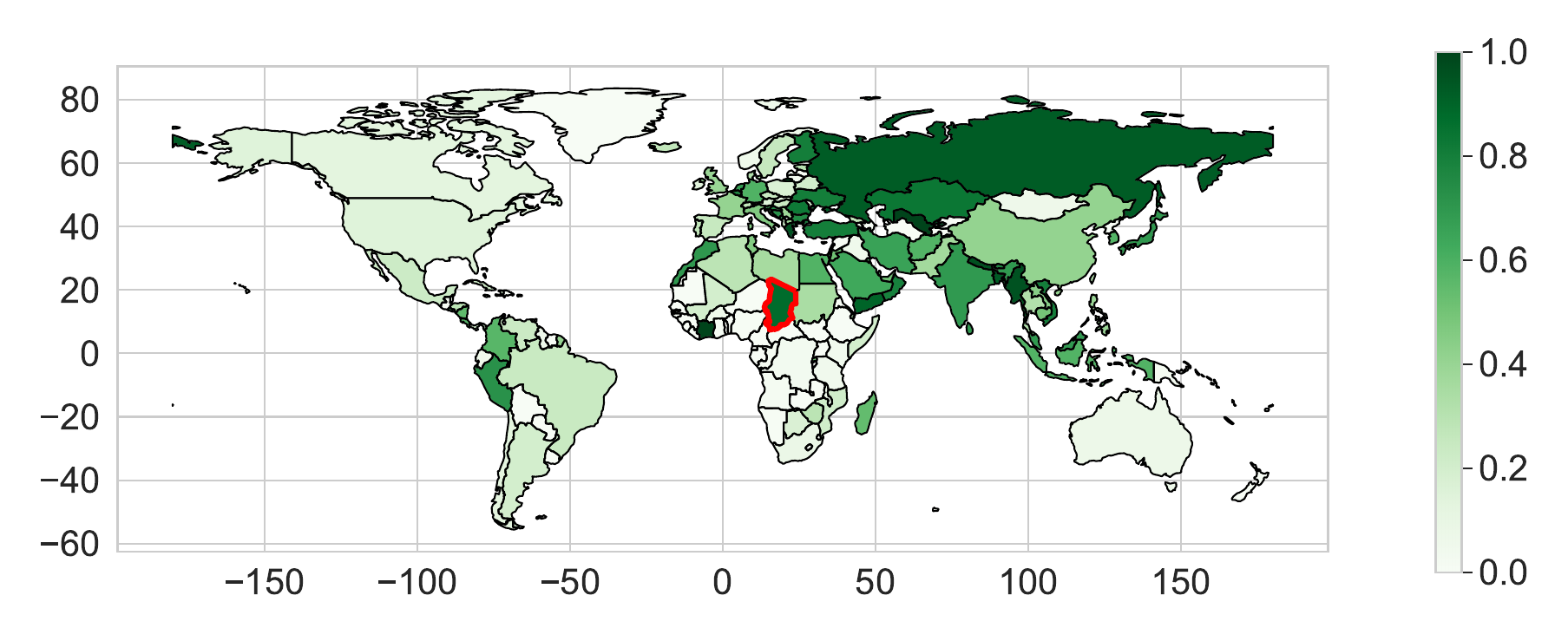}
        \caption{The normalized Inter-AS impact for countries due to sea rise (with a threshold of 1 meter or lower)}
        \label{fig:searise_inter_impact}
    \end{subfigure}
    \label{fig:searise_impact_intra_inter_new}
    \caption{The normalized interconnectivity impacts due to sea level rise. The results for each country are normalized based on the number of intra-AS and inter-AS links observed respectively.}
\end{figure}

\parait{Analyze Interconnectivity Patterns:} To make informed decisions about network layer infrastructure, it's crucial to understand interconnectivity patterns. Therefore, we compute normalized counts for intra-AS and inter-AS links for each country using the Intra and inter-AS risk comparison functionality in the resilience analysis module (Section~\ref{resilience_analysis_methods}). The results for sea rise's intra and inter-AS impact are depicted in Figures~\ref{fig:searise_intra_impact} and \ref{fig:searise_inter_impact} respectively. \ashwin{Plots for the other disasters can be found in the Appendix~\ref{gauging_risk_profile_appendix}.}

\parait{Insights:} While Figures~\ref{fig:earthquake_countries_impact} and \ref{fig:searise_countries_impact} provide a country's risk profile, it's important to consider that a country's overall risk can be elevated due to its reliance on neighboring countries for network layer components through submarine cable infrastructure. A clear example of this is observed in Chad's vulnerability to sea level rise. Despite being land-locked, Chad's (highlighted in red in Figure~\ref{fig:searise_countries_impact}) risk factor is significantly higher because it depends on a limited number of landing points in neighboring nations for submarine cable connectivity.

Additionally, varying levels of intra and inter-AS impact are observed across different countries. Notably, countries relying on a neighboring nation's submarine infrastructure tend to have a higher prevalence of inter-AS connections. This suggests that the absence of domestic submarine cable infrastructure leads to increased dependence on inter-AS links for submarine connectivity. Chad, highlighted in red in Figure~\ref{fig:searise_intra_impact} and \ref{fig:searise_inter_impact}, serves as a prime example, with only its inter-AS links being affected. A similar phenomenon is seen in land-locked countries in multiple regions across the globe.

\subsubsection{Examining Impact of Thresholds for Disasters}

\begin{figure}[ht]
    \centering
    \begin{subfigure}{0.48\textwidth}
        \centering
        \includegraphics[width=\textwidth]{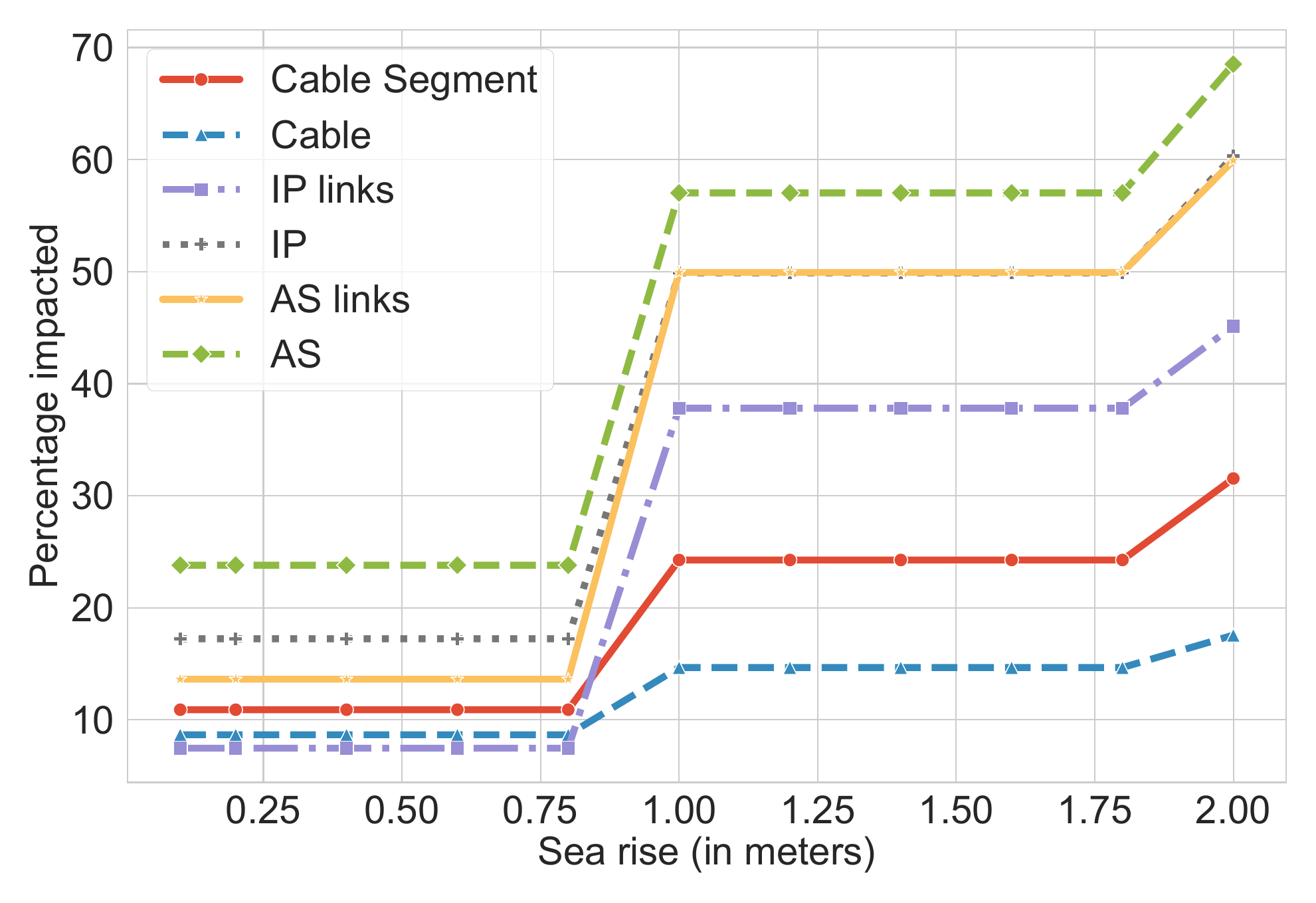}
        \caption{The maximum percentage of infrastructure at risk at various layers due to sea level rise at different heights}
        \label{fig:searise_range_impact}
    \end{subfigure}
    \hfill
    \begin{subfigure}{0.47\textwidth}
        \centering
        \includegraphics[width=\textwidth]{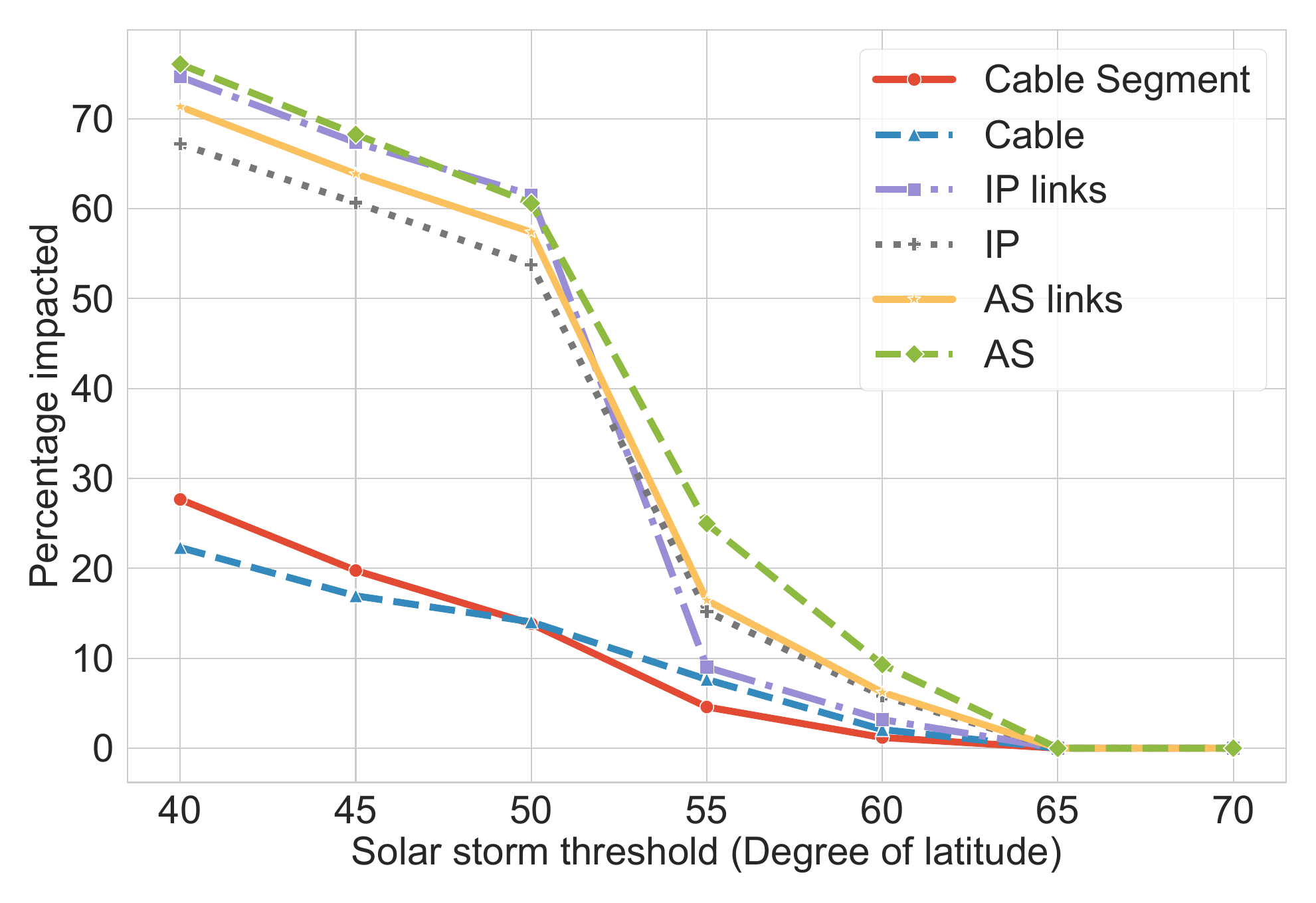}
        \caption{The maximum percentage of infrastructure at risk at various layers due to solar storms at different latitude thresholds}
        \label{fig:solarstorm_range_impact}
    \end{subfigure}
    \label{fig:range_impact}
    \caption{The maximum percentage of infrastructure at risk at various layers due to various disasters at multiple ranges.}
\end{figure}

While earthquakes and hurricanes tend to be regional disasters affecting specific areas at a time, sea-level rise and solar storms have the capacity to impact multiple regions simultaneously. Therefore, assessing their effects on various layers across multiple ranges is essential. Using the probabilistic failure impact capability (\S~\ref{additional_capabilities}) of Xaminer, we investigate the impact at various network components at different thresholds of impact. You can find the results of this experiment for sea-level rise and solar storms in Figures~\ref{fig:searise_range_impact} and \ref{fig:solarstorm_range_impact}, respectively.

\parait{Insights and Use Cases:} This analysis presents several interesting observations. For specific sea-level rise thresholds, the most substantial changes in percentage impact across all layers occurred between a rise of 0.8 to 1 meter in sea level, and the impact remains constant till 0.8 meter in sea level rise. Similarly, with solar storms, the most significant impact changes occurred between latitudes 50 and 55$^{\circ}$, with zero impact observed above 65$^{\circ}$ latitude. These specific ranges, 0.8 to 1 meter for sea-level rise and 50 to 55$^{\circ}$ latitudes for solar storms, represent critical transition zones between low and high impact ranges.

\parait{Capability:} This experimental setup, featuring a fixed event model but a range of thresholds, showcases Xaminer's versatility in accommodating various thresholds for any given event model and visualizing the results.

\subsection{Analysis with Multiple Disasters and Probability Distributions}

While understanding the effects of individual events at a regional or global scale has its advantages, for better resilience analysis, it is also essential to identify the risk profile by (i) exploring regional failure trends under varying probabilities and (ii) with multiple disaster events.

\subsubsection{Exploring Regional Trends with Varying Probabilities}

\begin{figure}
    \centering
    \includegraphics[width=\columnwidth]{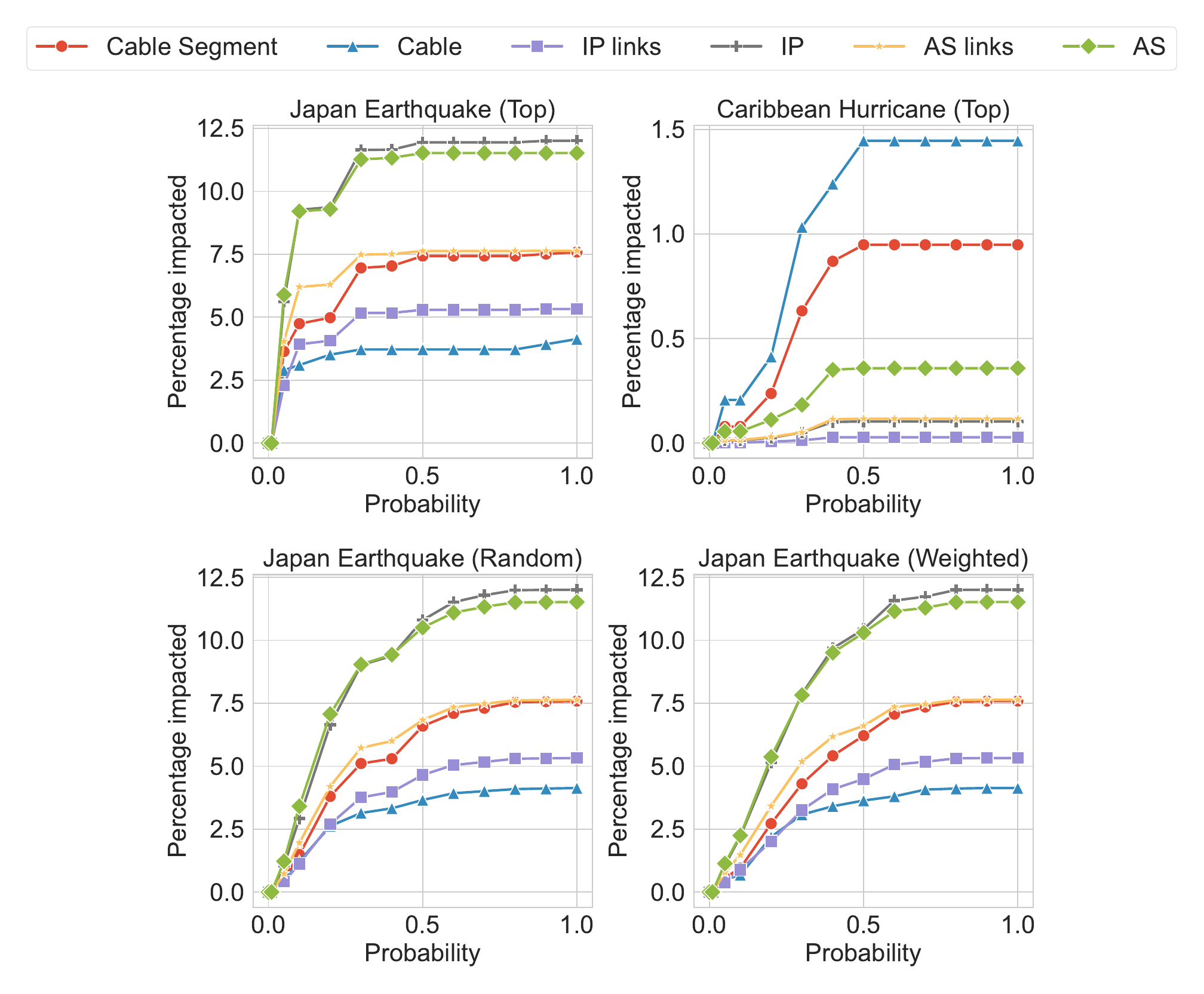}
    \caption{The percentage of infrastructure at risk at various layers in specific regions at various probabilities of failure. The title for each plot represents the region-disaster type with the sampling strategy used.}
    \label{trends_probabilities_regions}
\end{figure}

Understanding the localized impact of failure events with varying probabilities is crucial and to uncover these trends, Xaminer leverages its cross-layer impact analysis feature (\S~\ref{resilience_analysis_methods}) and probabilistic failure impact capability (\S~\ref{additional_capabilities}). This enables us to analyze how different components in both the physical and network layers respond to varying failure distributions within specific regions. Our experiments include four regions for the earthquake model and four for the hurricane model, utilizing three sampling strategies: random, top-n, and weighted. We examine probabilities of failure ranging from 1\% to 100\%. For sampling strategies (random and weighted), we conduct ten runs and average the results. The outcomes for two regions, including Japan, which features all three sampling distributions, and the Caribbean with the top sampling distribution are presented in Figure~\ref{trends_probabilities_regions}. \ashwin{Outcomes for other sampling distributions for the Caribbean and other regions can be found in Appendix~\ref{varying_probabilities_appendix}.}

\parait{Insights and Use Cases:} The results, as evident in the case of Japan and other regions (Figure~\ref{trends_probabilities_regions}), reveal that one component (IPs for Japan and cables for the Caribbean) consistently bears a higher impact compared to others, even at low probabilities of failure, regardless of the sampling method. This can contrast with global patterns observed for the same disaster type (as discussed in~\ref{assessing_max_risk}), highlighting unique regional trends influenced by both disaster type and location. Interestingly, we noticed that using the top-n sampling distribution, around a 50\% probability of failure, resulted in over 80\% of maximum impact for the components across all layers.

\parait{Capabilities:} This experimental setup, exploring diverse failure probabilities across different sampling methods and regions, underscores Xaminer's ability to assess cross-layer impacts at various regional granularities. Notably, we find Japan, a single country, the British Isles and the Caribbean as groups of countries, and the Pacific Northwest comprising states from two countries showcases Xaminer's capacity to assess cross-layer impact at different regional levels, allowing for detailed, customized analysis.

\begin{figure}[!ht]
    \centering
    \begin{subfigure}{0.35\textwidth}
        \centering
        \includegraphics[width=\textwidth]{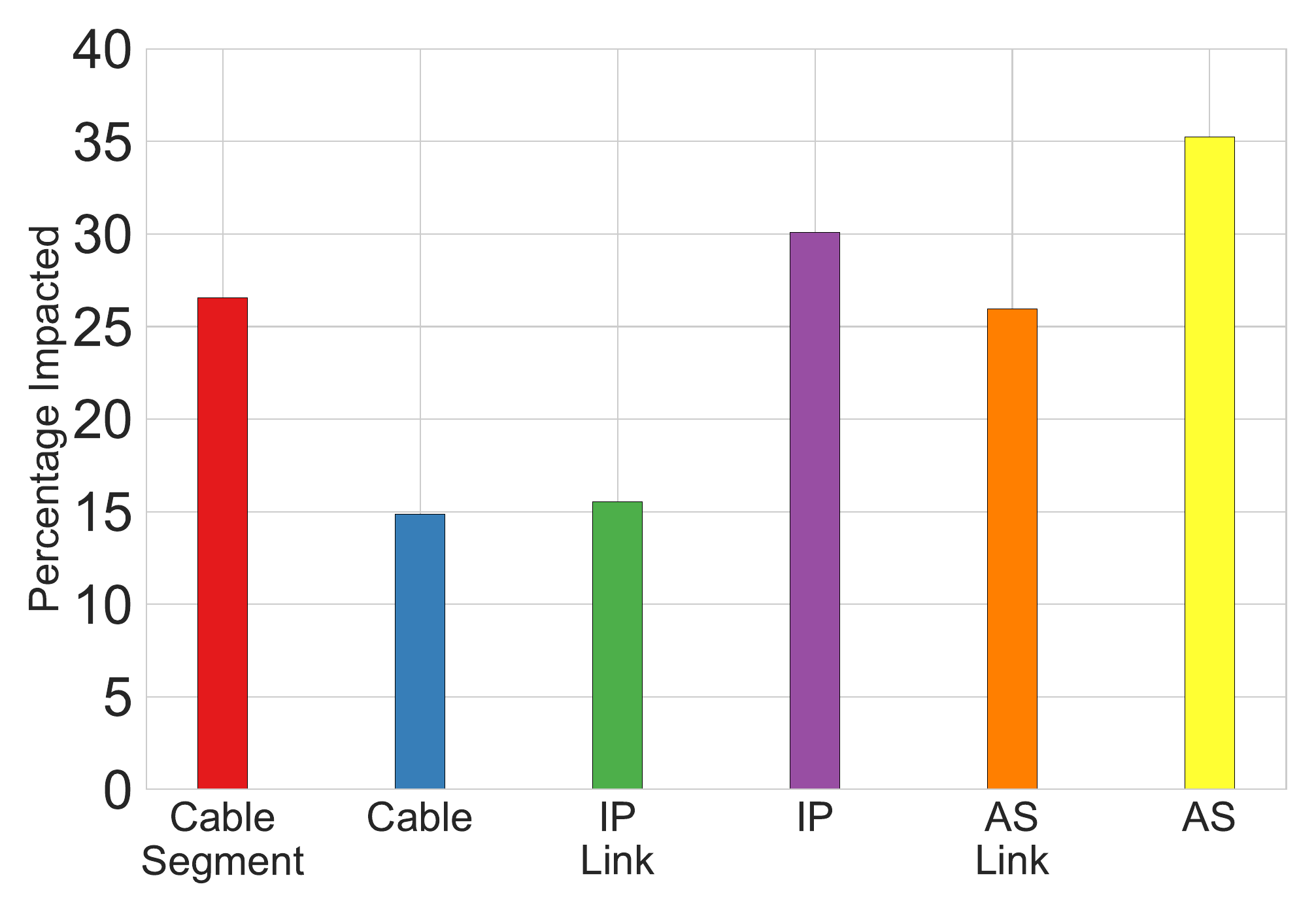}
        \caption{The maximum percentage of infrastructure at risk at various layers due to multiple disasters.}
        \label{fig:disaster_reasonable_impact}
    \end{subfigure}
    \hfill
    \begin{subfigure}{0.63\textwidth}
        \centering
        \includegraphics[width=\textwidth]{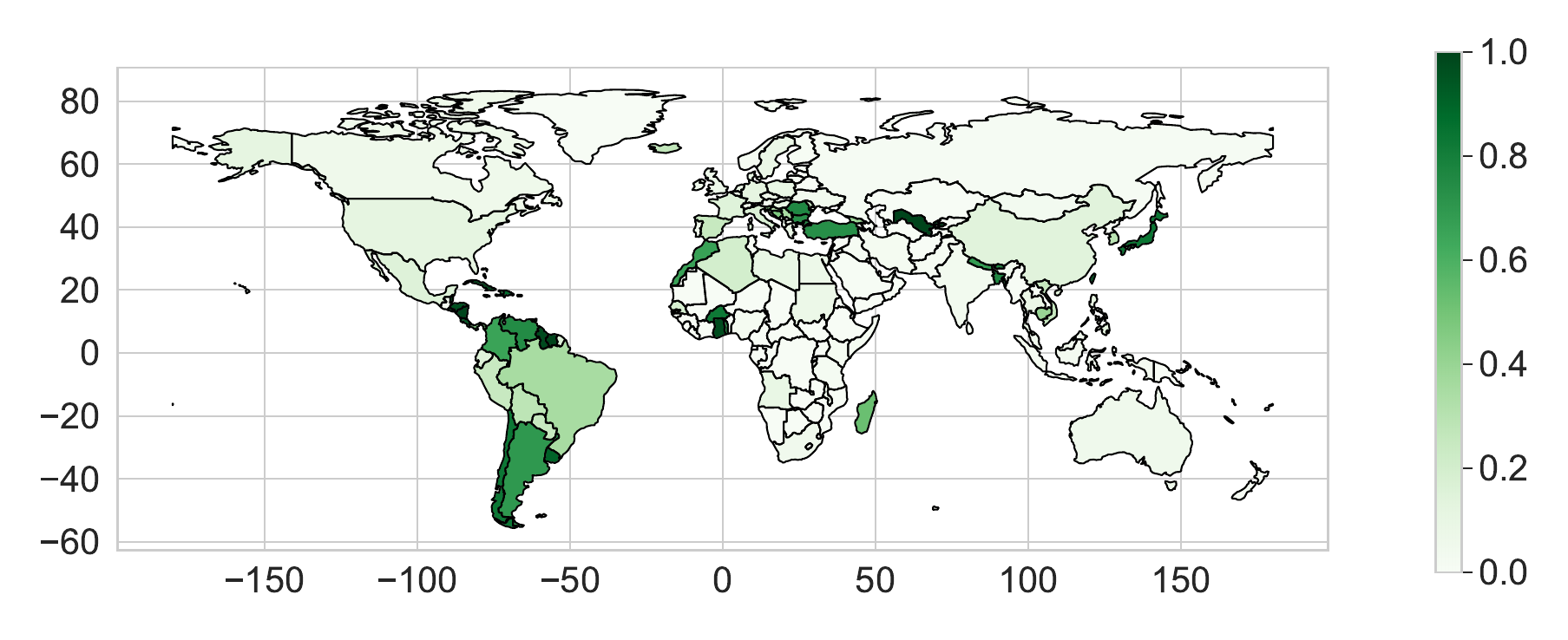}
        \caption{The risk profile for countries due to multiple disasters}
        \label{fig:mult_disaster_countries_impact}
    \end{subfigure}
    \caption{Plots showing the cross-layer impact due to multiple disasters}
    \label{fig:mult_disaster_results}
\end{figure}

\begin{figure}[ht]
    \centering
    \begin{subfigure}{\textwidth}
        \centering
        \includegraphics[width=0.8\textwidth]{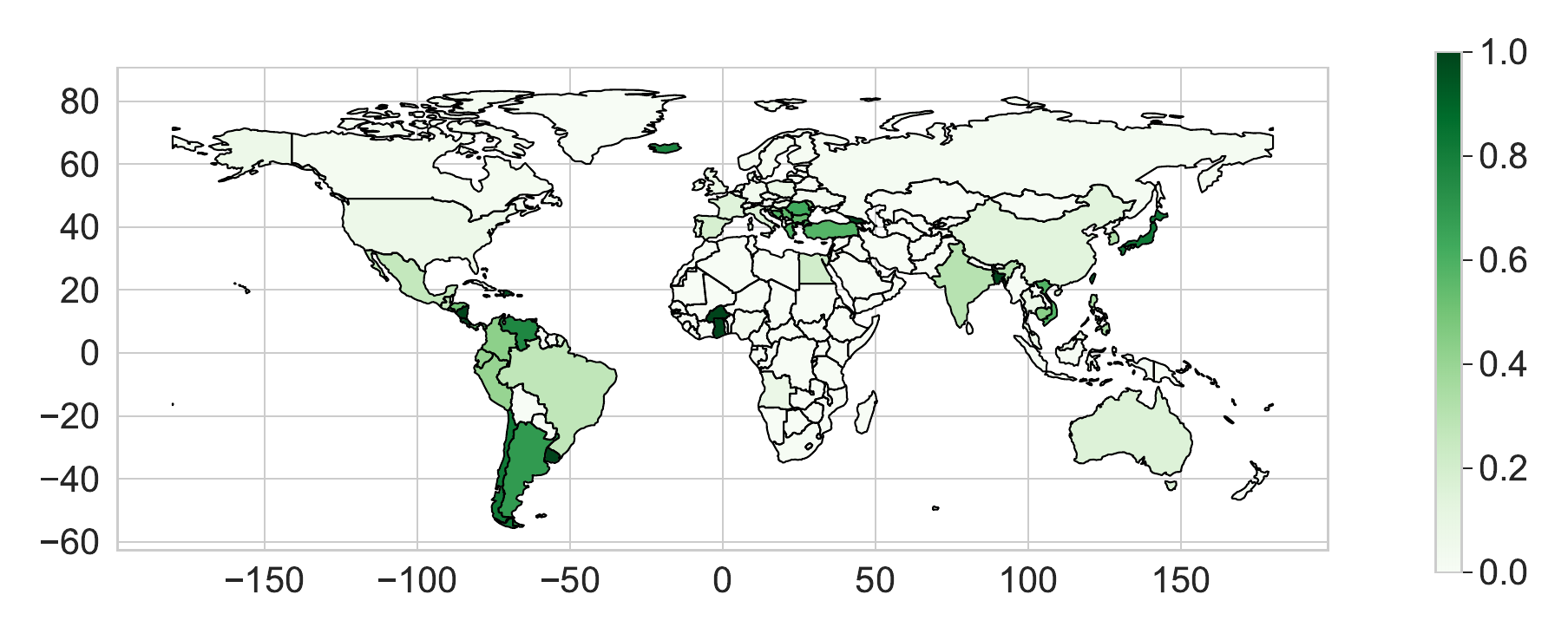}
        \caption{The normalized Intra-AS impact for countries due to multiple disasters}
        \label{fig:mult_dis_intra_impact_res}
    \end{subfigure}
    \begin{subfigure}{\textwidth}
        \centering
        \includegraphics[width=0.8\textwidth]{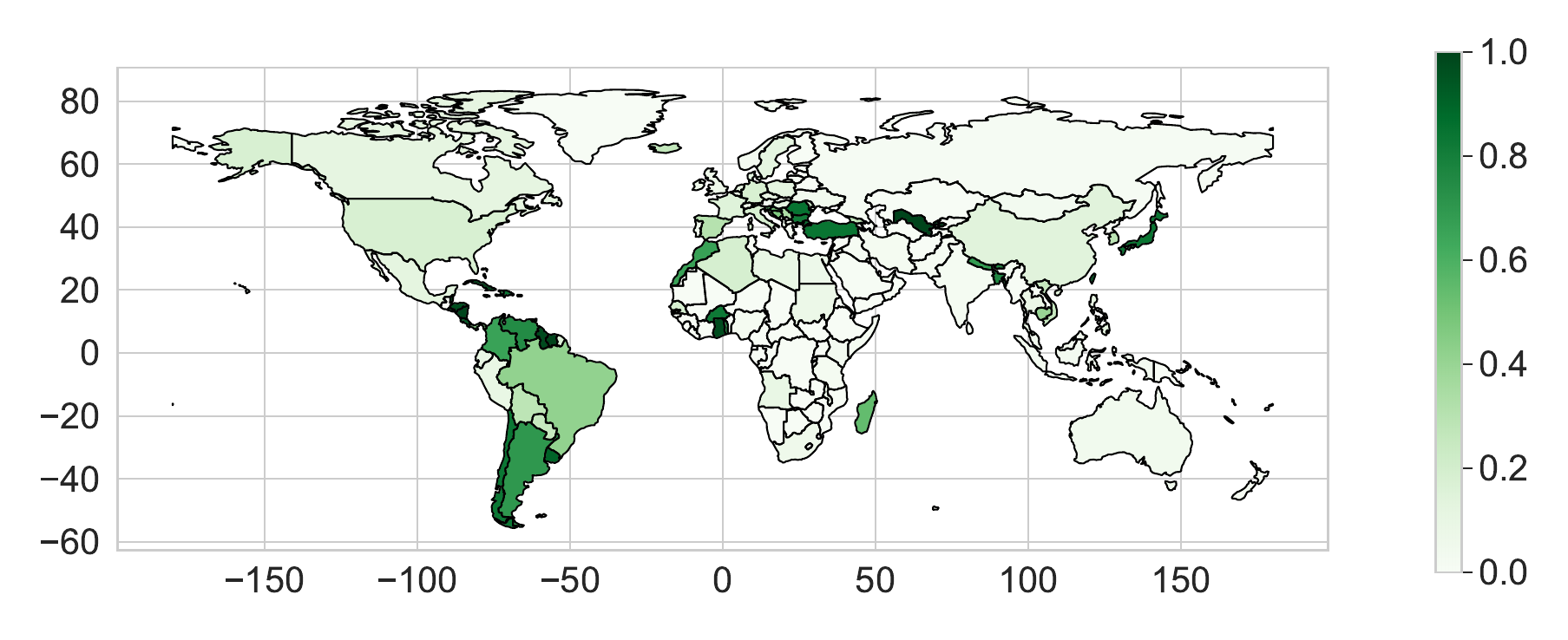}
        \caption{The normalized Inter-AS impact for countries due to multiple disasters}
        \label{fig:mult_dis_inter_impact_res}
    \end{subfigure}
    \caption{The normalized interconnectivity impacts due to multiple disasters. The results for each country are normalized based on the number of intra-AS and inter-AS links observed respectively.}
    \label{fig:mult_dis_impact_intra_inter_res}
\end{figure}

\subsubsection{Practical Assessment of Combined Disaster Effects}

In previous sections, we focused on assessing maximum infrastructure risk in extreme disaster scenarios. Now, we turn to a more practical assessment involving multiple disasters. In this experiment, we consider a 0.2-meter sea-level rise and a 5\% failure scenario for earthquakes, hurricanes, and solar storms, selecting the top 5\% of the most impactful points for each disaster to provide estimates for near-future impact. We evaluate overall impact, including the percentage of impact at each layer, the creation of country risk profiles, and an analysis of intra and inter-AS patterns using the Multiple Failure Events Analysis capability (as described in \S~\ref{additional_capabilities}). It's essential to note that these estimates showcase the tool's utility rather than predict actual impact. The results of this experiment can be found in Figures~\ref{fig:mult_disaster_results} and \ref{fig:mult_dis_impact_intra_inter_res}.

\parait{Insights and Use Cases:} Despite considering only the top 5\% of regions for multiple disasters, infrastructure components show substantial impact, with at least 15\% impact for each component. Additionally, we observed that, with this framework, regions including parts of South America, the Balkans region in Europe, and countries like Japan, Nepal, and Bangladesh in Asia are notably impacted (Figure~\ref{fig:mult_disaster_countries_impact}). In line with the previous result, land-locked countries like Nepal experience a significantly higher impact on inter-AS connectivity compared to intra-AS connectivity (Figure~\ref{fig:mult_dis_impact_intra_inter_res}).

\parait{Capabilities:} This experimental setup, involving analyzing the impact of multiple disaster models are varied failure distribution, demonstrates a combination of the multiple features and capabilities within Xaminer.

\subsection{Event Independent Cross-Layer Analysis}

Many research questions such as modeling the Internet infrastructure require an understanding of the distribution patterns and key trends. Having access to these patterns helps researchers and stakeholders with data-driven decision-making and modeling. To assist with this, Xaminer provides (i) the distribution patterns for aggregation of network layer against physical layer components, (ii) the inter-AS and intra-AS visualization, and (iii) correlation patterns amongst countries.

\subsubsection{Examining Distribution Trends}

While the previous sections have analyzed the cross-layer impacts of various events at multiple granularities, it is essential to identify the overall distribution patterns for the cross-layer map to build better models. In this segment, we examine the distribution patterns for IP endpoints within countries and ASes against the physical layer components, which are the cable segments, cables, landing stations, and P-countries. First, we present a visual representation of the IP endpoints in a country's (N-Country) dependence on other countries' submarine cable infrastructure (P-Country) is depicted in Figure~\ref{fig:world_distribution}. Next Figure~\ref{fig:country_cdf} and \ref{fig:as_cdf} present the CDF plots for the number of physical infrastructure components that IP endpoints within a country (N-Country) and AS use respectively.

\begin{figure}[ht]
    \centering
    \begin{subfigure}{0.6\textwidth}
        \centering
        \includegraphics[width=\textwidth, height=0.5\textwidth]{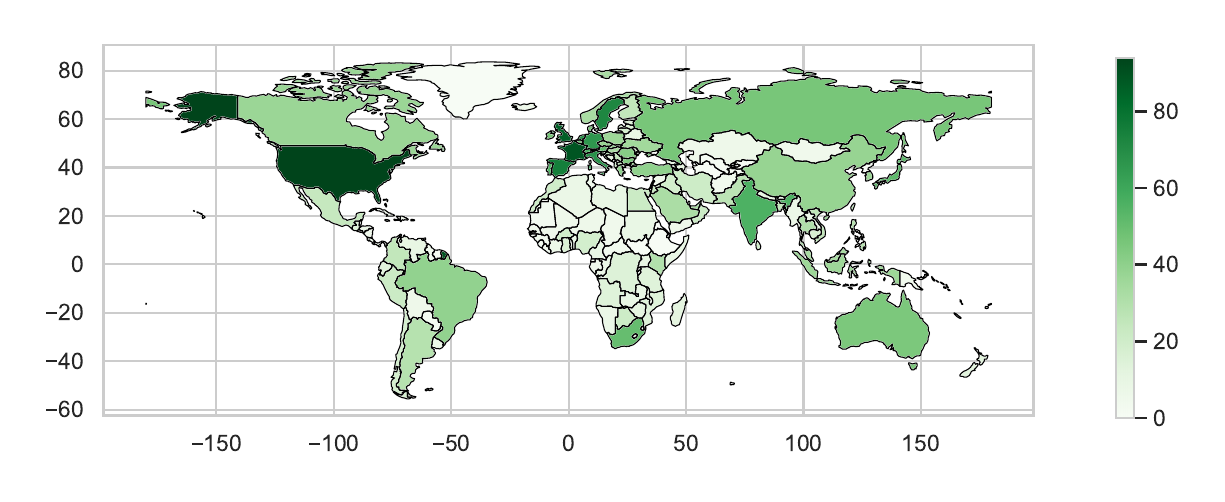}
        \caption{ }
        \label{fig:world_distribution}
    \end{subfigure}
    \hfill
    \begin{subfigure}{0.35\textwidth}
        \centering
        \includegraphics[width=\textwidth]{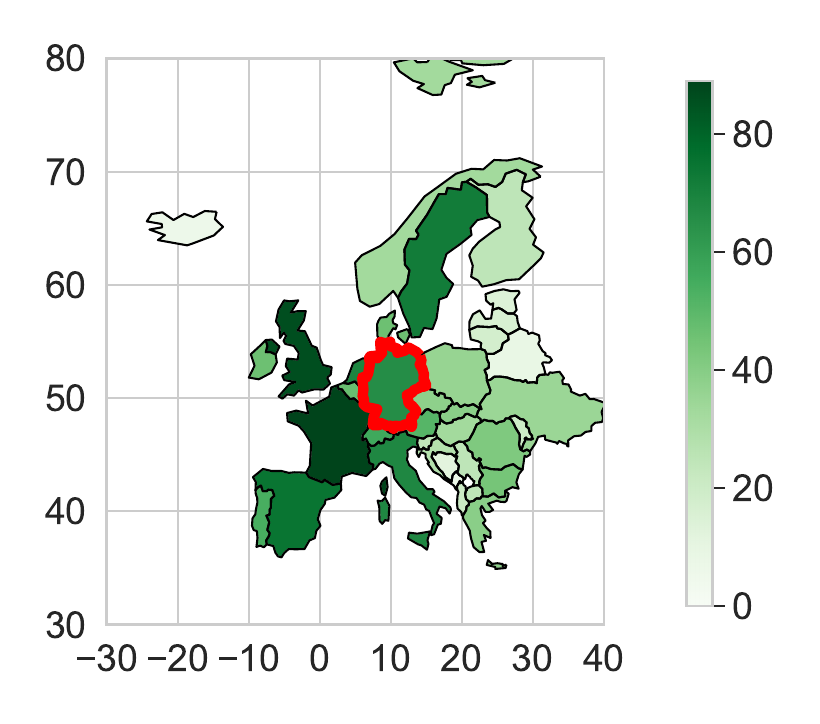}
        \caption{ }
        \label{fig:europe_distribution}
    \end{subfigure}
    \label{fig:distributions}
    \caption{Countries colored by the number of countries they are connected to by the Submarine Cable Infrastructure. Note that in Figure (b), Germany (highlighted with red borders) is connected to 65 countries by the submarine cable infrastructure, despite only housing cables connected to < 15 countries. This higher number revealed due to the use of a cross-layer map, is due to reliance on neighboring country's infrastructure to be connected to other nations.}
\end{figure}

\begin{figure}[ht]
    \centering
    \begin{subfigure}{0.48\textwidth}
        \centering
        \includegraphics[width=\textwidth]{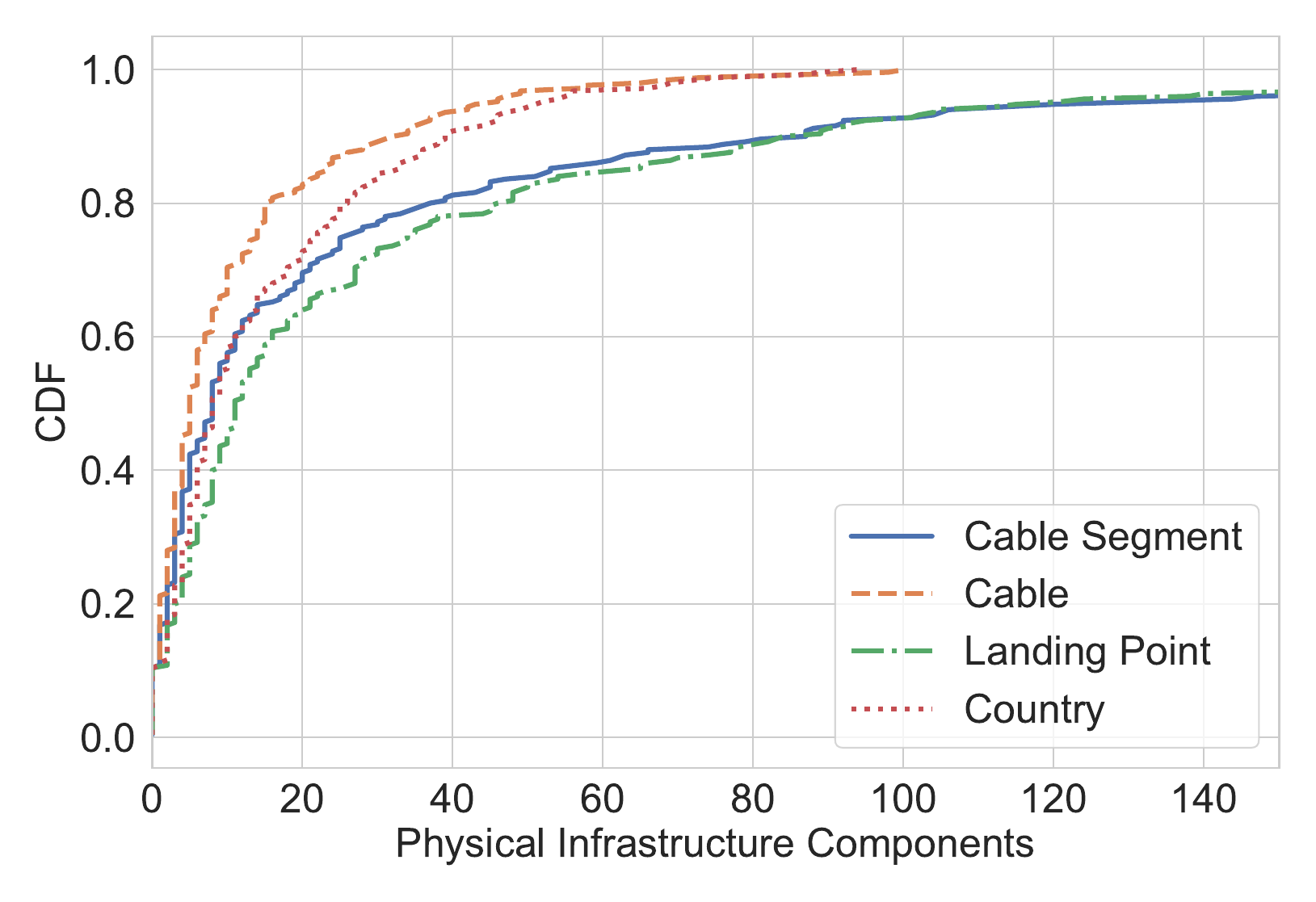}
        \caption{The CDF plot of the number of physical infrastructure components the IP endpoints within a country use.}
        \label{fig:country_cdf}
    \end{subfigure}
    \hfill
    \begin{subfigure}{0.47\textwidth}
        \centering
        \includegraphics[width=\textwidth]{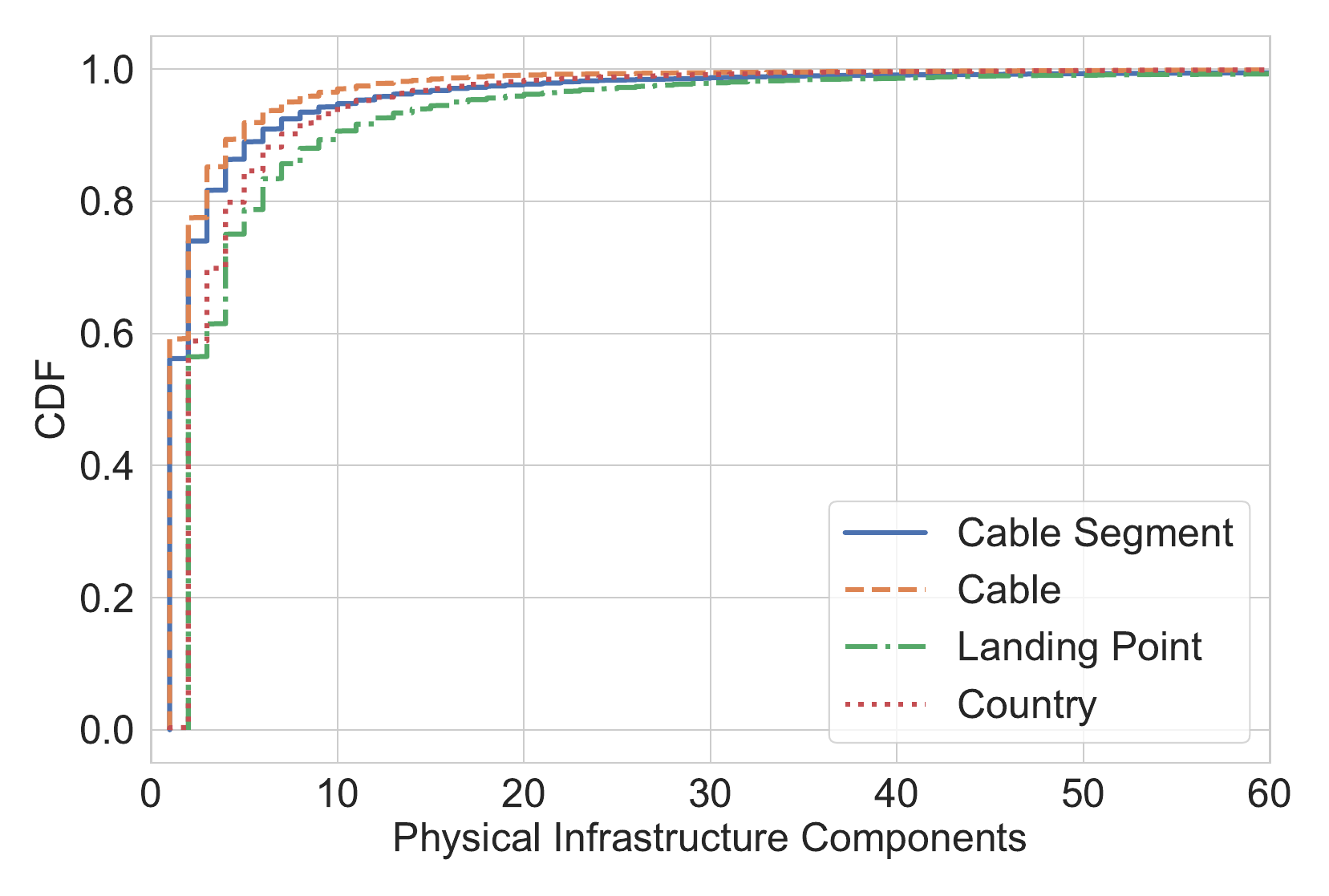}
        \caption{The CDF plot of the number of physical infrastructure components the IP endpoints within an AS use. }
        \label{fig:as_cdf}
    \end{subfigure}
    \caption{The CDF plots for the network layer infrastructure (aggregated by the country and AS) against the physical layer components (aggregated by Cable segments, Cables, Landing Points, and Countries)}
    \label{fig:cdfs}
\end{figure}

\parait{Insights:} In Figure~\ref{fig:world_distribution}, we observe that in each continent, there are just a handful of nations whose IP endpoints act as a gateway for international connectivity. Additionally, of the top 10 most connected nations, 8 are in Europe, 1 in North America (US), and 1 in Asia (India). A higher number indicates that a country is more resilient to singular cable failures and damages as alternative backup IP paths exist. The CDFs in Figure~\ref{fig:cdfs} indicate that (i) most ASes have a limited presence in the physical infrastructure, with more than 80\% of ASes using $\leq$ 5 physical infrastructure components, and (ii) the presence of a few countries and ASes in a large fraction of the physical infrastructure components indicated by the long tail in the plots.

\parait{Capabilities:} Figure~\ref{fig:europe_distribution} presents an instance where utilizing the Xaminer tool helped identify Germany's (highlighted with red borders) reliance on 65 countries' cable infrastructure for network layer connectivity. By just relying on the Submarine Cable Map provided by Telegeography~\cite{submarine_map}, Germany would be mapped to be connected to $<$ 15 countries, which would lead to incorrect inferences when analyzing the impact of events like natural disasters or outages.

\subsubsection{Visualizing Interconnectivity Patterns}

\begin{figure}
    \centering
    \includegraphics[width=\columnwidth]{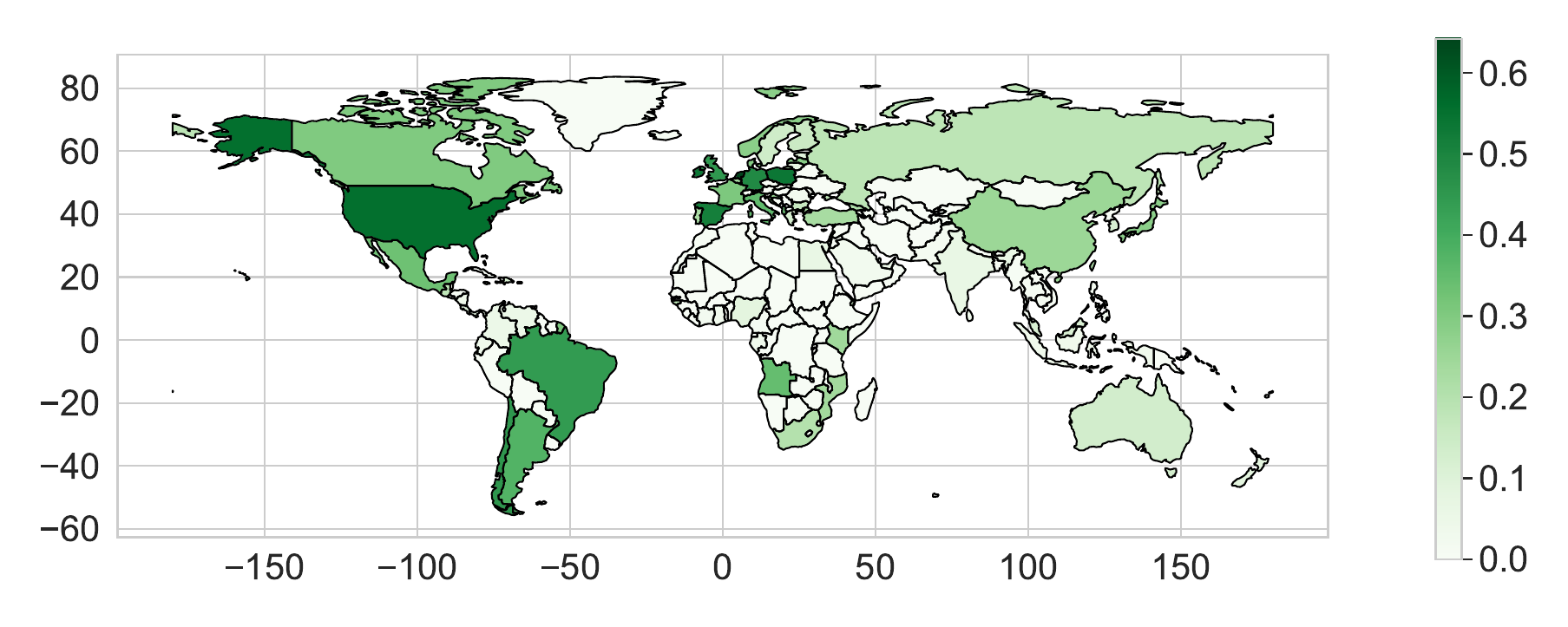}
    \caption{Countries (with submarine cable infrastructure) colored based on the percentage of Intra-AS links. 50\% or 0.5 indicates an equal use of Inter-AS and Intra-AS links, with higher than 0.5 indicating more Intra-AS links}
    \label{country_inter_connectivity_fig}
\end{figure}

In this section, we delve into the interconnectivity patterns of physical infrastructure components aggregated by country (P-Country). These insights are instrumental for resource allocation and strategic infrastructure development. To assess these patterns, we calculate the fraction of links using each country's submarine cables that fall within the Intra-AS category. The results are visualized in Figure~\ref{country_inter_connectivity_fig}.

\parait{Insights:} In Figure~\ref{country_inter_connectivity_fig}, we make intriguing observations. Notably, only five major nations—the US, Spain, Ireland, Netherlands, and Poland—exhibit a higher prevalence of intra-AS usage among their cables. This trend can be attributed to two key factors: First, the presence of Tier-1 Internet Service Providers (ISPs) fostering intra-AS links between the US and Europe. Second, the exchange of data center traffic between the US and Europe, notably in Ireland. Interestingly, we note that the Western regions, including the Americas and Europe, demonstrate a greater presence of intra-AS links compared to the Eastern regions. This difference may be influenced by the factors mentioned earlier.

\subsubsection{Correlation Trends}

In this section, we delve into the correlation trends exhibited by countries aggregated based on the network layer. To achieve this, we use the country correlation trends functionality from the resilience analysis module (\S~\ref{additional_capabilities}). The resulting insights are presented in Figure~\ref{country_cluster_corr} and the countries clustered within each cluster are detailed in Appendix~\ref{corr_trends_appendix}.

\begin{figure}
    \centering
    \includegraphics[width=\columnwidth]{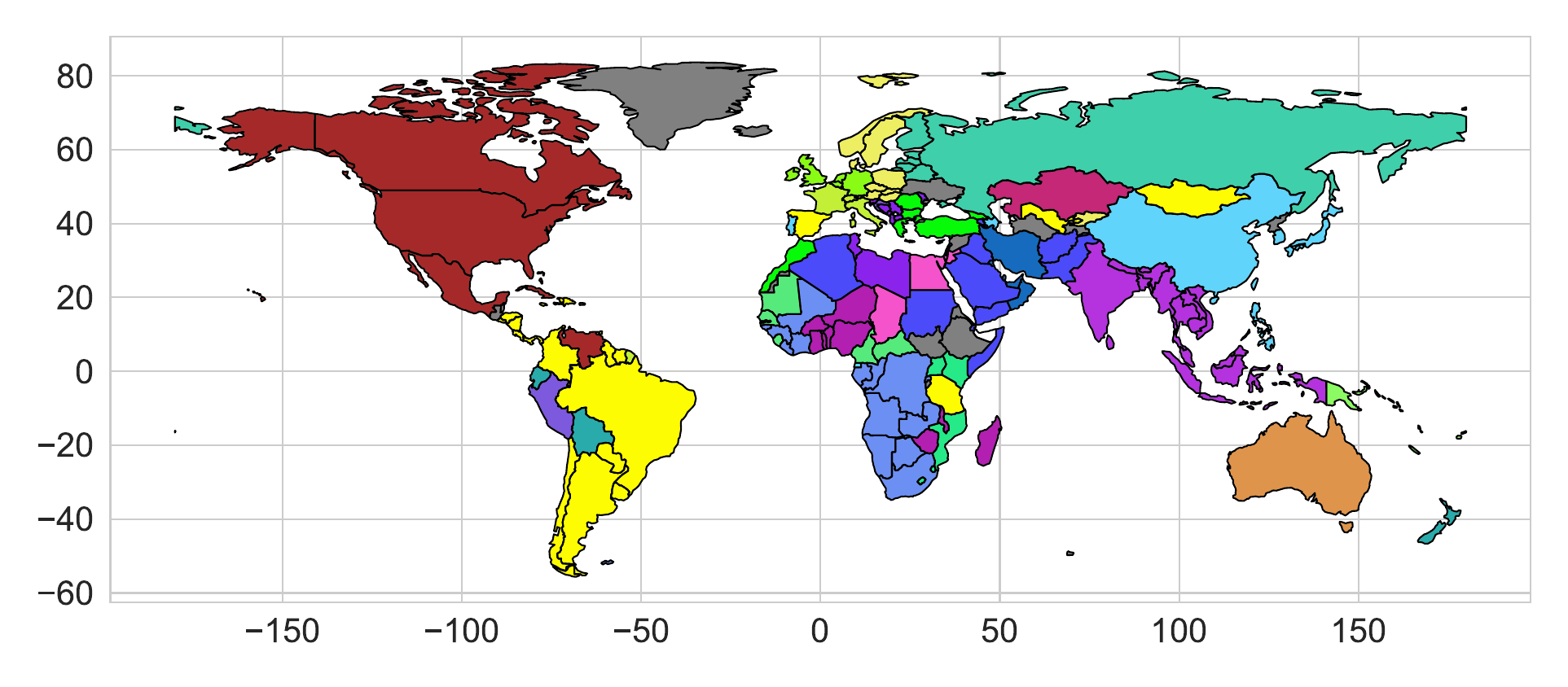}
    \caption{Countries clustered based on their correlation distributions. We observe that the regional blocs are the critical factor for a country's cluster assignment.}
    \label{country_cluster_corr}
\end{figure}

\parait{Insights and Use Cases:} Remarkably, our analysis from Figure~\ref{country_cluster_corr} reveals that most clusters correspond to regional blocs. This pattern persists even when countries within a cluster may be connected to entirely different sets of countries. For instance, India and Indonesia, despite their disparate lists of connected countries, exhibit strikingly similar correlation distributions. Even with a reduction in the distance threshold for clustering (leading to an increase in the number of clusters), most countries remain grouped within the close sub-divisions of regional blocs. This finding underscores the influence of geopolitics, proximity, and international relations on a country's connectivity patterns. A few anomalies in the clustering process are worth noting: (i) the grey cluster represents parts of Africa and Asia. However, upon closer examination, it becomes evident that all the countries in this cluster lack sufficient data, and (ii) the yellow cluster includes South America, Spain, Tanzania, and Mongolia. Nevertheless, these countries fragment into unique regional clusters when the distance threshold is lowered.

These unique trends and patterns derived from distribution data, interconnectivity analysis, and correlation provide essential features for modeling general behavior within the cross-layer submarine cable infrastructure. Such a model is invaluable for long-term infrastructure planning and resilience analysis.

\subsection{Comparison with Real-World Failures}

Given the prevalence of partial or complete cable failures due to damages at landing points and cuts in cable segments, characterizing the impact of such events on the global Internet is essential. \ashwin{We validate the country-level impact analysis of Xaminer using traffic losses reported by monitoring platforms---Cloudflare~\cite{Cloudflare}, IODA~\cite{IODA}, and NetBlocks~\cite{netblocks}.} 

\ashwin{Using Xaminer, we compute the risk profile (fraction of IP links lost) for all countries using three real-world events---SeaMeWe-5 cable disruption, Egypt cable cut, and Yemen cable failure. Despite the limitations of the risk profile metric, it serves as a proxy for traffic loss, and we discuss its implications based on the observed inconsistencies from the three events.}

\ashwin{Since Xaminer uses the Nautilus cross-layer map, which can assign an IP link to multiple cables, particularly in cases involving parallel cables, we expand our analysis to two scenarios: (i)\textit{Top}, where the risk is computed by considering only the top cable segment prediction for each IP link, and (ii)\textit{Weighted}, where IP links mapped to parallel cables are assigned an equal weight. For instance, if an IP link is mapped to four parallel cables, the \textit{Top} method considers only the top cable segment, while the \textit{Weighted} method considers each cable segment with a weight of 0.25. Now we discuss the results of these three events in detail.}

\parait{SeaMeWe-5 Cable Disruption:} \ashwin{ In November 2022, a terrestrial cable cut (in Egypt) on the SeaMeWe-5 cable caused outages in parts of Asia and Africa. We compare the impact estimated by Nautilus with observations from real-world monitoring platforms in Table~\ref{tab:seamewe5_cable_cut}. Xaminer uses the impacted segments to evaluate the risk profile (a cable cut in Egypt is likely to impact the segment between France and Pakistan but is unlikely to affect the segment between Singapore and Indonesia). Table~\ref{tab:seamewe5_cable_cut} shows that, although Xaminer estimates the fraction of links lost and the monitoring platforms evaluate the fraction of traffic lost, Xaminer correctly estimates the impact for Pakistan, Somalia, and Yemen. However, Xaminer inaccurately estimates traffic loss for Djibouti, Eritrea, and Indonesia for the following reasons. A lack of data in Nautilus for Eritrea causes the estimated risk to be 0\%. For Indonesia, a large disparity is observed due to two factors: (i) disproportionate links on the Pacific end compared to the Eurasian end in Nautilus data, and (ii) the potential for high traffic from Europe not being identified due to lack of traffic capacity information for identified links. Notably, while NetBlocks identified traffic drops in all 6 countries, IODA reported an impact for Djibouti and Somalia, and Cloudflare radar showed no significant impact for these countries, revealing tool-specific inconsistencies likely stemming from varied vantage points in traffic data collection. This could be the cause for the disparity in results for Djibouti.}

\begin{table}[]
    \parbox{.47\textwidth}{
    \centering
    \begin{tabular}{ c c c c }
    \hline
    \textbf{Country} & \textbf{Top} & \textbf{Wghtd} & \textbf{Val} \\
    \hline
    Pakistan & 12.75\% & 14\% & 19\% \\
    Djibouti & 4\% & 18.75\% & 54\% \\
    Somalia & 6.75\% & 30\% & 25\% \\
    Indonesia & 2\% & 4\% & 55\% \\
    Yemen & 40\% & 25\% & 32\% \\
    Eritrea & 0\% & 0\% & 45\% \\
    \hline
    \end{tabular}
    \vspace{2mm}
    \caption{\ashwin{The risk profile identified for various countries by Xaminer using the Top and Weighted (Wghtd) mechanisms for the SeaMeWe-5 cable disruption. The validation column (Val) indicates the traffic loss reported by NetBlocks for this event~\cite{seamewe5_damage_pakistan}. For these countries, there was no significant traffic impact seen from IODA, Cloudflare radar, or both.}}
    \label{tab:seamewe5_cable_cut}
    }
    \hfill
    \parbox{.47\textwidth}{
    \centering
    \begin{tabular}{ c c c c }
    \hline
    \textbf{Country} & \textbf{Top} & \textbf{Wghtd} & \textbf{Val} \\
    \hline
    Somalia & 73\% & 72.5\% & 85\% \\
    Pakistan & 5\% & 20.25\% & 25\% \\
    Ethiopia & 0\% & 0\% & 90\% \\
    Djibouti & 33.5\% & 35\% & 29\% \\
    Saudi Arabia & 7.5\% & 9.25\% & 16\% \\
    Egypt & 15\% & 3\% & 3\% \\
    \hline
    \end{tabular}
    \vspace{2mm}
    \caption{\ashwin{The risk profile identified for various countries by Xaminer using the Top and Weighted (Wghtd) mechanisms for a terrestrial cable cut in Egypt. The validation column (Val) indicates the traffic loss reported as outages for Somalia, Pakistan, and Ethiopia~\cite{egypt_cable_cut} or identified for Djibouti, Saudi Arabia, and Egypt~\cite{Cloudflare_djibouti, Cloudflare_saudi, Cloudflare_egypt} by Cloudflare radar.}}
    \label{tab:egypt_cable_cut}
    }
    \vspace{-8mm}
\end{table}

\parait{Egypt Cable Cut:}  \ashwin{In June 2022, a terrestrial cable cut in Egypt caused significant network disruptions across Africa and Asia, resulting in an 85\% traffic reduction in Somalia, a 25\% reduction in Pakistan, and impacts in Djibouti and Saudi Arabia~\cite{egypt_cable_cut}. Similar to the SeaMeWe-5 failure, we identify cable segments likely affected by the Egypt cut and use Xaminer to compute risk profiles for countries (Table~\ref{tab:egypt_cable_cut}). Notably, Xaminer's risk closely aligns with Cloudflare's reported traffic loss for all countries except Ethiopia, where Nautilus lacked data, resulting in a 0\% risk. High variance in risk scores for Pakistan is observed due to multiple cables along the same path. Although Cloudflare doesn't report outages for Djibouti, Saudi Arabia, and Egypt, immediate traffic drops were observed at the onset of the failure, recovering within 1 to 1.5 hours due to traffic rerouting through alternative routes.}

\parait{Yemen Falcon Cable Failure:} \ashwin{ In January 2022, the Falcon cable, a key source of international connectivity for Yemen, suffered a complete capacity drop due to an airstrike in Al-Hudaydah~\cite{yemen_cable_cut}. Simulating this event, we designate all cable segments with an endpoint in Al-Hudaydah as failed and use Xaminer to compute risk profiles. Xaminer identifies a 70\% drop (Top method) and a 45\% drop (Weighted method) in IP links for Yemen. The high variance can be attributed to the sparse Nautilus data in Yemen and the presence of multiple parallel cables in the region. Traffic loss in other countries was negligible and hence is not detailed.}

\parait{Insights and Use Cases:} These results underscore the direct correlation between submarine cable link failures and their far-reaching impacts on a country's Internet traffic. It's important to note that, in our analysis, Xaminer considers only the definitely submarine links. As a result, some discrepancies were observed in certain countries, which could be further mitigated with improved cross-layer mapping and link bandwidth details. With Xaminer's ability to quantify these impacts, cable repair teams can gain a more efficient means of prioritizing segments, and network operators are better equipped to devise effective rerouting strategies.

\parait{Capability:} This experimental setup, exploring various combinations of cable segment, cable, and landing point failures, showcases Xaminer's ability to assess cross-layer impacts on submarine cable infrastructure at a highly granular level.

\section{Discussion}~\label{discussion}

\vspace{-2mm}

Having conducted an extensive analysis of the cross-layer impact of diverse natural disasters in both regional and global contexts, we now direct our attention to recognizing the inherent limitations associated with our data sources and Xaminer. While improving the accuracy of underlying datasets is not the goal of Xaminer, its design enables better models and maps (as they become available) to be easily plugged into Xaminer.

\begin{figure}
    \centering
    \includegraphics[width=0.65\columnwidth]{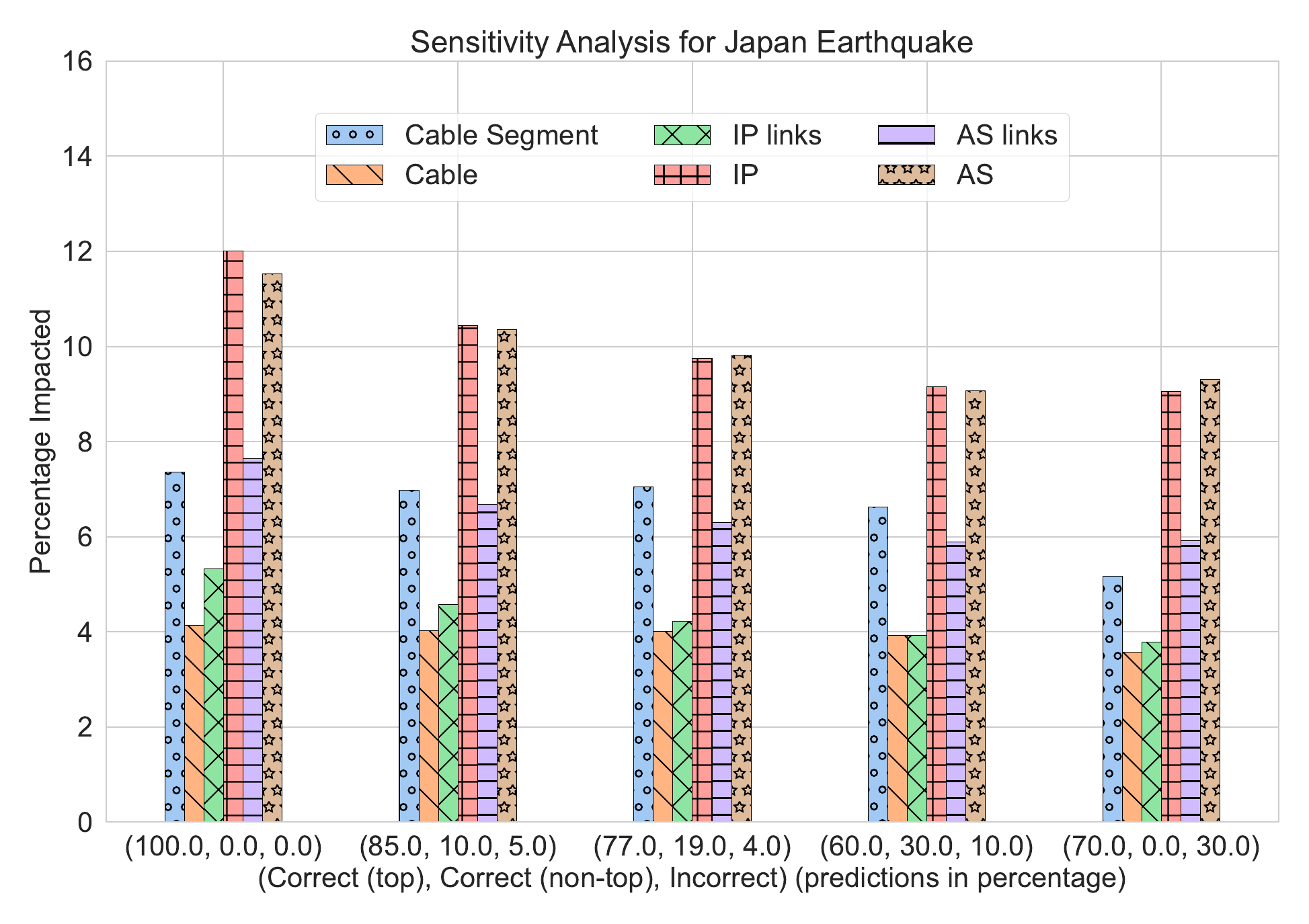}
    \caption{\ashwin{The maximum percentage of infrastructure at risk at varied levels of inaccuracies with Nautilus data. The horizontal axis depicts a 3-tuple: the percentage of IP links with correct top cable segment, correct non-top (secondary) cable segment, and incorrect cable segment predictions by Nautilus, respectively. Notably, significant Nautilus data inaccuracies do not proportionally reduce risk in Xaminer's analysis. (77, 19, 4) value corresponds to Nautilus' observed result in its validation experiment.}}
    \label{fig:japan_sensitivity_analysis}
    \vspace{-5mm}
\end{figure}

\parait{Inaccuracies with cross-layer map:} \ashwin{Nautilus, the cross-layer map, exhibits limitations stemming from multiple inaccurate and incomplete sources. Despite efforts to mitigate errors, Nautilus predicts multiple cables per IP link and may generate inaccurate mappings, especially when considering MPLS tunnels. The validation experiments in Nautilus further underscore some of its limitations.}

\ashwin{To gauge the sensitivity of Xaminer's analysis to cross-layer mapping errors, we conduct a sensitivity analysis. As analyses are based on the top cable prediction, potential errors include cases where Nautilus correctly predicts the cable segment but not as the top (secondary) choice and instances where Nautilus fails to predict the correct cable segment. Lacking ground-truth information, we simulate error scenarios by randomly selecting a fraction of IP links in the impacted region to have one of these errors. To simulate secondary choice errors for an IP link, we randomly select a secondary cable segment prediction as the ground truth. To simulate scenarios where Nautilus fails to predict the correct cable segment for an IP link, we apply stricter constraints by excluding the impact of such links. This assumption, based on the possibility that the true cable segment might not be in the impacted regions, thus serves as a lower bound for the results.}

\ashwin{Figure~\ref{fig:japan_sensitivity_analysis} contains the averaged results from the sensitivity analysis across 10 rounds for various combinations of the two errors with varying levels of added errors. Despite the artificially introduced inaccuracies, the impact evaluation shows similar trends. Since parallel cables have a high likelihood of being jointly impacted by a failure event due to their geographical proximity, non-top errors (where Nautilus maps a link to the wrong candidate in a set of parallel cables) have minimal impact. We see similar patterns for other regions and disaster models (additional plots are presented in Appendix~\ref{sensitivity_analysis_appendix}).}

\parait{Measurement infrastructure bias:} \ashwin{Xaminer's analysis indirectly relies on traceroute data sourced from RIPE Atlas and CAIDA, introducing bias towards Europe and America, with coverage gaps in Asia and Africa. Despite Nautilus' attempt to mitigate this bias, it still persists. Initiatives like MIRA~\cite{MIRA} aim to fill data gaps, particularly in regions like Africa, presenting opportunities to enhance cross-layer maps. While recognizing these limitations, our experiments included well-represented regions like Japan and the Pacific Northwest that revealed consistent patterns. Furthermore, Xaminer's adaptable design can easily integrate improved cross-layer maps (as they become available), continually enhancing the precision and reliability of the analyses.}

\parait{Cable endpoint considerations:} \ashwin{The analyses presented focused solely on the impact at cable endpoints, omitting considerations for the intermediate path between these points. This decision is based on the nature of the three evaluated disasters (sea rise, hurricanes, and solar storms), where the impact is primarily determined by cable endpoints and not affected by the cable's remaining length. Only earthquakes (and volcanic eruptions – not amongst the evaluated disasters) have the potential to cause significant disruptions along the cable's length. However, the currently available earthquake dataset covers inland and regions closer to the coast, making it inadequate for estimating the impact on the cable's length. Xaminer's results are therefore based on the endpoint (ie., landing stations) impact alone, disregarding the cable segment's length.}

\ashwin{If disaster data becomes available for locations in the ocean, Xaminer can estimate the impact using a simple straight-line assumption or line strings-based cable data from publicly available submarine cable maps. Given the sensitive nature of submarine cable paths, precise cable path information is unlikely to be publicly available. However, since earthquakes in the ocean bed have a large impact radius, these approximations (straight-line or line strings) are likely to capture the impact effectively.}

\parait{Limitations in the risk profile metrics:} \ashwin{While analyzing past failure events, we utilize the risk profile as a proxy for traffic loss, acknowledging its potential divergence from true traffic loss. The actual risk is dependent on the fraction of traffic carried by each IP link, which is currently not available. This is due to all underlying datasets relying on traceroute measurements, which lack data on bandwidth or traffic fraction per IP link. Moreover, the visible links in traceroutes are influenced by BGP routing policies, load balancing, and failover configurations, making it challenging to distinguish between active and backup links. While an argument can be made to use the frequency of links in the traceroute measurements as a proxy, this would be erroneous due to incomplete views and associated biases in traceroute measurements and vantage points.}

\parait{Enhancing impacted cable segments identification:} \ashwin{In evaluating failure events, we assume that if cable segment endpoints can only be connected while passing through the failure region, the cable segment is affected. For example, in the SeaMeWe-5 cut in Egypt, we assume that the France to Pakistan link is vulnerable.  While generally valid, the internal fiber optic structure may create scenarios where this assumption fails. An alternative approach involves collecting traceroute measurements before, during, and after a failure event, identifying cable segments where most links disappear during failure while being present before and after. However, this method has limitations due to traceroute coverage, potentially leading to incomplete information and incorrect inferences for some cable segments. Hence we leave uncovering the internal interdependencies within a cable to generate more accurate estimates as a direction for future work.}

\section{Conclusion}

In this paper, we introduce Xaminer, the first cross-layer resilience analysis tool that significantly enhances our understanding of the relationship between physical and network failures on both a global and regional scale. Xaminer's unique capabilities allow us to quantify the potential infrastructure impact of various natural disasters, offering risk assessments at varying levels of granularity, from entire countries and regions down to individual cable segments. Its adaptability ensures that Xaminer can accommodate future improvements in cross-layer mapping and emerging data sources. Through extensive experiments, we demonstrate Xaminer's capabilities in dissecting the cross-layer impacts of diverse disaster scenarios, encompassing events such as earthquakes and hurricanes. Its flexibility enables us to address a wide array of event combinations across different regions, unveiling the distinct patterns and consequences of each disaster type. Xaminer can serve as an indispensable tool in informing our ongoing mission to enhance the robustness and resilience of the Internet.
\bibliographystyle{ACM-Reference-Format}
\bibliography{main}

\appendix

\section{Ethics}

This work does not raise any ethical concerns.

\begin{figure}
    \centering
    \includegraphics[width=0.85\columnwidth]{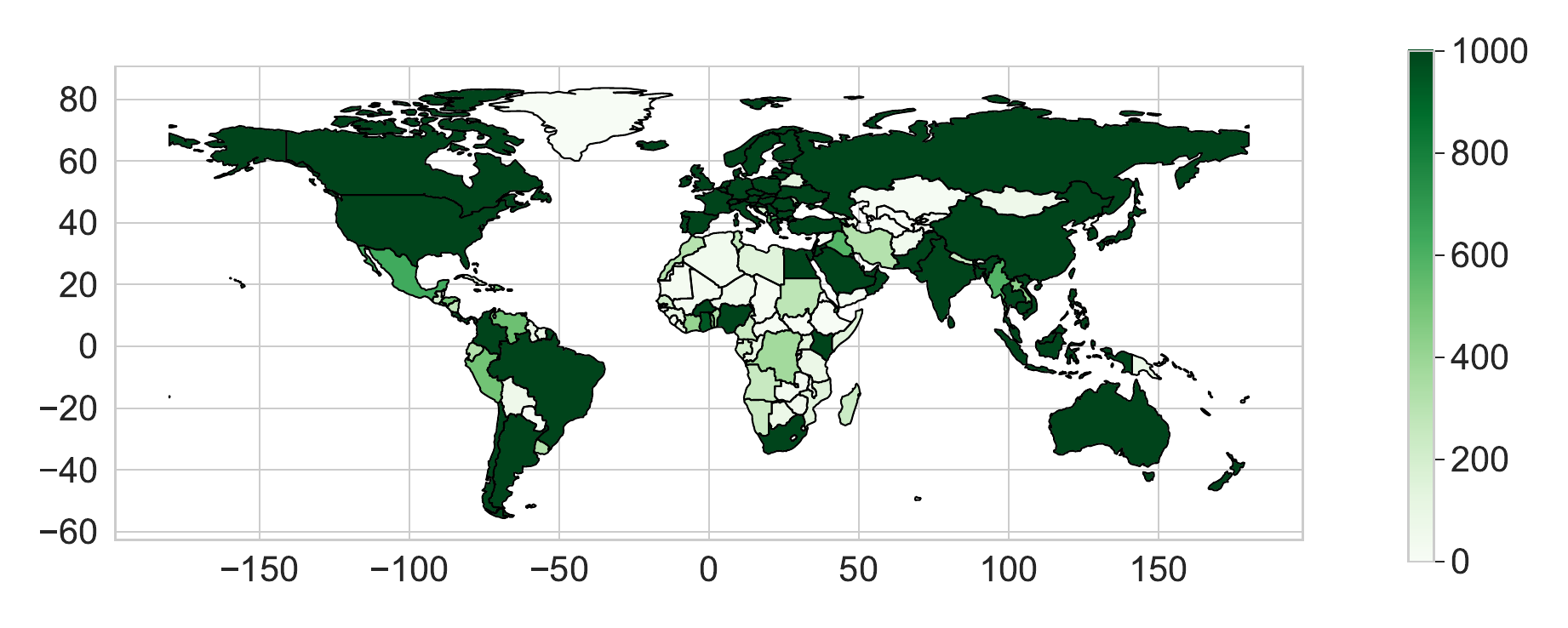}
    \caption{Countries colored by the number of IP links with an endpoint in the country. All countries with more than 1000 links are shown with the same dark green color.}
    \label{fig:countries_colored}
\end{figure}

\section{Additional Results}

\ashwin{In this section, we present the additional results from our analyses of the cross-layer impact on submarine cables.}

\subsection{Identifying Maximum Infrastructure at Risk Across Geographic Regions} \label{region_maximal_impact_appendix}

\begin{figure}
    \centering
    \includegraphics[width=0.85\columnwidth]{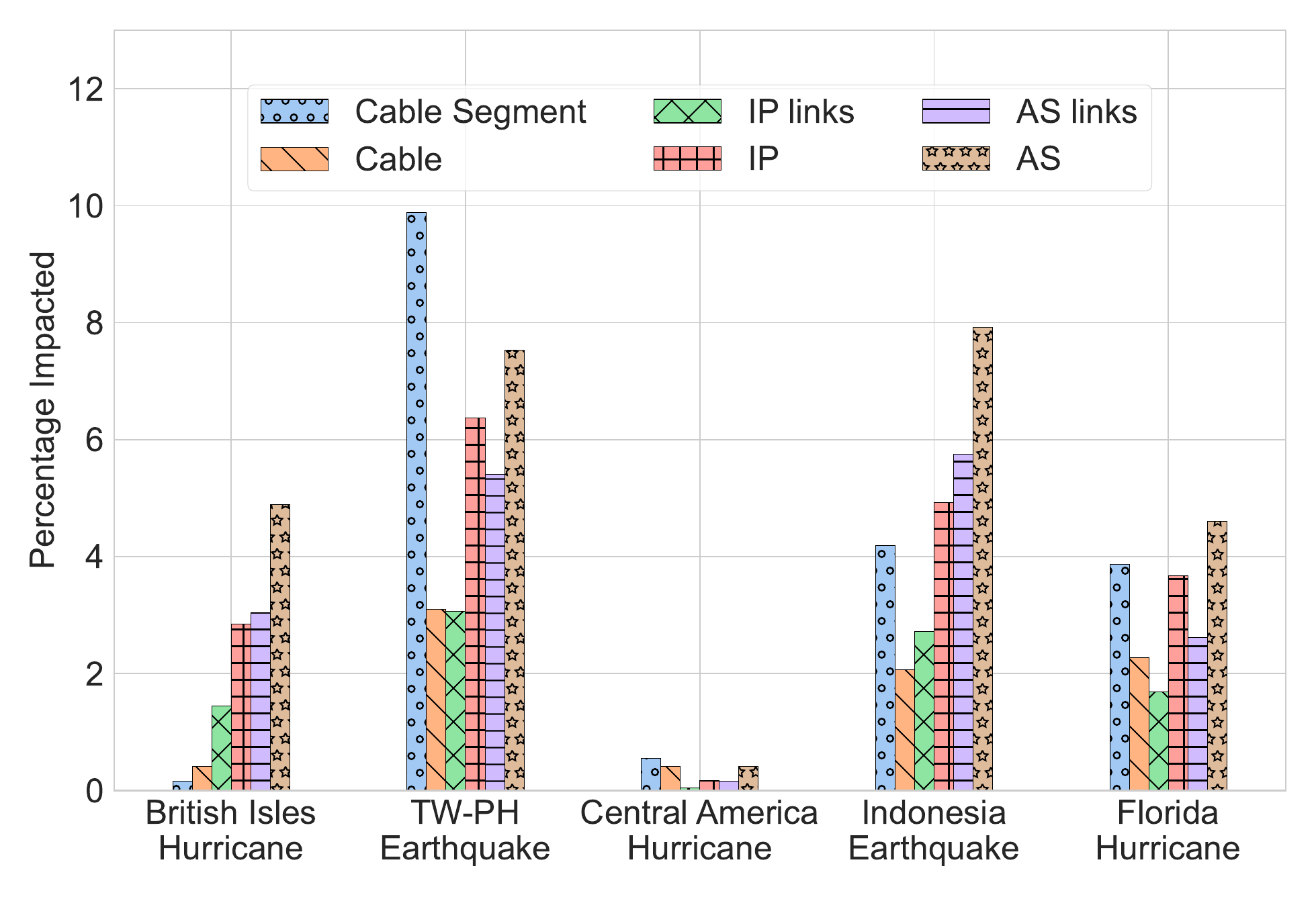}
    \caption{The maximum percentage of infrastructure at risk across layers due to various disasters at a regional level.}
    \label{fig:regional_maximal_impact_additional}
\end{figure}

\ashwin{To assess the maximum infrastructure at risk across different geographical regions, we leverage the cross-layer impact analysis feature detailed in our resilience analysis module (\S~\ref{resilience_analysis_methods}). In addition to results for Japan, the Caribbean, and the Pacific Northwest discussed earlier, we look at five more regions susceptible to earthquakes and hurricanes: (i) Hurricanes in the British Isles, (ii) Earthquakes in the Taiwan and Philippines region, (iii) Hurricanes in Central America, (iv) Earthquakes in Indonesia, and (v) Hurricanes in Florida. The graphical representation of the impact of these disasters on the various layers is shown in Figure~\ref{fig:regional_maximal_impact_additional}. Similar to Japan, the British Isles region has a limited number of cables, and the network layer components bear a higher risk compared to the Taiwan and Philippines regions, where the physical layer components face maximal susceptibility. Florida and Central America bear similar impacts across both physical and network layer components, while Indonesia bears a higher risk, particularly for ASes.}

\subsection{Gauging A Country's Risk Profile} \label{gauging_risk_profile_appendix}

\begin{figure}[ht]
    \centering
    \begin{subfigure}{\textwidth}
        \centering
        \includegraphics[width=0.8\textwidth]{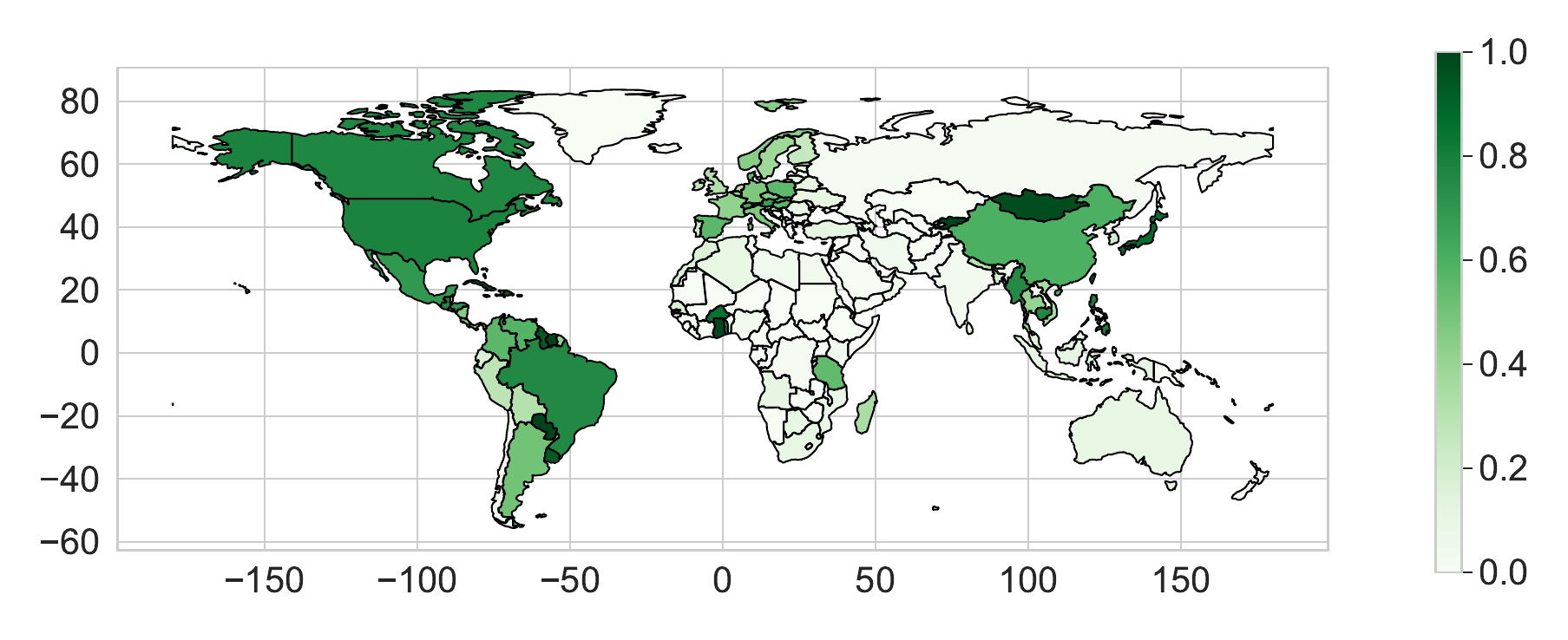}
        \caption{The risk profile for countries due to hurricanes (with a threshold of 64 knots or higher)}
        \label{fig:hurricane_countries_impact}
    \end{subfigure}

    \begin{subfigure}{\textwidth}
        \centering
        \includegraphics[width=0.8\textwidth]{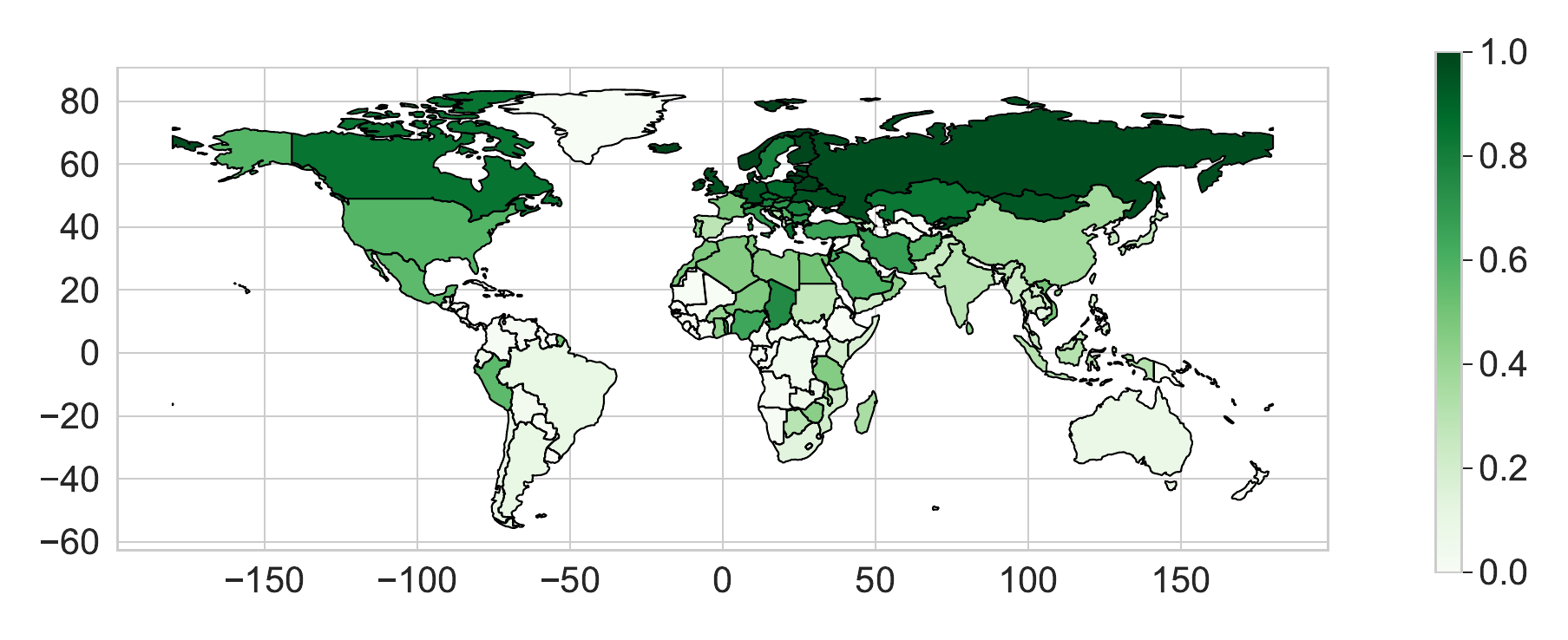}
        \caption{The risk profile for countries due to sea level rise (with a threshold of 50 latitude or higher)}
        \label{fig:solarstorm_countries_impact}
    \end{subfigure}
    \label{fig:countries_impact_maximal_additional_v2}
    \caption{The risk profiles for countries due to various disaster settings. The results for each country are normalized based on the number of IPs within the country.}
\end{figure}

\textit{Assess Country Risk Profiles:} \ashwin{To assess the normalized impact of disasters on a given country, we used the risk profile metric in the resilience analysis module (Section~\ref{resilience_analysis_methods}). Figures~\ref{fig:hurricane_countries_impact} and~\ref{fig:solarstorm_countries_impact} contain the results for hurricanes and solar storms respectively. Similar to earthquakes and sea level rise, countries directly affected by these events (hurricanes and solar storms) also have a cascading impact on those countries dependent on them. For instance, Paraguay, a country in South America, faces severe hurricane impact at the network layer despite being land-locked due to its reliance on a limited number of neighboring countries that are directly impacted by hurricanes.}

\begin{figure}[ht]
    \centering
    \begin{subfigure}{\textwidth}
        \centering
        \includegraphics[width=\textwidth]{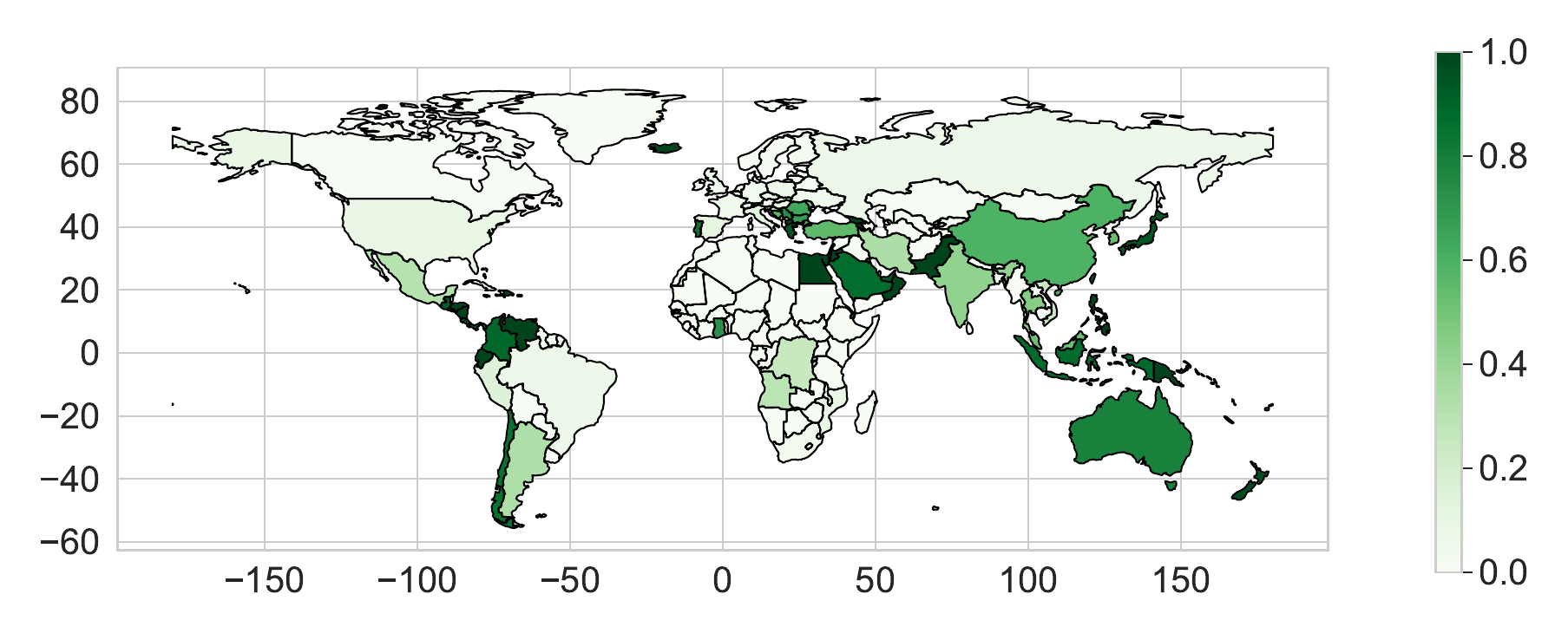}
        \caption{The normalized Intra-AS impact for countries due to earthquake (with a threshold of 6 PGA or higher)}
        \label{fig:earthquake_intra_impact}
    \end{subfigure}
    \begin{subfigure}{\textwidth}
        \centering
        \includegraphics[width=\textwidth]{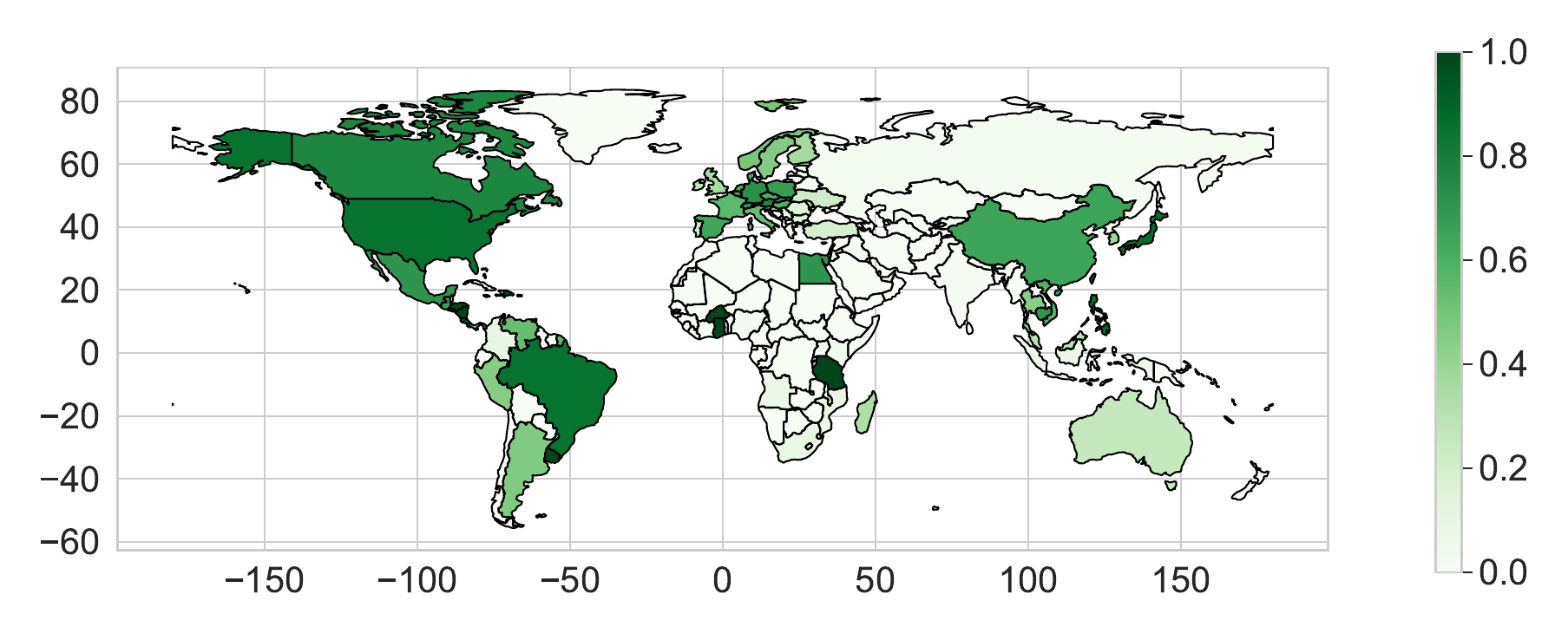}
        \caption{The normalized Intra-AS impact for countries due to hurricane (with a threshold of 64 knots or higher)}
        \label{fig:hurricane_intra_impact}
    \end{subfigure}
    \begin{subfigure}{\textwidth}
        \centering
        \includegraphics[width=\textwidth]{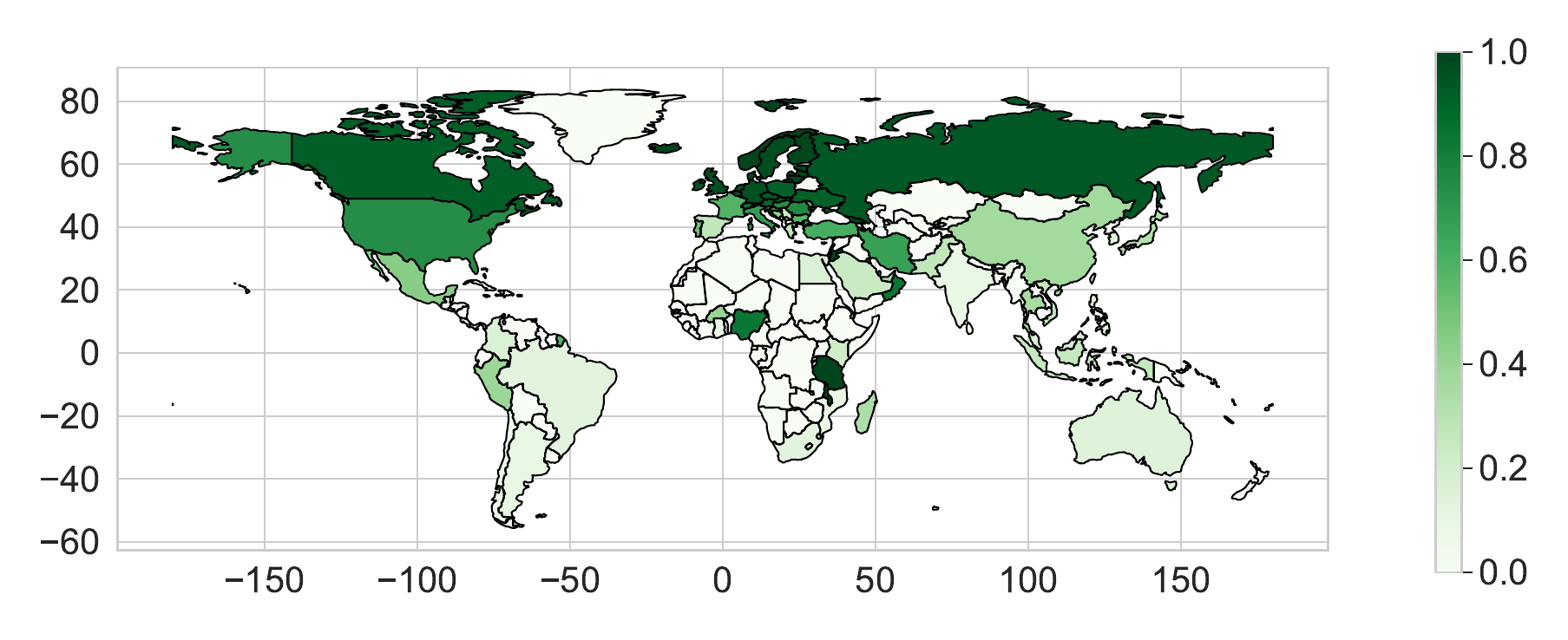}
        \caption{The normalized Intra-AS impact for countries due to Solarstorms (with a threshold of 50 latitude or higher)}
        \label{fig:solarstorm_intra_impact}
    \end{subfigure}
    \caption{The normalized intra-AS impacts due to earthquakes, hurricanes, and solar storms. The results for each country are normalized based on the number of intra-AS links observed.}
    \label{fig:disasters_impact_intra_new}
\end{figure}

\begin{figure}[ht]
    \centering
    \begin{subfigure}{\textwidth}
        \centering
        \includegraphics[width=\textwidth]{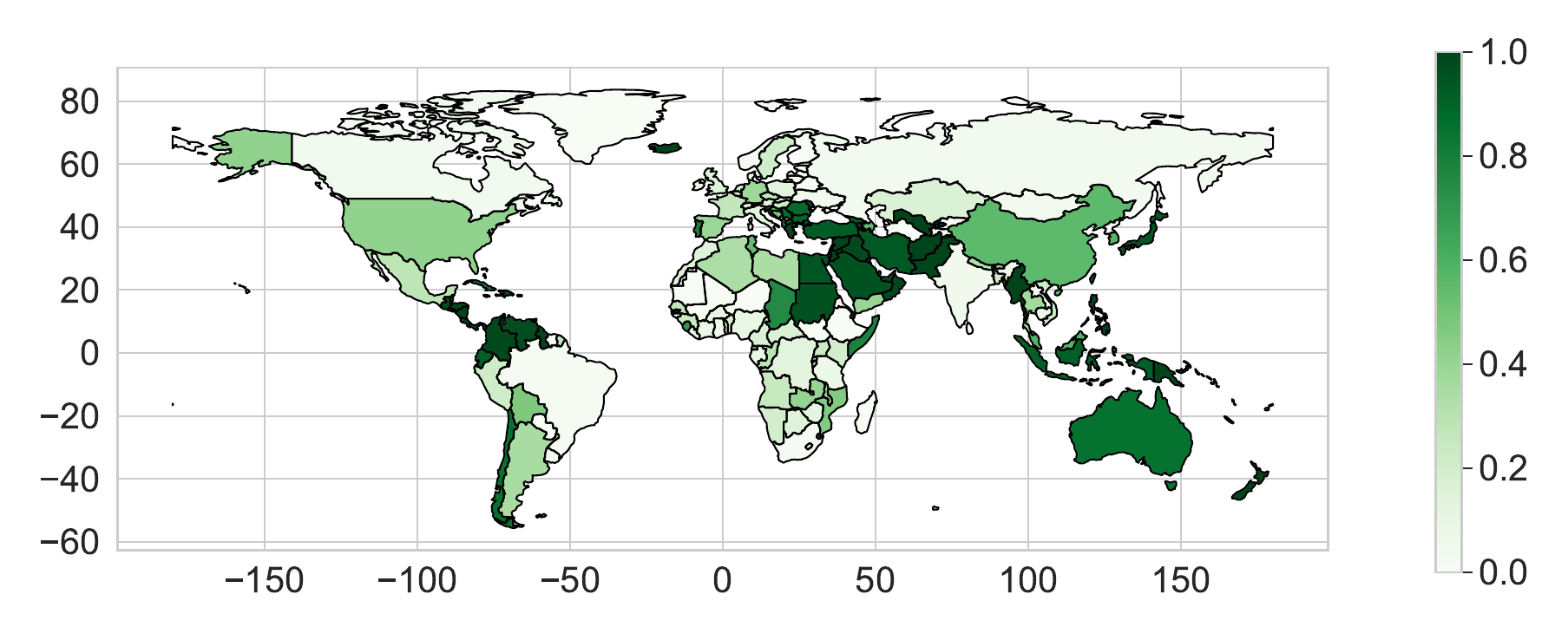}
        \caption{The normalized Inter-AS impact for countries due to earthquake (with a threshold of 6 PGA or higher)}
        \label{fig:earthquake_inter_impact}
    \end{subfigure}
    \begin{subfigure}{\textwidth}
        \centering
        \includegraphics[width=\textwidth]{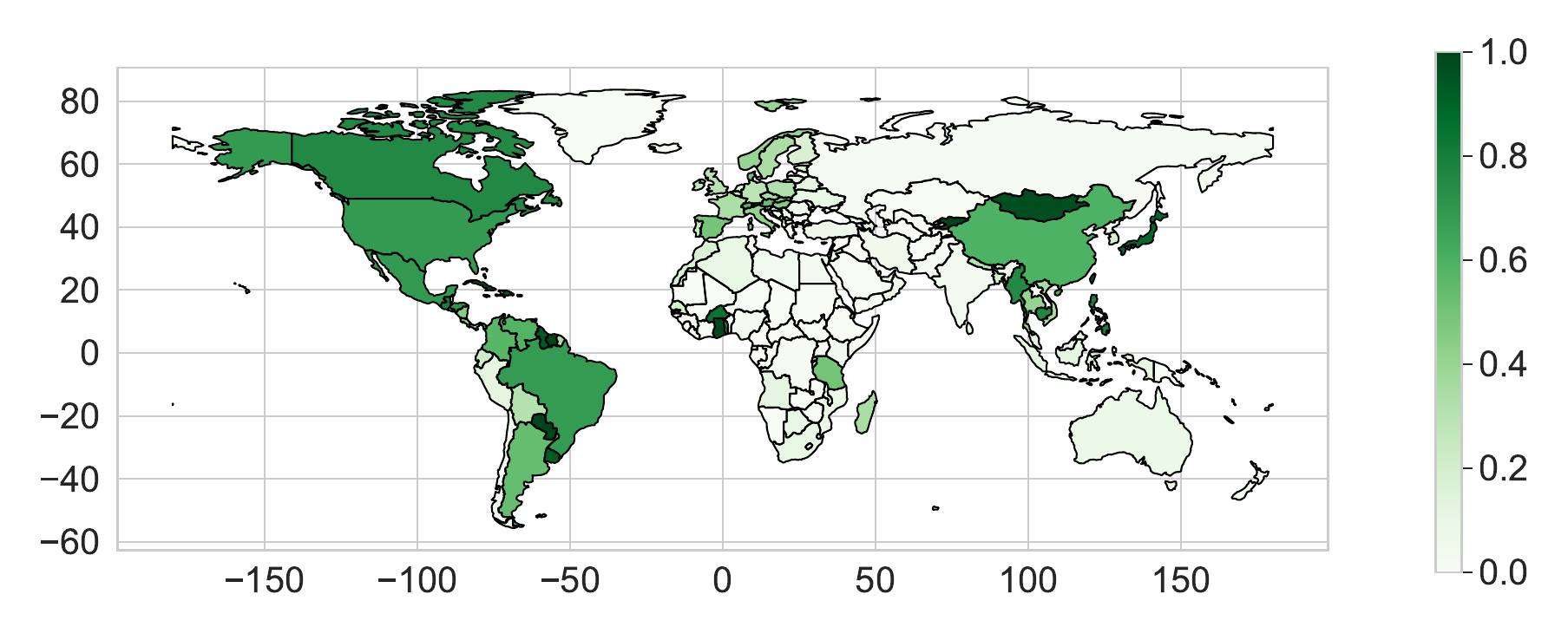}
        \caption{The normalized Inter-AS impact for countries due to hurricane (with a threshold of 64 knots or higher)}
        \label{fig:hurricane_inter_impact}
    \end{subfigure}
    \begin{subfigure}{\textwidth}
        \centering
        \includegraphics[width=\textwidth]{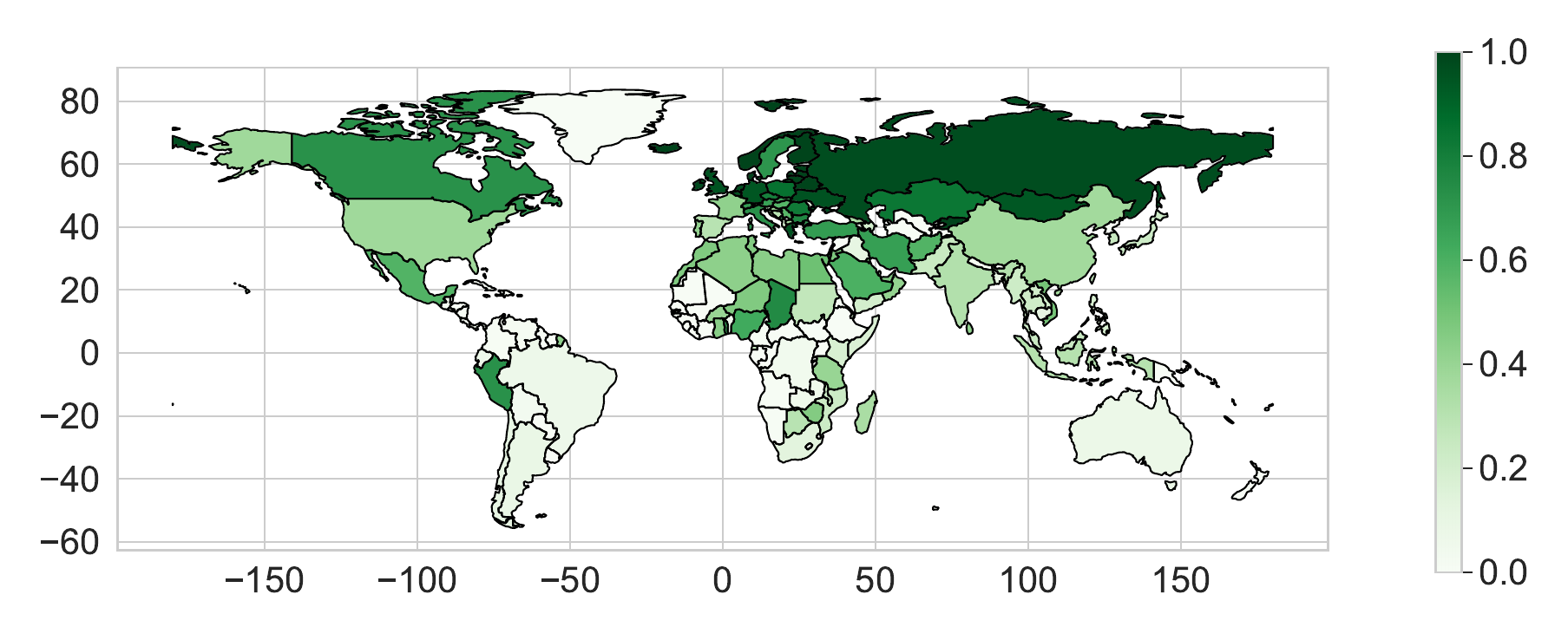}
        \caption{The normalized Inter-AS impact for countries due to Solarstorms (with a threshold of 50 latitude or higher)}
        \label{fig:solarstorm_inter_impact}
    \end{subfigure}
    \caption{The normalized inter-AS impacts due to earthquakes, hurricanes, and solar storms. The results for each country are normalized based on the number of inter-AS links observed.}
    \label{fig:disasters_impact_inter_new}
\end{figure}

\parait{Analyze Interconnectivity Patterns:} \ashwin{We compute the normalized count of intra-AS and inter-AS links for each N-Country using the intra- and inter-AS risk comparison functionality in the resilience analysis module (Section~\ref{resilience_analysis_methods}). The results for earthquakes, hurricanes, and solar storms intra and inter-AS impact are depicted in Figures~\ref{fig:disasters_impact_intra_new} and~\ref{fig:disasters_impact_inter_new} respectively. Similar to the earlier observation with Chad for sea level rise, Paraguay experiences a higher impact on inter-AS links, compared to the intra-AS links, due to its reliance on a limited number of landing points in neighboring nations for submarine cable connectivity.}

\begin{figure}
    \centering
    \includegraphics[width=\columnwidth]{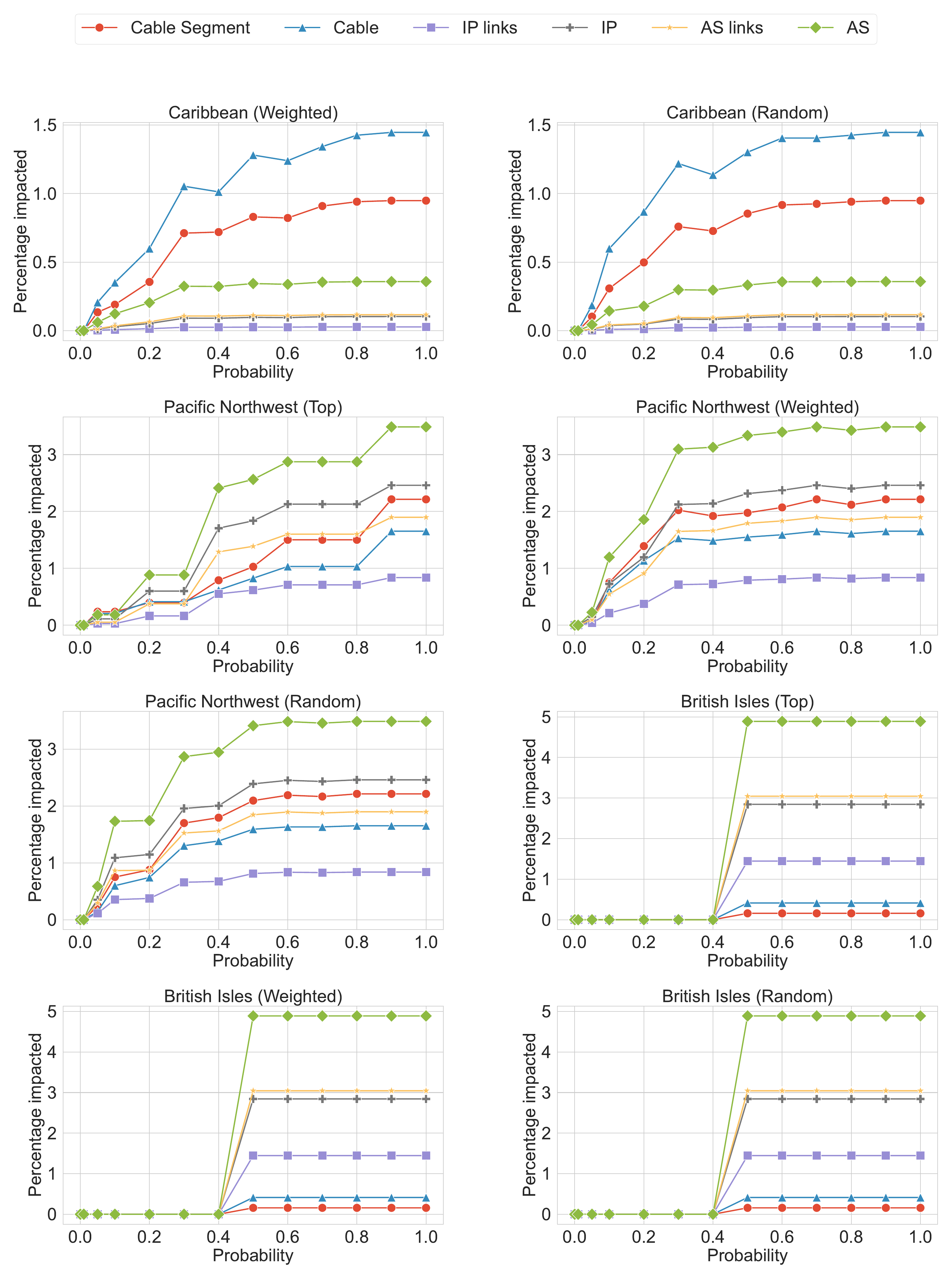}
    \caption{The percentage of infrastructure at risk at various layers in specific regions at various probabilities of failure. The title for each plot represents the region-disaster type with the sampling strategy used.}
    \label{trends_probabilities_regions_additional_1}
\end{figure}

\begin{figure}
    \centering
    \includegraphics[width=\columnwidth]{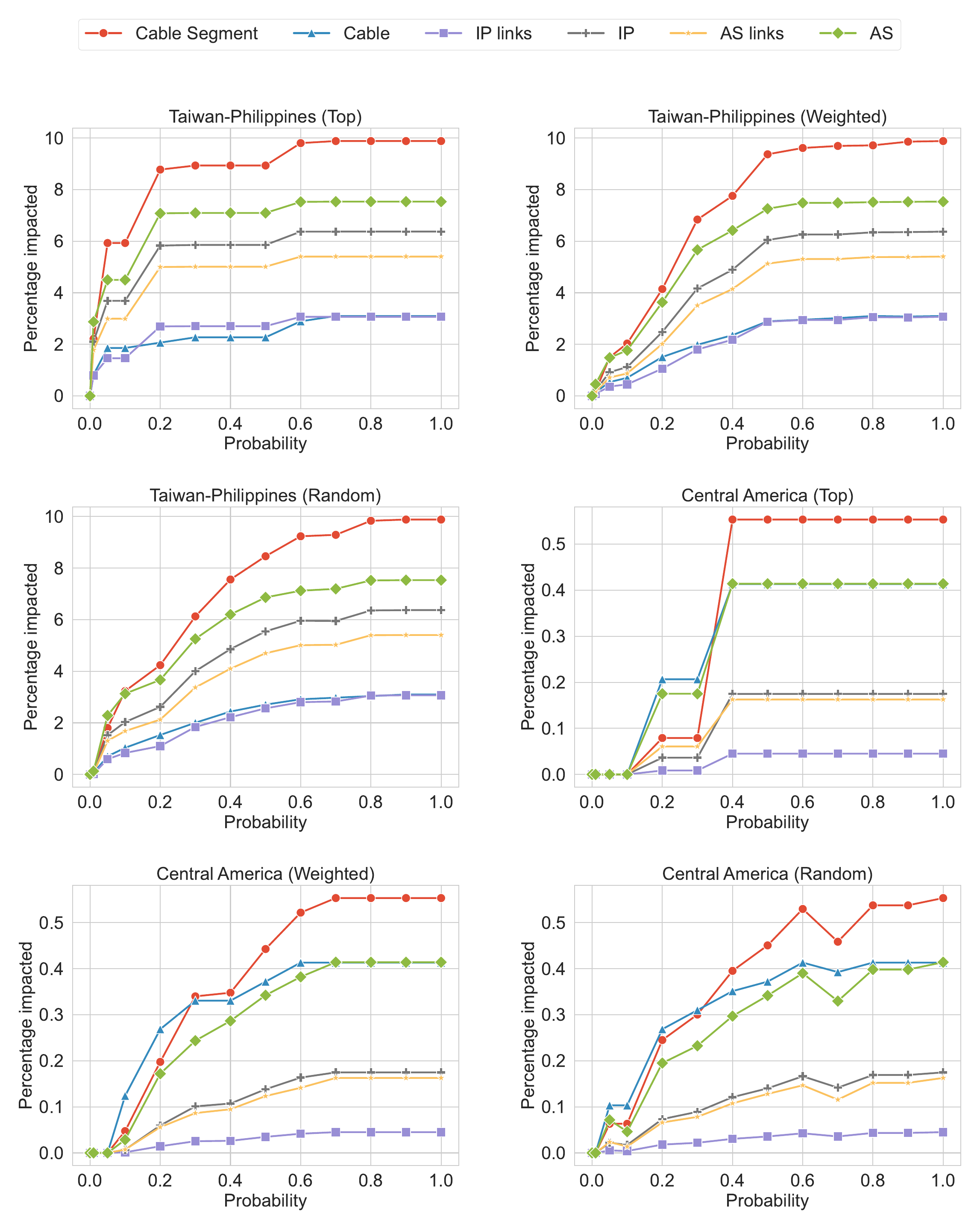}
    \caption{The percentage of infrastructure at risk at various layers in specific regions at various probabilities of failure. The title for each plot represents the region-disaster type with the sampling strategy used.}
    \label{trends_probabilities_regions_additional_2}
\end{figure}

\begin{figure}
    \centering
    \includegraphics[width=\columnwidth]{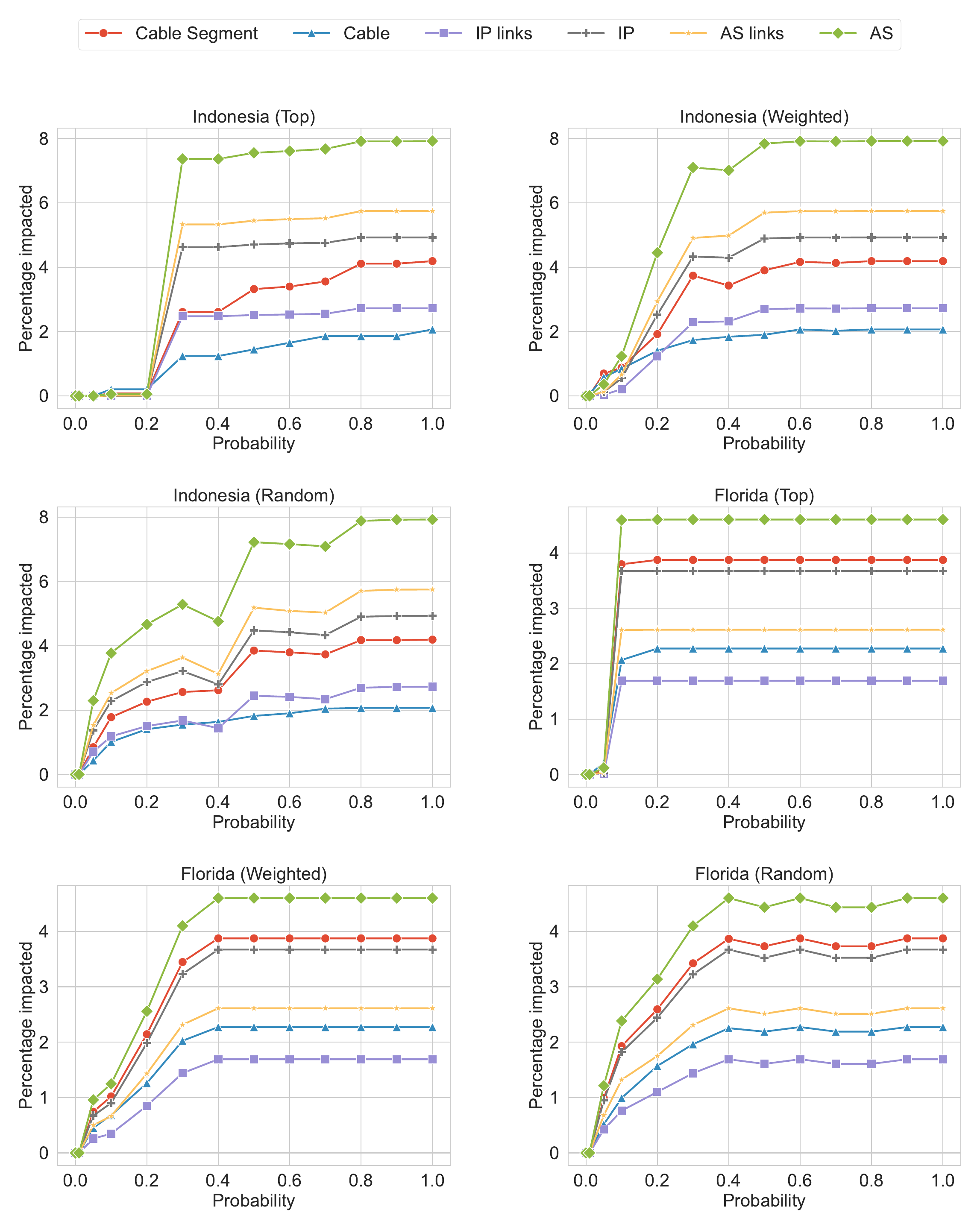}
    \caption{The percentage of infrastructure at risk at various layers in specific regions at various probabilities of failure. The title for each plot represents the region-disaster type with the sampling strategy used.}
    \label{trends_probabilities_regions_additional_3}
\end{figure}

\subsection{Exploring Regional Trends with Varying Probabilities} \label{varying_probabilities_appendix}

\ashwin{To analyze the effects of different failure probabilities, we utilize the cross-layer impact analysis functionality (\S~\ref{resilience_analysis_methods}) and the probabilistic failure impact capability (\S~\ref{additional_capabilities}).We employ three sampling distributions: random, top-n, and weighted. For the random and weighted distributions, we conduct ten experimental runs and report the averages. Figures \ref{trends_probabilities_regions_additional_1}, \ref{trends_probabilities_regions_additional_2}, and \ref{trends_probabilities_regions_additional_3} show the impact of varying failure probabilities using these three sampling distributions in different geographic regions and for various disasters.} 

\ashwin{Similar to the findings in Japan, we consistently observe that a specific component bears a higher risk compared to other physical infrastructure components. For example, this component is cables in the Caribbean, ASes in the Pacific Northwest, the British Isles, Indonesia, and Florida, and cable segments in the Taiwan-Philippines region and Central America. Moreover, akin to the previous results from Japan, a 50\% probability of failure leads to over 80\% of maximum impact across all layers for these components.}

\clearpage

\begin{figure}[ht]
    \centering
    \begin{subfigure}{0.45\textwidth}
        \centering
        \includegraphics[width=\textwidth]{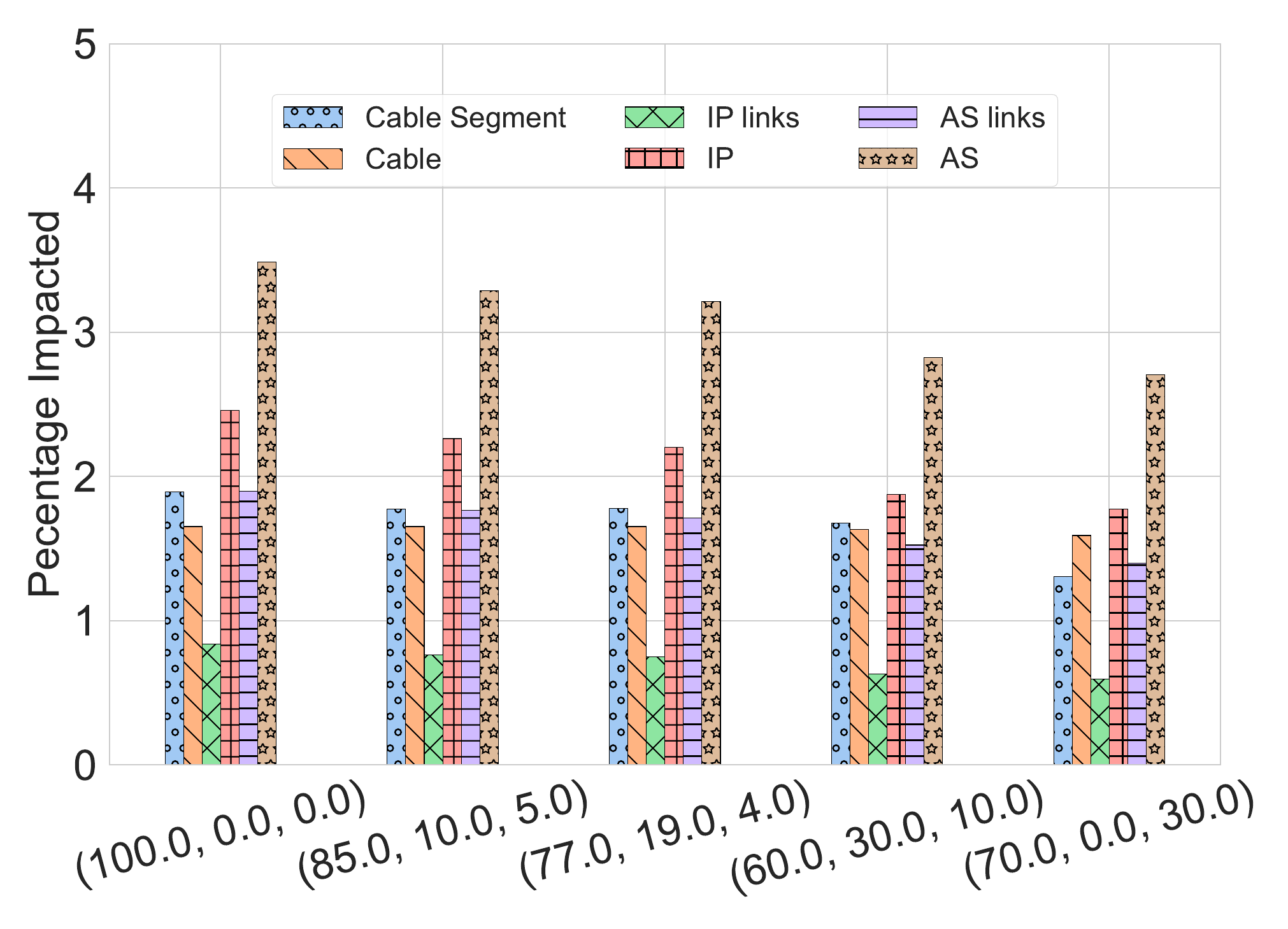}
        \caption{Sensitivity Analysis for the Pacific Northwest due to earthquakes}
        \label{fig:pnw_sensitivity}
    \end{subfigure}
    \hfill
    \begin{subfigure}{0.45\textwidth}
        \centering
        \includegraphics[width=\textwidth]{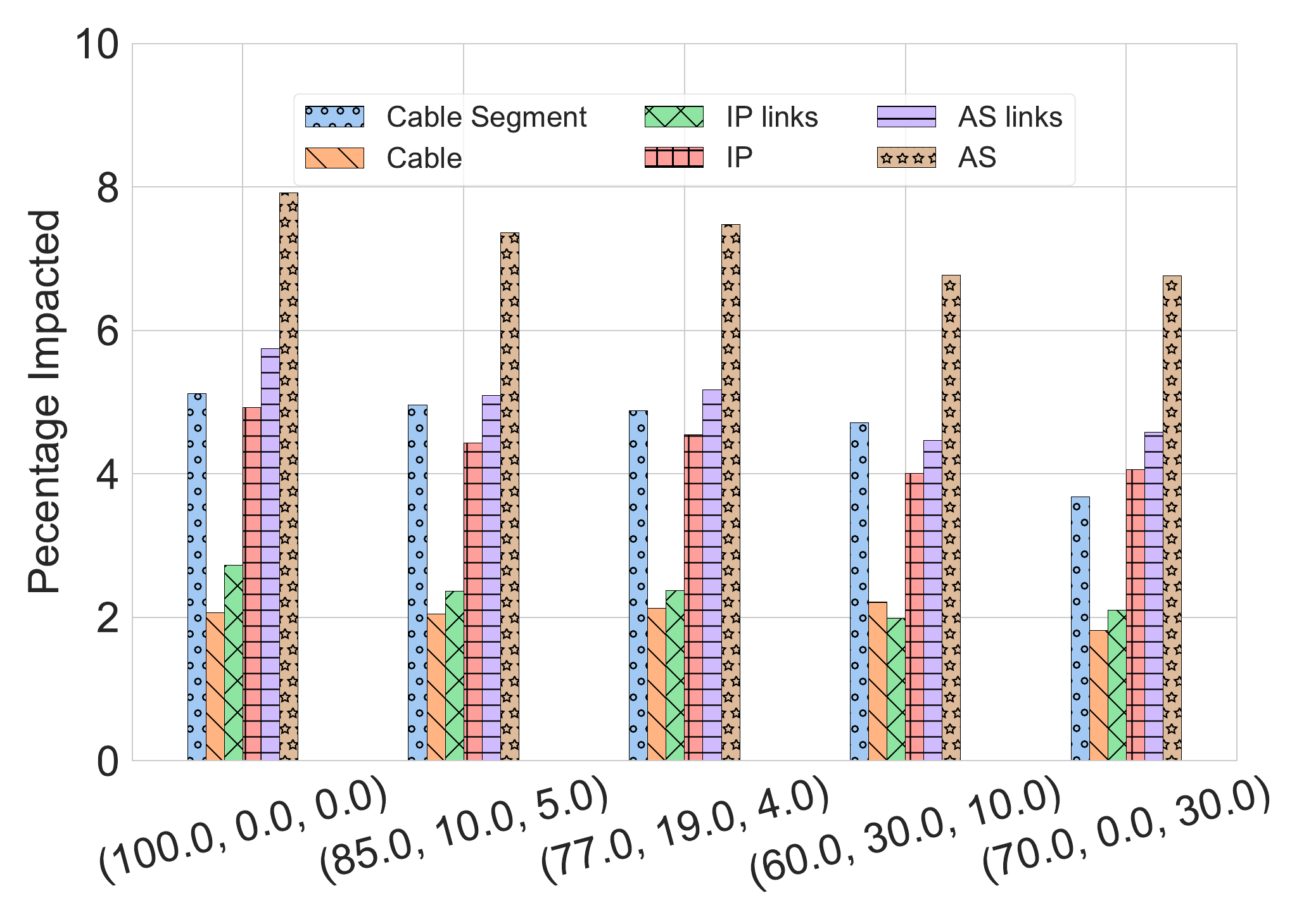}
        \caption{Sensitivity Analysis for Indonesia due to earthquakes}
    \end{subfigure}
    \begin{subfigure}{0.45\textwidth}
        \centering
        \includegraphics[width=\textwidth]{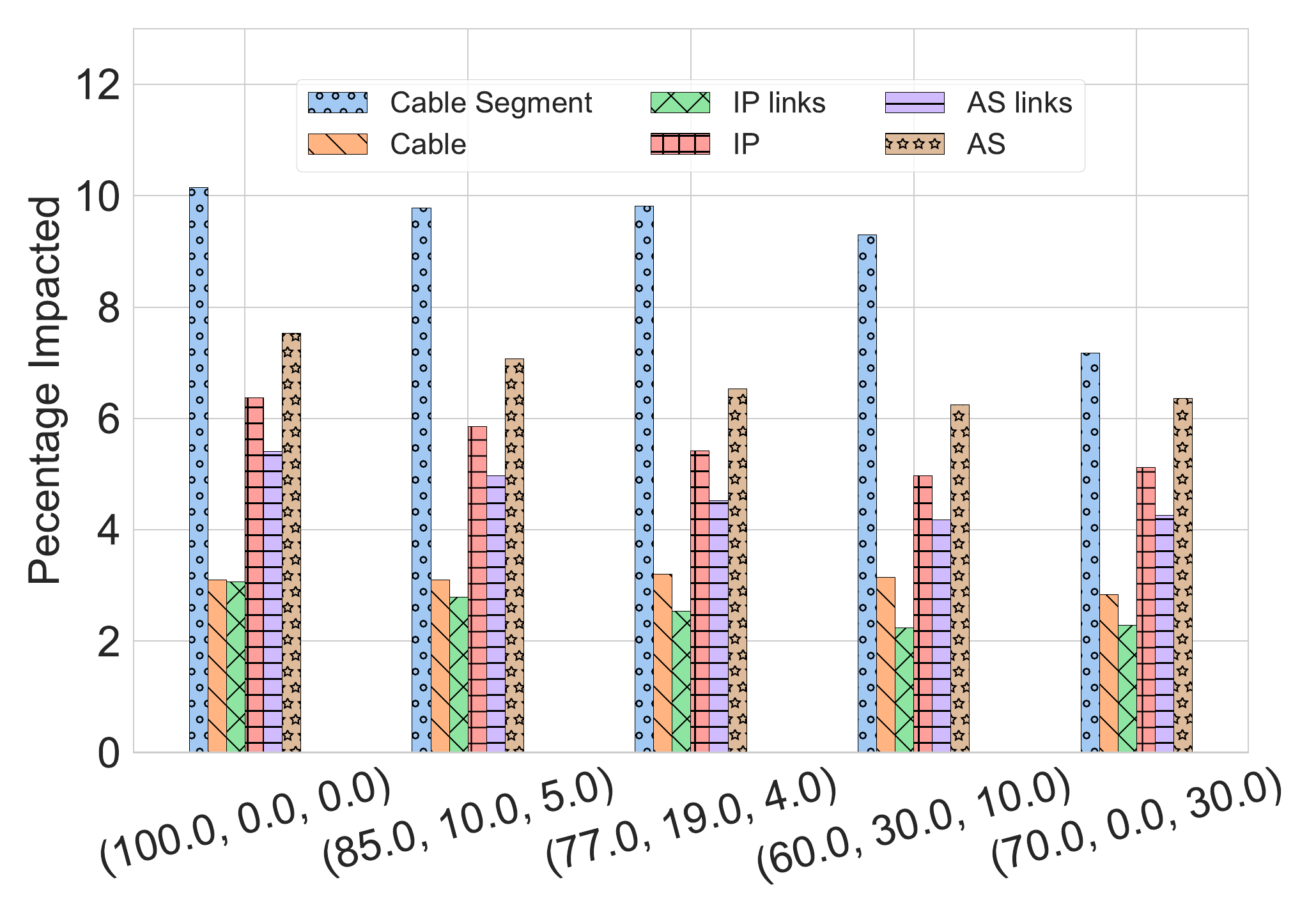}
        \caption{Sensitivity Analysis for Taiwan-Philippines due to earthquakes}
    \end{subfigure}
    \hfill
    \begin{subfigure}{0.45\textwidth}
        \centering
        \includegraphics[width=\textwidth]{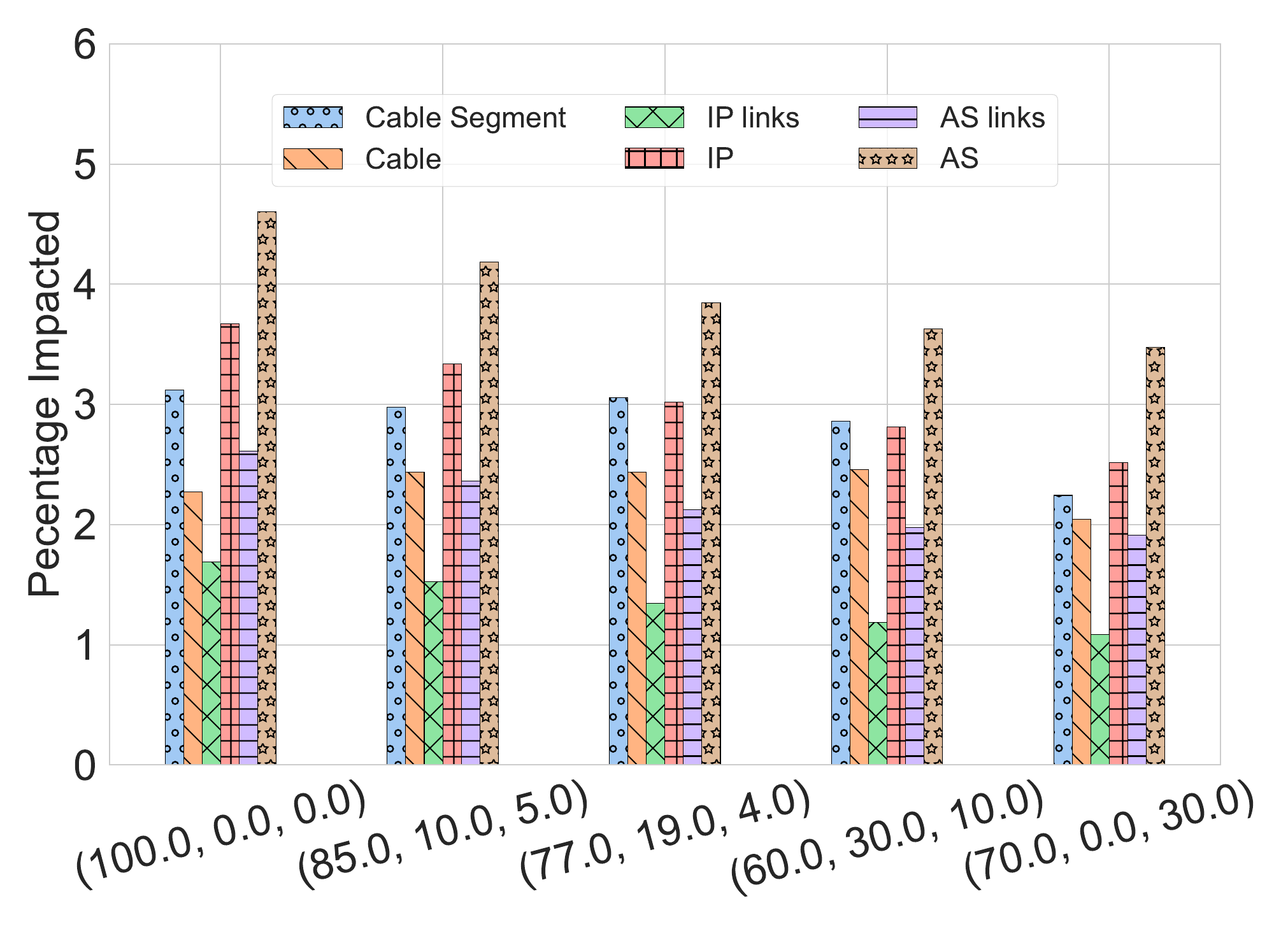}
        \caption{Sensitivity Analysis for Florida due to hurricanes}
    \end{subfigure}
    \begin{subfigure}{0.45\textwidth}
        \centering
        \includegraphics[width=\textwidth]{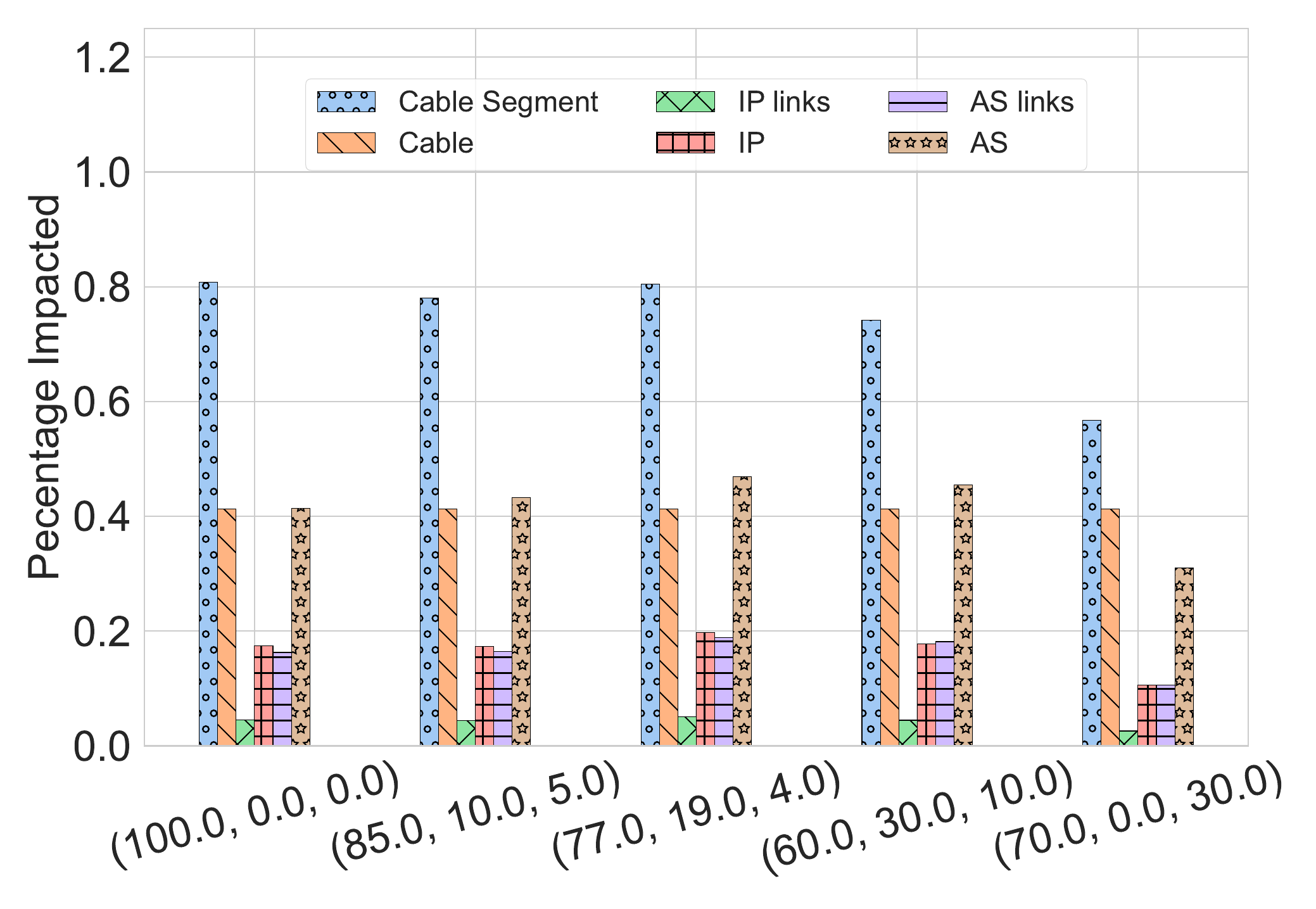}
        \caption{Sensitivity Analysis for Central America due to hurricanes}
    \end{subfigure}
    \hfill
    \begin{subfigure}{0.45\textwidth}
        \centering
        \includegraphics[width=\textwidth]{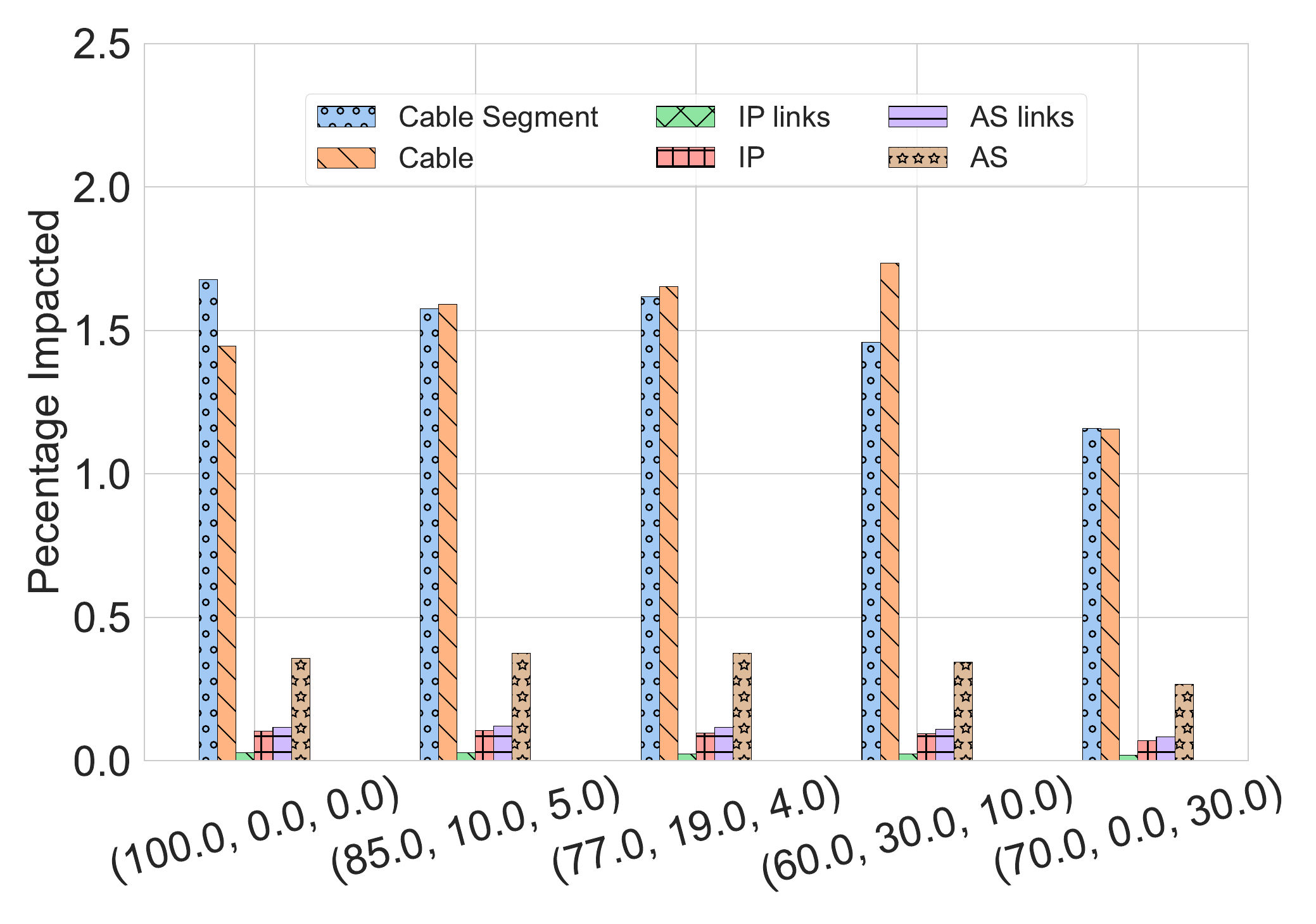}
        \caption{Sensitivity Analysis for the Caribbean due to hurricanes}
        \label{fig:crb_sensitivity}
    \end{subfigure}
    \caption{The maximum percentage of infrastructure at risk at varied levels of inaccuracies with Nautilus data. The horizontal axis depicts a 3-tuple: the percentage of IP links with correct top cable segment, correct non-top (secondary) cable segment, and incorrect cable segment predictions by Nautilus, respectively. Notably, significant Nautilus data inaccuracies do not proportionally reduce risk in Xaminer's analysis. (77, 19, 4) value corresponds to Nautilus' observed result in its validation experiment.}
    \label{fig:sensitivity_analysis_all}
\end{figure}

\begin{figure}[ht]
    \centering
    \begin{subfigure}{0.45\textwidth}
        \centering
        \includegraphics[width=\textwidth]{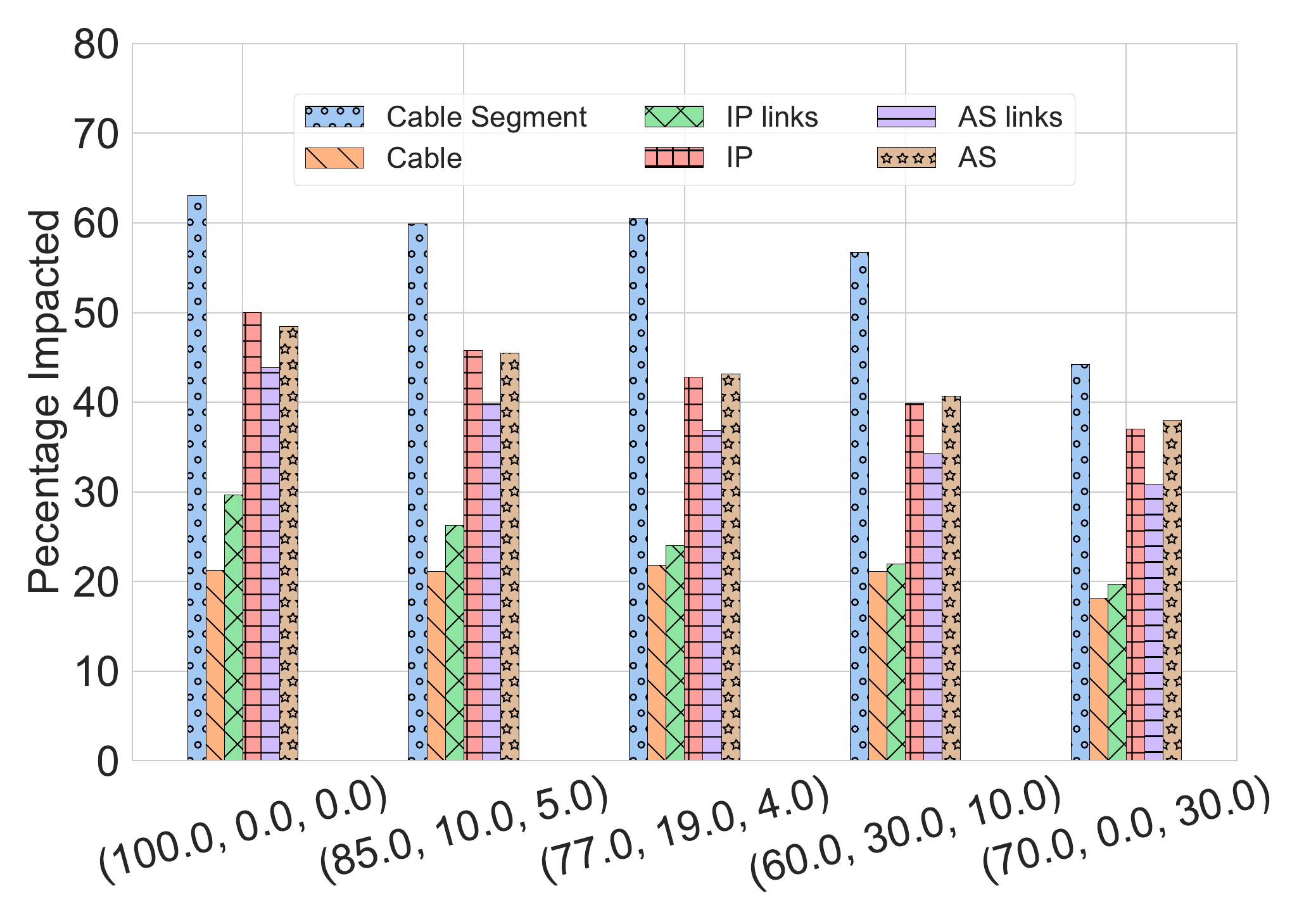}
        \caption{Sensitivity Analysis for Earthquakes (global)}
    \end{subfigure}
    \hfill
    \begin{subfigure}{0.45\textwidth}
        \centering
        \includegraphics[width=\textwidth]{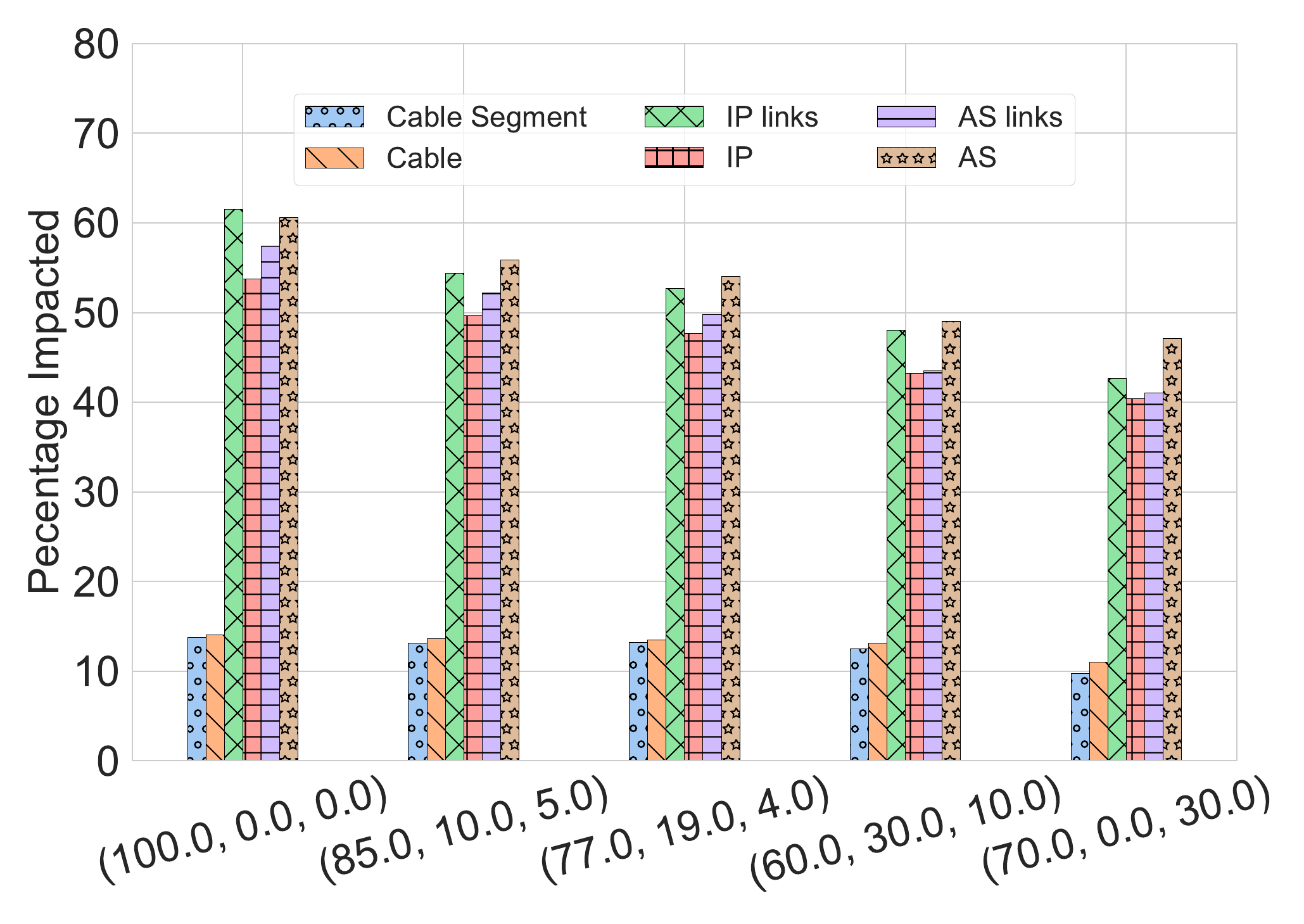}
        \caption{Sensitivity Analysis for Solar storms (global)}
    \end{subfigure}
    \caption{The maximum percentage of infrastructure at risk at varied levels of inaccuracies with Nautilus data globally. The horizontal axis depicts a 3-tuple: the percentage of IP links with correct top cable segment, correct non-top (secondary) cable segment, and incorrect cable segment predictions by Nautilus, respectively. Notably, significant Nautilus data inaccuracies do not proportionally reduce risk in Xaminer's analysis. (77, 19, 4) value corresponds to Nautilus' observed result in its validation experiment.}
    \label{fig:sensitivity_analysis_all_global}
\end{figure}

\subsection{Sensitivity Analysis} \label{sensitivity_analysis_appendix}

\ashwin{To assess the impact of errors in Nautilus mapping on Xaminer analyses, we conduct a sensitivity analysis, considering potential errors in top cable predictions. These errors encompass cases where Nautilus correctly predicts the cable segment but not as the top choice and instances where Nautilus fails to predict the correct cable segment. Simulating these errors, we randomly select a fraction of IP links in the impacted region and introduce errors. Results in Figures~\ref{fig:sensitivity_analysis_all} and~\ref{fig:sensitivity_analysis_all_global} averaged across 10 rounds indicate that, despite Nautilus inaccuracies, the trends remain similar. For example, in the Pacific Northwest region (Figure~\ref{fig:pnw_sensitivity}), the impact reduction due to earthquakes shows a limited change in affected components, even when a significant number of links are considered erroneous. It is worth highlighting that the impact due to secondary choice errors is lower than due to inaccurate prediction errors. Finally, the peculiar observation of higher risk despite higher inaccuracies at (60, 30, 10) configuration with the Caribbean region due to hurricanes (Figure~\ref{fig:crb_sensitivity}), can be attributed to the randomness associated with the secondary cable choice selection.}

\subsection{Correlation Trends} \label{corr_trends_appendix}

\ashwin{The countries comprising each cluster in Figure~\ref{country_cluster_corr} are listed in Table~\ref{tab:cluster_country}. In Figure~\ref{country_cluster_corr}, the shaded countries in grey belong to cluster 24, while those highlighted in yellow belong to cluster 4.}

\clearpage

\begin{table}
    \centering
    \small
    \begin{tabularx}{\linewidth}{c X}
    \hline
    \textbf{Cluster} & \textbf{List of Countries} \\
    \hline
    1 & 'Aruba', 'Anguilla', 'Bahamas', 'Canada', 'Cuba', 'Curaçao', 'Mexico', 'Montserrat', 'Puerto Rico', \\
      & 'Turks and Caicos Islands', 'United States', 'Venezuela, Bolivarian Republic of',  \\
      & 'Virgin Islands, British', 'Virgin Islands, U.S.' \\
    2 & 'Antigua and Barbuda', 'Barbados', 'Dominica', 'Guadeloupe', 'Saint Lucia',  \\
      & 'Saint Martin (French part)', 'Martinique', 'Saint Vincent and the Grenadines' \\
    3 & 'Australia', 'Nauru', 'French Polynesia' \\
    4 & 'Argentina', 'American Samoa', 'Bonaire, Sint Eustatius and Saba', 'Bermuda', 'Brazil', 'Chile', \\
      & 'Colombia', 'Costa Rica', 'Cayman Islands', 'Dominican Republic', 'Spain',  \\
      & 'Micronesia, Federated States of', 'Grenada', 'French Guiana', 'Guam', 'Guyana', 'Honduras',  \\
      & 'Haiti', 'Jamaica', 'Mongolia',  'Northern Mariana Islands', 'Nicaragua', 'Panama',  \\
      & 'Paraguay', 'El Salvador', 'Suriname', 'Sint Maarten (Dutch part)', 'Trinidad and Tobago',  \\
      & 'Tanzania, United Republic of', 'Uruguay', 'Uzbekistan' \\
    5 & 'Austria', 'Czechia', 'Denmark', 'Hungary', 'Kyrgyzstan', 'Norway', 'Poland', 'Slovakia', 'Sweden' \\
    6 & 'Andorra', 'Switzerland', 'France', 'Italy' \\
    7 & 'Belgium', 'Germany', 'United Kingdom', 'Ireland', 'Netherlands' \\
    8 & 'Fiji', 'New Caledonia', 'Papua New Guinea', 'Solomon Islands', 'Tonga', 'Vanuatu' \\
    9 & 'Bulgaria', 'Cyprus', 'Georgia', 'Greece', 'Liechtenstein', 'Luxembourg', 'Morocco', 'Romania', 'Turkey' \\
    10 & 'Central African Republic', 'Cameroon', 'Cabo Verde', 'Gambia', 'Mauritania', 'Senegal', 'Sierra Leone' \\
    11 & 'Kenya', 'Lesotho', 'Mozambique', 'Mauritius', 'Réunion', 'Eswatini', 'Uganda' \\
    12 & 'Åland Islands', 'Belarus', 'Estonia', 'Finland', 'Lithuania', 'Latvia', 'Russian Federation', 'San Marino' \\
    13 & 'Bolivia, Plurinational State of', 'Ecuador', 'Saint Kitts and Nevis', 'Niue', 'New Zealand', 'Samoa' \\
    14 & 'Azerbaijan', 'China', 'Equatorial Guinea', 'Japan', 'Korea, Republic of', 'Philippines', 'Portugal', \\
      & 'Taiwan, Province of China' \\
    15 & 'United Arab Emirates', 'Bahrain', 'Iran, Islamic Republic of', 'Kuwait', 'Oman', 'Qatar' \\
    16 & 'Angola', 'Botswana', "Côte d'Ivoire", 'Congo, The Democratic Republic of the', 'Congo', \\
      & 'Falkland Islands (Malvinas)', 'Gabon', 'Guinea', 'Guinea-Bissau', 'Liberia', 'Mali', 'Namibia', \\
      & 'Rwanda', 'Sao Tome and Principe', 'South Africa', 'Zambia' \\
    17 & 'Afghanistan', 'Armenia', 'Djibouti', 'Algeria', 'Iraq', 'Lebanon', 'Monaco', 'Pakistan', \\
      & 'Saudi Arabia', 'Sudan', 'Somalia', 'Yemen' \\
    18 & 'Croatia', 'Peru', 'Holy See (Vatican City State)' \\
    19 & 'Albania', 'Bosnia and Herzegovina', 'Libya', 'Moldova, Republic of', 'North Macedonia', 'Malta', \\
      & 'Montenegro', 'Serbia', 'Slovenia', 'Tunisia' \\
    20 & 'Bangladesh', 'Bhutan', 'Indonesia', 'India', 'Cambodia', "Lao People's Democratic Republic", \\
      & 'Sri Lanka', 'Maldives', 'Myanmar', 'Malaysia', 'Nepal', 'Singapore', 'Thailand', 'Timor-Leste', \\
      & 'Viet Nam', 'Wallis and Futuna' \\
    21 & 'Benin', 'Burkina Faso', 'Guernsey', 'Ghana', 'Gibraltar', 'Isle of Man', 'Jersey', 'Madagascar', \\
      & 'Malawi', 'Niger', 'Nigeria', 'Togo', 'Zimbabwe' \\
    22 & 'Egypt', 'Israel', 'Jordan', 'Palestine, State of', 'Chad' \\
    23 & 'Hong Kong', 'Kazakhstan', 'Macao' \\
    24 & 'Antarctica', 'French Southern Territories', 'Burundi', 'Saint Barthélemy', 'Belize',  \\
      & 'Brunei Darussalam', 'Bouvet Island', 'Cocos (Keeling) Islands', 'Cook Islands', 'Comoros',  \\
      & 'Christmas Island', 'Eritrea', 'Western Sahara', 'Ethiopia', 'Faroe Islands', 'Greenland', \\
      & 'Guatemala', 'Heard Island and McDonald Islands', 'British Indian Ocean Territory', 'Iceland',   \\
      & 'Kiribati', 'Marshall Islands', 'Mayotte', 'Norfolk Island', 'Pitcairn', 'Palau',   \\
      & "Korea, Democratic People's Republic of", 'South Georgia and the South Sandwich Islands',   \\
      & 'Saint Helena, Ascension and Tristan da Cunha', 'Svalbard and Jan Mayen',  \\
      & 'Saint Pierre and Miquelon', 'South Sudan', 'Seychelles', 'Syrian Arab Republic', 'Tajikistan', 'Tokelau', 'Turkmenistan', \\
    \hline
\end{tabularx}
\caption{Clusters with the associated list of countries}
\label{tab:cluster_country}
\end{table}

\end{document}